\documentclass[aps,pre,twocolumn]{revtex4}
\usepackage{epsfig}
\usepackage{citesort}
\usepackage{amssymb}

\begin{document}

\newcommand{\vn}[1]{\mbox{\boldmath$#1$}}
\newcommand{\bsigma}{\vn{\sigma}}
\newcommand{\bvarepsilon}{\vn{\epsilon}}
\newcommand{\bepsilon}{\vn{\varepsilon}}
\newcommand{\bkappa}{\vn{\kappa}}
\newcommand{\tr}{{\rm tr}}
\newcommand{\dev}{{\rm dev}}
\newcommand{\grad}{\rm grad\,}
\renewcommand{\div}{\rm div\,}
\renewcommand{\matrix}[1]{{\bf{#1}}}

\title{A discrete model for long time sintering}

\author{S.~Luding$^{1,2}$, 
        Karsten Manetsberger$^{3}$, Johannes M\"ullers$^{4}$} 
\affiliation{
        (1) Institut f\"ur Computeranwendungen 1, Universit\"at Stuttgart \\
            Pfaffenwaldring 27, 70569 Stuttgart, Germany\\
        (2) Particle Technology, DelftChemTech, TU Delft,\\
            Julianalaan 136, 2628 BL Delft, The Netherlands\\
        (3) Daimler Chrysler,  Messtechnik und Modelltechnik, \\
            Postfach 2360, 89013 Ulm, Germany\\
        (4) Daimler Chrysler, Forschung und Technologie, \\
            Simulationstechnik/Theorie, 88039 Friedrichshafen, Germany\\
}

\begin{abstract}
A discrete model for the sintering of polydisperse, inhomogeneous arrays of cylinders
is presented with empirical contact force-laws, taking into account plastic deformations,
cohesion, temperature dependence (melting), and long-time effects.
Samples are prepared under constant isotropic load, and are sintered for different
sintering times.  Increasing both external load and sintering time leads to a 
stronger, stiffer sample after cooling down. The material behavior is 
interpreted from both microscopic and macroscopic points of view. 

Among the interesting results is the observation, that the coordination 
number, even though it has the tendency to increase, sometimes slightly 
decreases, whereas the density continuously increases during
sintering -- this is interpreted as an indicator of reorganization effects
in the packing.
Another result of this study is the finding, that 
strongly attractive contacts occur during cool-down of the sample and
leave a sintered block of material with almost equally strong 
attractive and repulsive contact forces.
\end{abstract}

\maketitle

\tableofcontents

\section{Introduction}

  In mechanics and physics exist two ways to describe and model a 
  particulate, inhomogeneous material like powder, ceramics or concrete.
  The first approach is based on continuum theory and relies on 
  empirical assumptions about the macroscopic material behavior
  \cite{schwedes68,vermeer01}. For a summary of recent progress
  on the macroscopic modeling of sintering processes see Refs.\ 
  \cite{jagota90b,jagota95,olevsky98,bellehumeur98,zachariah99,riedel01,manetsberger01} 
  and the references therein.
  This ansatz is complemented by a ``microscopic'' description
  of the material on the particle or grain level where the particles
  and their interactions are modeled one by one 
  \cite{herrmann98,delo99,vermeer01}.
  The former involves stress, strain and plastic yield conditions
  \cite{redanz01}, whereas the latter deals with local force-deformation 
  laws for each contact \cite{jagota90}.  The macroscopic approach neglects the
  microstructure due to its nature and often isotropy is assumed 
  \cite{jagota90b}. 
  In contrast to network models or models with fixed topology 
  \cite{delo99,heyliger01,redanz01} we will follow the path of microscopic, 
  dynamic modeling, where no assumptions about either geometry, topology,
  homogeneoity, or isotropy of the powder packing are involved.  
  However, we have to assume certain contact force-laws and, furthermore, 
  restrict ourselves to two dimensions.
  
  In order to test the approach, a model system in a two-dimensional 
  box filled with cohesive-frictional disks of different sizes, see Sec.\
  \ref{sec:model}, is examined by means of a ``microscopic'' discrete element 
  method (DEM).  
  The microscopic interaction model for plastic deformations, friction,
  and cohesion is discussed and time-, temperature-, and history-dependent 
  behavior is introduced and rationalized in Sec.\ \ref{sec:DEMmodel}.  
  The long time sintering of a block of material is simulated 
  and the results are discussed in Sec.\ \ref{sec:results}.

\section{Model System}
\label{sec:model}

\begin{figure}[htb]
  \begin{center}
   \epsfig{file=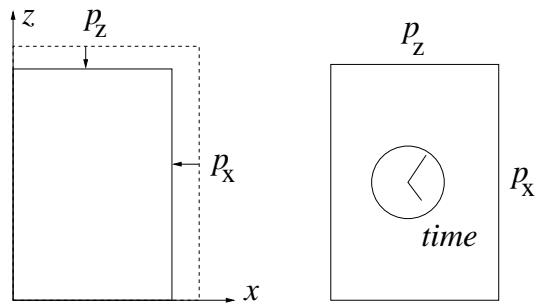,width=7.0cm}
  \end{center}
  \caption{(Left) Preparation of the sample with pressure on both sides.
           (Right) After the system is relaxed the evolution in time of the 
           sample with time is examined.}
  \label{fig:schemes}
\end{figure}

One possibility to obtain information about the material behavior 
is to perform elementary tests in the laboratory.  However, because
it is difficult to observe what is going on inside the material,
an alternative way is the simulations with the discrete element model (DEM) 
\cite{cundall79,bashir91,baars96,thornton98b,thornton98c,herrmann98,vermeer01}.
The numerical ``experiment'' chosen is a bi-axial box set-up, 
see Fig.\ \ref{fig:schemes},
where the left and bottom walls are fixed, and a stress- or strain-controlled
deformation is applied to the other walls.  In the future, quantitative 
agreement between experiment and simulation has to be achieved, however,
this issue is far from the scope of this paper.

Stress control means that the wall is subject to a predefined external
pressure that, in equilibrium, is cancelled by the stress, which the material
exerts on the wall.  In a typical ``experiment'', the particles are rapidly 
compressed with constant, isotropic pressure and then left alone for some
time and at constant temperature, typically close below the melting point.  

The stress-controlled motion of the walls is described by the 
differential equation
\begin{equation}
m_{\rm w} \ddot x(t) = F_{\rm w}(t)-p_{\rm w} z(t) - \gamma_{\rm w} \dot x(t) ~,
\label{eq:mxt}
\end{equation}
where $m_{\rm w}$ is the mass of the wall.  Values of $m_{\rm w}$ large
as compared to the sample mass lead to slow adaption and vibrations, whereas
small values allow for a rapid adaption to the actual
situation.  Three forces are active: (i) the force $F_{\rm w}(t)$ due
to the bulk material, (ii) the force $-p_{\rm w}z(t)$ due to the external
pressure, and (iii) a viscous force which damps the motion
of the wall so that oscillations are reduced.  

Inside the system, $N$ disks with radii $a_i$ ($i=1$, $\ldots$, $N$) 
and height $h$ are placed.  The radii are drawn from a homogeneous 
distribution with mean $a_0$ and relative width $w_0$ so that 
$a_i/a_0 \in [1-w_0, 1+w_0]$.  The particle-particle interactions
and the parameters involved are discussed in the next section.

\section{Discrete Particle Model}
\label{sec:DEMmodel}

The elementary units of granular materials are mesoscopic grains
which deform under stress, possibly yield, change their properties 
with time, and can behave different for different temperatures.  
Since the realistic modeling 
of the deformations of the particles is much too complicated to allow
for a subsequent many-particle simulation,
we relate the interaction force to the overlap $\delta$ of two
particles, see Fig.~\ref{fig:force}. In the absence of long-range
forces, an interaction takes place only if particles are in contact and
thus $\delta>0$.  In that case, the forces are split into a normal and a
tangential component denoted by $n$ and $t$, respectively.

If all forces $\vn f_i$ acting on the particle $i$, either from other
particles, from boundaries or from external forces, are known, the
problem is reduced to the integration of Newton's equations of motion
for the translational and rotational degrees of freedom
\begin{equation}
m_i \frac{{\rm d}^2}{{\rm d} t^2} \vn r_i = \vn f_i ~, 
{\rm ~~~ and ~~~}
I_i \frac{{\rm d}^2}{{\rm d} t^2} \vn \varphi_i = \vn t_i ~,
\end{equation}
with the mass $m_i$ of particle $i$, its position $\vn r_i$
the total force $\vn f_i = \sum_c \vn f^c_i$\,,
its moment of inertia $I_i$, its angular velocity $\vn \omega_i =
{\rm d}\vn \varphi_i/{\rm d}t$, the total torque
$\vn t_i = \sum_c \vn l^c_i \times \vn f^c_i$, and
the center-contact vector $\vn l^c_i$.
The integration of the equations of motion is performed with a standard 
molecular dynamics Verlet algorithm together with Verlet-table neighborhood 
search \cite{allen87}.

\subsection{Normal Contact Model}
\label{sec:forcenormal}

Two particles $i$ and $j$ at positions $\vn{r}_i$ and $\vn{r}_j$, with
radii $a_i$ and $a_j$, interact only if they are in contact so that 
their overlap 
\begin{equation}
\delta = (a_i + a_j) - (\vn r_i - \vn r_j) \cdot \vn n
\end{equation}
is positive, $\delta > 0$, with the unit vector 
$\vn n = \vn n_{ij}= (\vn r_i - \vn r_j) / |\vn r_i - \vn r_j|$ 
pointing from $j$ to $i$.  The force on particle $i$, from 
particle $j$ can be written as $\vn{f}_{ij} = f^n_{ij} \vn{n}
+  f^t_{ij} \vn{t}$, with $\vn n$ perpendicular to $\vn t$.  
In this subsection, the normal forces are discussed.

\subsubsection{Short time contact model}

\begin{figure}[tbh]
  \begin{center}
    \hfill
    \epsfig{file=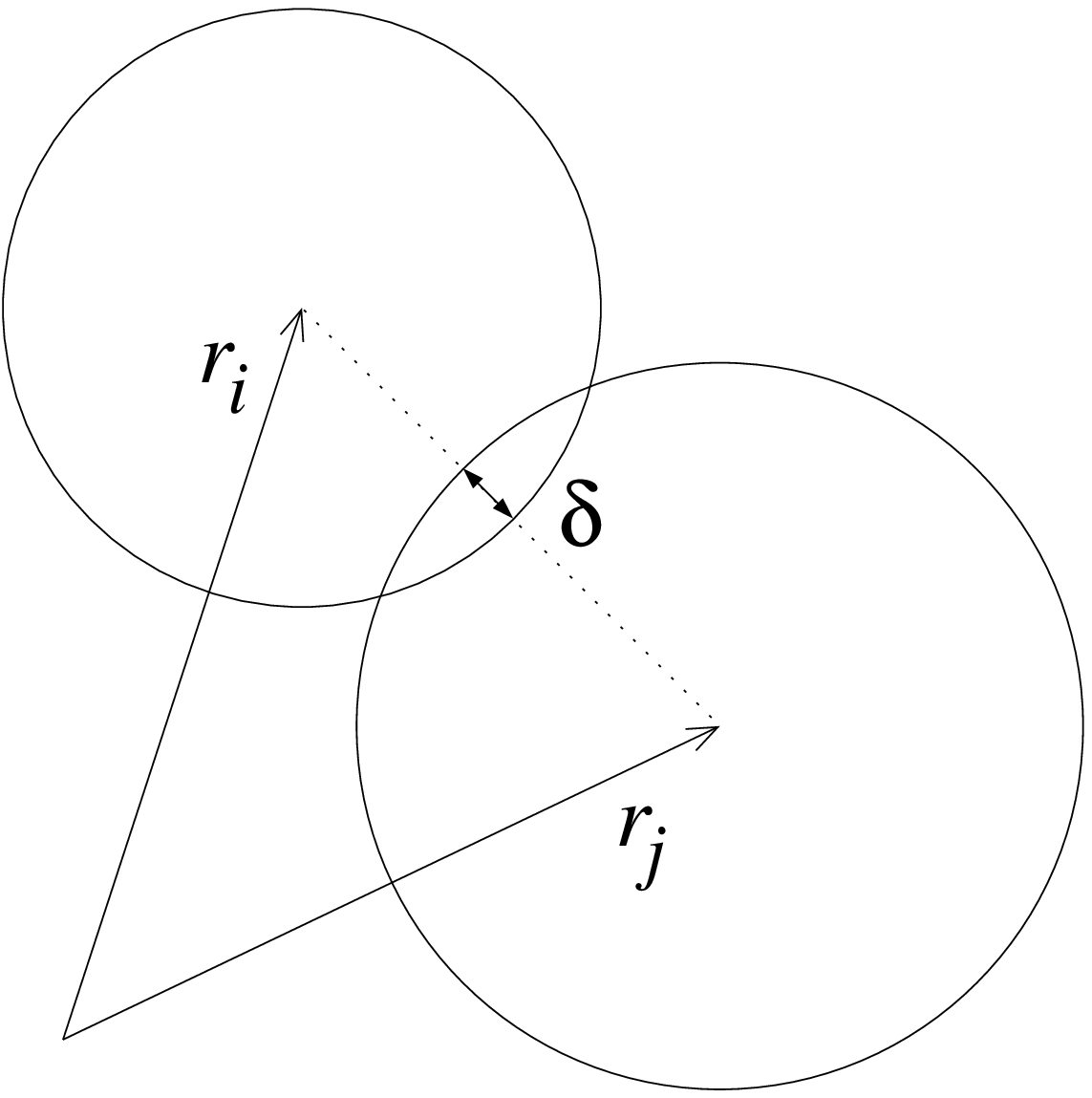,width=2.9cm} \hfill
    \epsfig{file=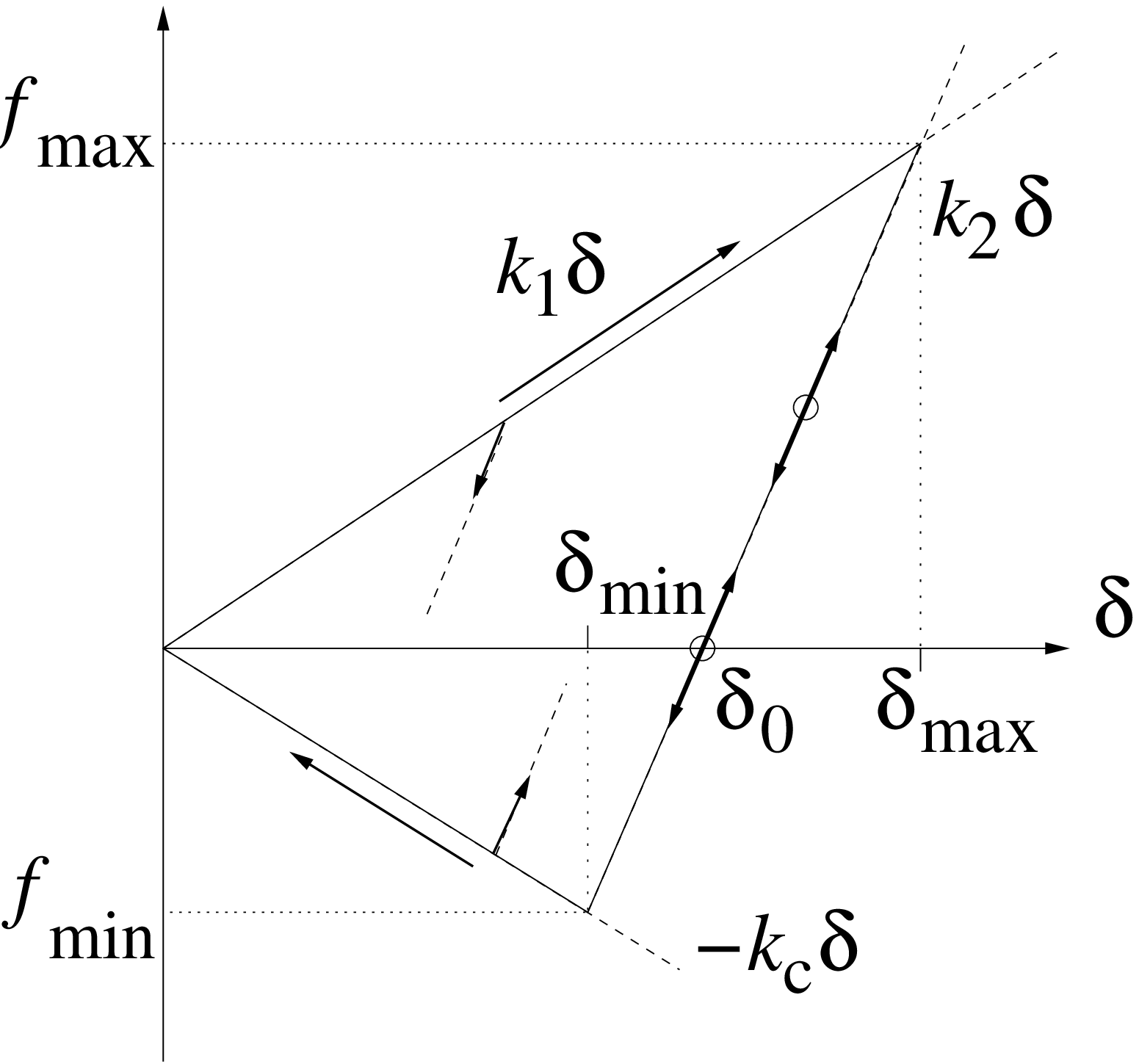,width=5.6cm}
    \hfill
  \end{center}
  \caption{(Left) Two particle contact with overlap $\delta$.
           (Right) Force law for two springs
           with stiffness $k_1$ and $k_2$ for initial loading and 
           subsequent un/reloading, respectively.  Attractive
           forces are possible due to the cohesion strength $k_c$. }
  \label{fig:force}
\end{figure}

As first step, we discuss the time- and temperature-independent
behavior of the contact forces between a pair of particles.
For this, we modify and extend the linear hysteretic spring model 
\cite{walton86,luding98c,tomas00}.  It is the simplest version 
of some more complicated nonlinear-hysteretic force laws 
\cite{walton86,zhu91,sadd93}, which reflects the fact that
at the contact point, plastic (permanent) deformations may take place.
The repulsive (hysteretic) force can be written as
\begin{equation}
f_{ij} = \left \{ 
\begin{array}{lll}
k_1  \delta           & {\rm loading,~}       \\
k_2 (\delta-\delta_0) & {\rm un/reloading,~}  \\
-k_c \delta           & {\rm unloading,~}     \\
\end{array}
\right .
\label{eq:fhys}
\end{equation}
with $k_1 \le k_2$, see Fig.\ \ref{fig:force}.

During the initial loading the force increases linearly with
the overlap $\delta$, until the maximum overlap $\delta_{\rm max}$ 
is reached, which is kept in memory as a {\em history parameter}. 
The line with slope $k_1$ thus defines the maximum force possible
for a given $\delta$. During unloading the force drops,
on the line with slope $k_2$, from its value
at $\delta_{\rm max}$ down to zero at the force-free overlap 
$$\delta_0=(1-k_1/k_2)\delta_{\rm max} ~.$$ 
Reloading at any instant leads to an increase of the force along this
line, until the maximum force is reached; for still increasing $\delta$,
the force follows again the line with slope $k_1$ and $\delta_{\rm max}$
has to be adjusted accordingly.  
Unloading below $\delta_0$ leads to negative, attractive forces
until at the overlap
$$\delta_{\rm min}=\frac{k_2-k_1}{k_2+k_c}\delta_{\rm max} ~,$$
the minimum force 
$f_{\rm min} = -k_c \delta_{\rm min}$, 
i.e.~the maximum attractive 
force, is obtained as a function of the model parameters $k_1$, $k_2$, 
$k_c$, and the history parameter $\delta_{\rm max}$.  
Further unloading leads to attractive forces on the branch $f=-k_c \delta$.
\footnote{
The highest possible attractive force, for given $k_1$ and $k_2$,
is reached for $k_c \rightarrow \infty$, so that 
$f_{\rm min} = - (k_2-k_1) \delta_{\rm max}$. 
This would lead to a discontinuity at $\delta=0$ that is avoided 
by using finite $k_c$.}

The cone formed by the lines with slope $k_1$ and $-k_c$ defines 
the range of possible force values. 
If a force would fall outside the cone, it is forced to remain
on the limit lines.  Departure from these lines into the cone takes 
place in the case of unloading and reloading, respectively.  
Between these two extremes, unloading and reloading follow the same 
line with slope $k_2$. Possible equilibrium
states are indicated as circles in Fig.\ \ref{fig:force},
where the upper and lower circle
correspond to a pre-stressed and stress-free state, respectively.

\subsubsection{Viscous dissipation}

In the case of collisions of particles and for large deformations,
dissipation takes place due to the hysteretic nature of the force-law.
For small displacements around some equilibrium state, the
model does not contain dissipation. Therefore, in order to allow for 
stronger dissipation and thus faster relaxation, a viscous, velocity dependent 
dissipative force in normal direction,
\begin{equation}
f^{n{\rm ,d}}_{ij} = \gamma_0 \dot \delta ~,
\label{eq:fdiss0}
\end{equation}
is assumed with some damping coefficient $\gamma_0$.  
The half-period of a vibration around the equilibrium position, see 
Fig.\ \ref{fig:force}, can be computed for arbitrary values of
$k_1$ and $k_c$, as long as the overlap fulfills the condition
$\delta_{\rm min} < \delta < \delta_{\rm max}$. In this case, $k_2$ 
determines the stiffness and one obtains
a typical response time on the contact level \cite{luding97c},
\begin{equation}
t_c = {\frac{\pi}{\omega}} ~, {\rm ~~~with~~~}
\omega=\sqrt{\frac{k_2}{m_{12}}-\eta_0^{2}} ~,
\label{eq:tc12}
\end{equation}
the eigenfrequency of the contact, the rescaled damping coefficient
$\eta_0=\gamma_0/(2 m_{12})$, and the reduced mass $m_{12}=m_1 m_2 /
(m_1+m_2)$.
The time-step of the simulation $t_{\rm MD}$ has to be chosen such that 
$$t_{\rm MD} \approx t_c/50$$ for a proper integration of the equations
of motion,  so that we chose a typical time-step $t_{\rm MD}=\pi/(50 \omega)$
after the model parameters $k_2$ and $\gamma_0$ are specified
\footnote
{In this study, we will not discuss more advanced viscous forces 
\cite{svoboda94,riedel94}
(including time- and temperature dependencies) because this dissipative
force is mainly a means of dissipating surplus energy and not so much 
based on realistic assumptions about the behavior of the material
\cite{riedel94}. Also
in the limit of the fluid, a more realistic viscous force law would
be required, but since we will not reach this limit, we rather prefer
to keep the model as simple as possible.  Extensions of the present
model are always possible in the future.}.

\subsubsection{Stiffness increase with contact area}

In order to account for the fact that a larger contact surface leads to
a larger contact stiffness, the coefficient $k_2$ is made dependent on
the maximum overlap history parameter $\delta_{\rm max}$ (and thus on
the force-free overlap $\delta_0$), as long as the overlap is below
the threshold $\delta^{\rm fluid}$ that corresponds to the ``complete 
melting'' of the particles.  Complete melting is here the limit of an
incompressible liquid that is contained in the model, however, neither
discussed in detail nor verified for reasons of brevity.
 
{The overlap $\delta^{\rm fluid}$ corresponding to the stress-free fluid, 
is computed such that the volume fraction in the system equals unity.  
The volume fraction of a dense packing of rigid spheres is 
$$\nu_{\rm solid} = N \pi h a_0^2 / V_{\rm solid} = \pi /(2\sqrt{3}) ~.$$
If all the material would be melted, the volume fraction is
$$\nu_{\rm fluid} = N \pi h a_0^2 / V_{\rm fluid} = 1 ~,$$
which leads to the ratio of fluid and solid volumes
$$  V_{\rm fluid} / V_{\rm solid} 
  = a^2_{\rm fluid} / a^2_0 
  = \pi /(2\sqrt{3}) ~.$$
The minimum radius for the incompressible melt is 
$a_{\rm fluid} 
  = a_0-\delta^{\rm fluid} 
  = a_0 \sqrt{ V_{\rm fluid} / V_{\rm solid} }
  \approx 0.9523 \, a_0$ 
which corresponds to the maximum overlap 
$$\delta^{\rm fluid} \approx 0.0477 \, a_0 ~,$$
in two dimensions. In the following, we use $\delta^{\rm fluid}=0.20$
in order to magnify the range of possible non-fluid overlaps.
}

The stiffness is maximal in the fluid limit for 
$\delta_0 = \delta^{\rm fluid}$, which corresponds to 
$\delta_{\rm max} = \delta_{\rm max}^{\rm fluid} 
                  = k_2 \delta^{\rm fluid} / (k_2-k_1)$,
and varies between $k_1$ and $k_2$ for smaller overlaps, so that
\begin{equation}
k_2(\delta_{\rm max}) = \left \{ 
\begin{array}{lll}
k_2 & {\rm ~if~} \delta_{\rm max} \ge \delta_{\rm max}^{\rm fluid} \\
k_1 + (k_2-k_1)  \frac{\delta_{\rm max} }{ \delta_{\rm max}^{\rm fluid}}
    & {\rm ~if~} \delta_{\rm max} < \delta_{\rm max}^{\rm fluid} 
\end{array}
\label{eq:k2}
\right . ~.
\end{equation}
For large overlaps (in the fluid regime), the stiffness and the force
is thus only dependent on $k_2$, independent of $k_1$.  For smaller
overlaps both $k_1$ and $k_2$ affect the force together with the 
history of this contact.

\begin{figure}[htb]
  \begin{center}
    \epsfig{file=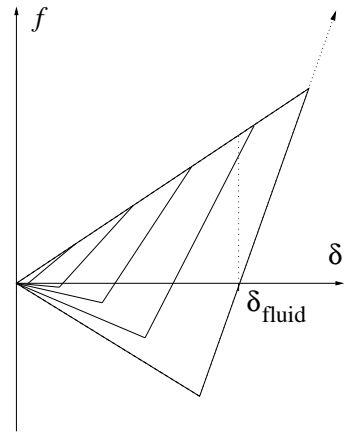,width=4.4cm}
  \end{center}
  \caption{Force law with varying stiffness $k_2(\delta_{\rm max})$, 
           according to Eq.\ (\protect\ref{eq:k2}). If $\delta_0$ becomes
           larger than $\delta^{\rm fluid}$, the stiffness remains
           equal to $k_2$ and the force remains on the corresponding
           liquid branch with slope $k_2$ (dotted line). The cohesion
           strength is maximal for the maximum contact strength
           and decreases with $k_2$, 
           see subsection\ \protect\ref{sec:cohesion} for details.}
  \label{fig:force2}
\end{figure}


The stiffness in the incompressible fluid should diverge, however,
for reasons of numerical stability, we have to limit the maximum stiffness
to $k_2$.  Larger stiffness values would require smaller time-steps
which would reduce the simulation efficiency.  Therefore, the model
does not really take the incompressibility of the liquid into account,
see Fig.\ \ref{fig:force2}.  In the fluid no plastic deformation
can take place so that $\delta_0$ is fixed and cannot be shifted to
larger force free overlaps -- in contrast to the hysteretic model described 
above. 

In summary, the hysteretic stiffness model takes into account
an increasing stiffness with increasing overlap.  The first loading
is plastic with low stiffness, and subsequent un- and reloading 
are stiffer because the material was initially compressed.  
As a consequence, also the maximum cohesive force
depends on the maximum compression which was experienced by the 
contact during its history. If the material is compressed so strong
that the liquid density is reached, the force-free overlap is equal
to the fluid equivalent overlap and the material behaves like a fluid.

\subsection{Density Temperature Dependence}

If a solid or a liquid (we assume a simple material here - not like water - 
where the density dependence on the temperature is continuous through the
phase transition) is heated, in general, its volume increases so that its
density decreases.
Therefore, we assume a temperature dependent density of the single
particles (disks with radius $a$ and height $h$):
\begin{equation}
\rho(T) = \frac{m}{\pi a^2 h} 
        = \rho(T_{\rm melt}) + \delta \rho_T \,(T_{\rm melt}-T) ~,
\end{equation}
with the density change per unit temperature $\delta \rho_T$.
This corresponds (in linear approximation) to a change of the 
particle radius 
\begin{equation}
a(T) = a(T_{\rm melt}) [1 - \delta a_T \,(T_{\rm melt}-T) ]~,
\label{eq:aT}
\end{equation}
with the relative change of the radius per unit temperature $\delta a_T$.  
This approximation can be used if the range of temperatures
is rather narrow and the changes per unit temperature are very small.

In the following, we use $\delta a_T = 10^{-4}$\,K$^{-1}$, so that
the particle radius is changed by 0.01 per-cent if the temperature is
changed by one Kelvin.  In the interesting range of temperatures
between a low temperature (80$^0$C) and some melting point (120$^0$C),
the radius changes by one per-cent, accordingly.

\subsection{Contact Temperature Dependence}

\begin{figure}[htb]
 \begin{center}
  \epsfig{file=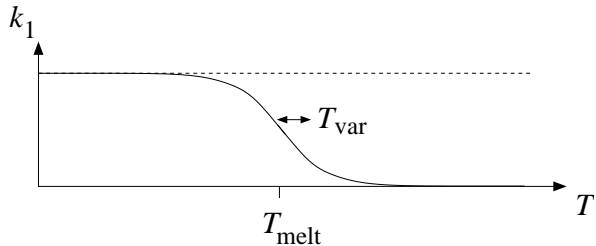,height=3.2cm}
 \end{center}
\caption{Schematic plot of the stiffness $k_1$ as a 
         function of the temperature.}
\label{fig:kTschem}
\end{figure}

For the temperature dependence, we focus on an inhomogeneous material with 
a melting temperature $T_{\rm melt}$ and assume that the material behaves 
static, as described above, if the temperature $T$ is much smaller than 
the melting temperature. The behavior of the stiffness $k_1$ is schematically 
shown in Fig.\ \ref{fig:kTschem} as a function of the temperature.

The material becomes softer when $T_{\rm melt}-T$ becomes
small and will lose all stiffness in the limit $T_{\rm melt}-T \ll 0$.
The temperature range in which the melting takes place is quantified by
$T_{\rm var}$.
In the transition regime
$|T_{\rm melt}-T| \approx T_{\rm var}$, the particles are significantly
softer than in the {\em cold} limit $T_{\rm melt}-T \gg T_{\rm var}$.  In the
{\em hot} and the liquid regime, $T - T_{\rm melt} \gg T_{\rm var}$, one has
$k_1 \rightarrow 0$ and the particles lose their nature,
however, the `incompressibility' is accounted for with a stiffness
$k_2$, as defined in Eq.\ (\ref{eq:k2}), and $\delta_0 = \delta^{\rm fluid}$
is fixed.  

\subsubsection{Increasing temperature}

When the temperature is increased to a rather large value, close to the
melting point, two particles under stress and in equilibrium due to 
compressive forces will lose stiffness and thus will deform stronger 
so that their overlap becomes larger.  Therefore, we assume for the 
stiffness coefficient
\begin{eqnarray}
k_{1}(T) 
  & = & \frac{k_1}{2}
    \left [ 1+ \tanh \left ( 
                       \frac{T_{\rm melt}-T}{T_{\rm var}} 
                    \right ) 
    \right ]   \nonumber \\
  & = & \frac{k_1}{2} \left [ 1 + \tanh (\tau) \right ] ~,
\label{eq:k1T}
\end{eqnarray}
with the dimensionless temperature difference $\tau$, and the typical range 
of considerable temperature dependency $T_{\rm var}$, where the temperature
has a strong effect on the stiffness of the particles, 
see Fig.\ \ref{fig:kTschem}.  If experimentally available in the future, 
the function $\tanh(x)$ can be replaced by any
other function $f(x)$ that decays from unity to zero at $x \approx 0$.  

When $k_1$ is reduced due to an increase in temperature ($+$), 
we assume that $\delta^+_{\rm max}$ remains constant,
so that one obtains a larger force-free overlap  
$\delta_0^+(T) = [1-k_1(T)/k_2(\delta^+_{\rm max})] \delta^+_{\rm max}$.
Thus the material volume shrinks due to sintering at the contact level.

\begin{figure}[htb]
  \begin{center}
    \epsfig{file=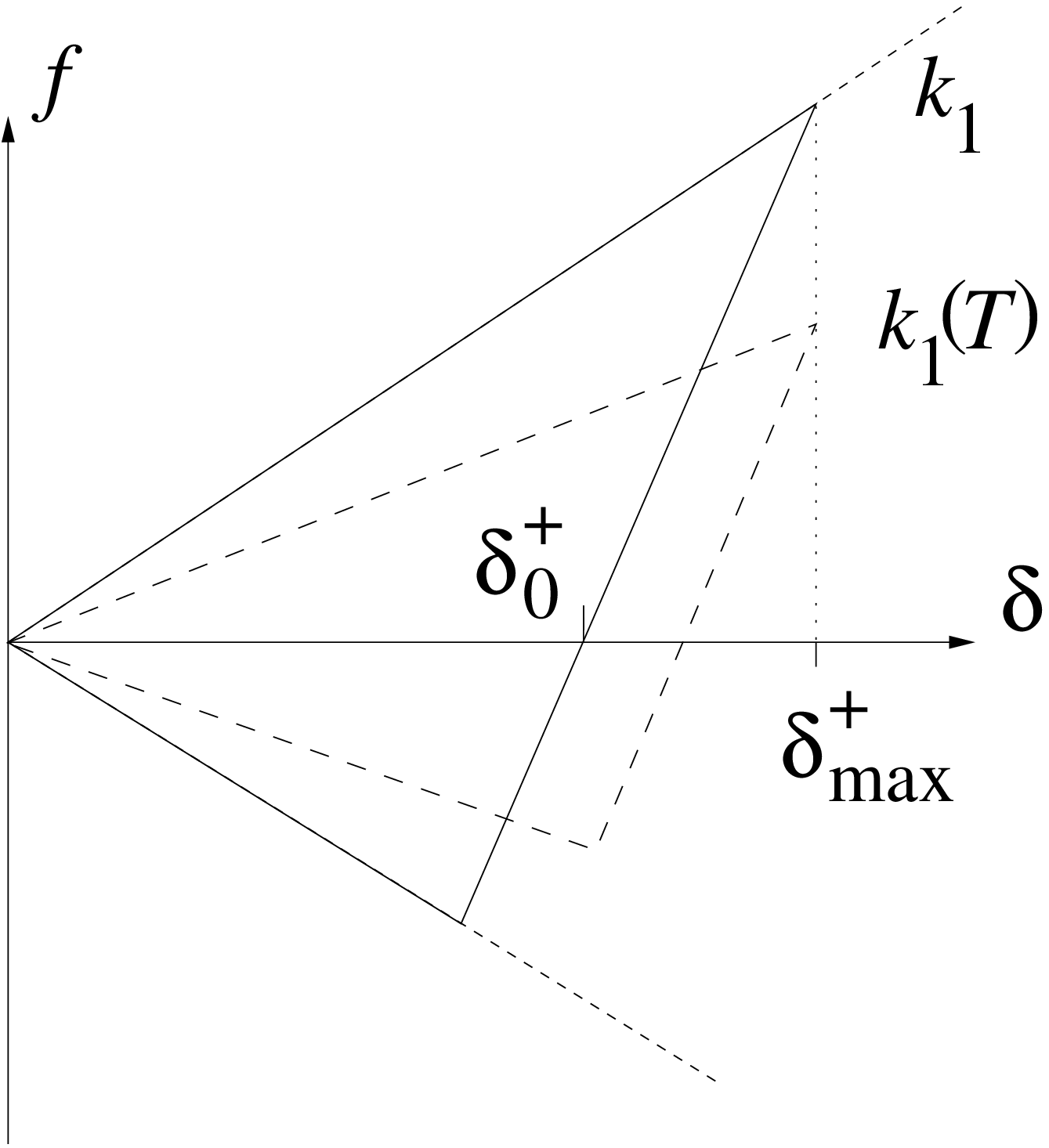,width=4.0cm} \hfill
    \epsfig{file=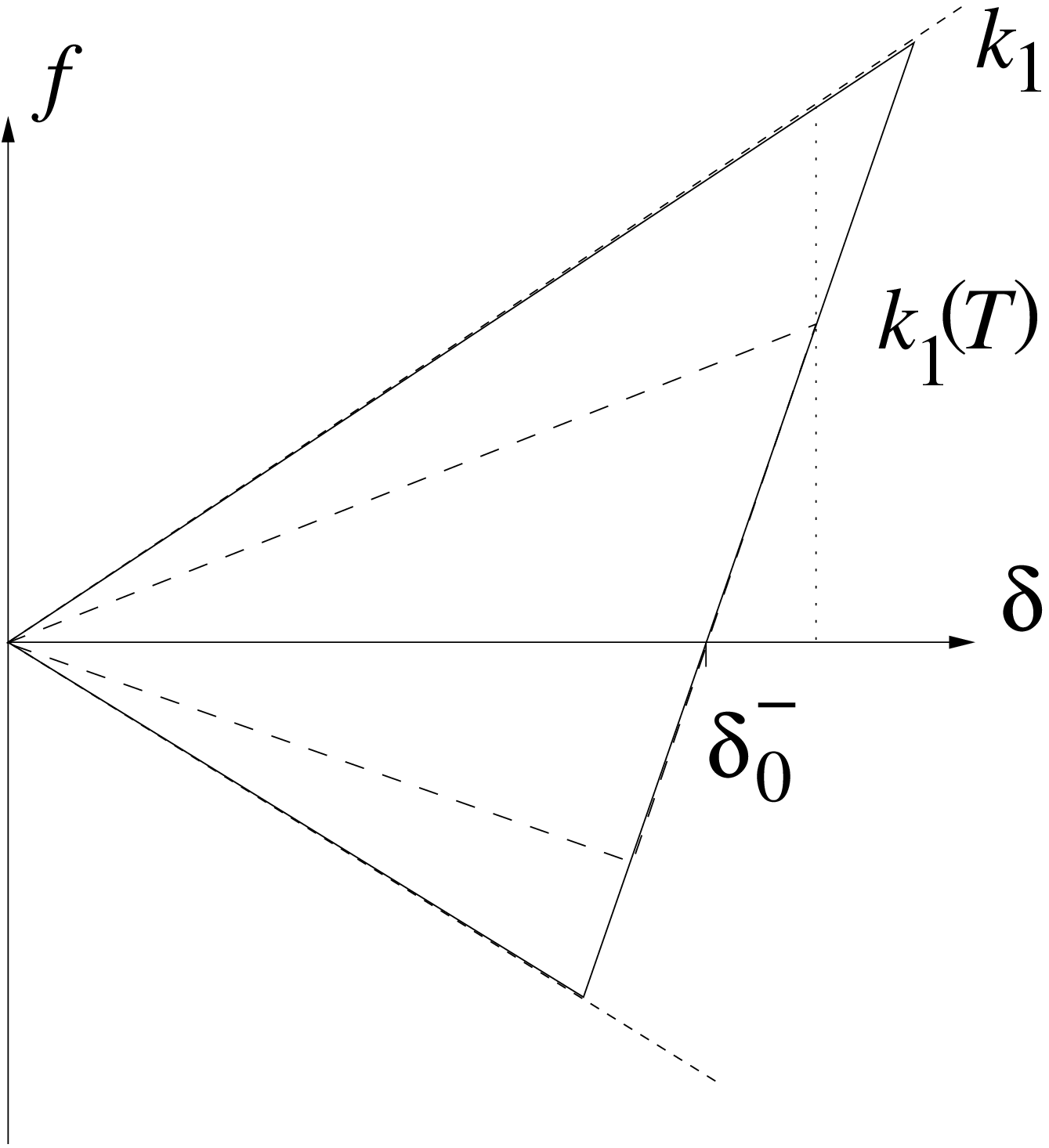,width=4.0cm}
  \end{center}
  \caption{Force laws for varying stiffness $k_1$, according to 
           Eq.\ (\protect\ref{eq:k1T}). 
           (Left) 
           If the temperature is increased, $k_1$ is reduced while 
           $\delta^+_{\rm max}$ remains constant (dashed line, stress-free 
           case). 
           (Right) 
           If the temperature is subsequently decreased, $k_1$ is increased
           while $\delta^-_0$ remains constant (solid line with slope $k_2$).
          }
  \label{fig:force3}
\end{figure}

Note that $k_2$ is not changed directly when $k_1(T)$ is decreased, 
see the left panel in Fig.\ \ref{fig:force3} or Eq.\ (\ref{eq:k2}).
The cohesion in this model, however, is directly affected by
a change of $k_1(T)$, see Eq.\ (\ref{eq:kc}).
In a pre-stressed situation, corresponding to a finite confining force 
at the contact, also $\delta_{\rm max}^+$ is shifted in order to
balance the confining force -- but only after
$k_1 \delta_{\rm max}^+$ became smaller than the confining force.

If, as an example, the material has a melting point
$T_{\rm melt}=120^0$C with a range of softening of $T_{\rm var}=10^0$C,
for a temperature of $T=118^0$C, the stiffness $k_1$ is reduced to 
$0.6$ of its cold-limit value, for a temperature of $122^0$C the
stiffness is $0.4$, and for a temperature of $130^0$C the
stiffness is only $0.12$. For a temperature of $T=160^0$C only
a stiffness of $3.10^{-4}$ remains.
The fact that there is a remaining stiffness
above the melting temperature can be attributed to the inhomogeneity
of the material, i.e.\ not all the material melts at the same temperature.

\subsubsection{Decreasing temperature}

If later in time the temperature is decreased again, $k_1(T)$ is
adjusted according to Eq.\ (\ref{eq:k1T}), but since the melted (sintered)
area around the contact point will not return to its previous state, 
we now assume that $\delta^-_{0}={\rm const.}$, so that the maximum 
overlap increases to the value 
$\delta^-_{\rm max}(T)=\delta^-_{0} / [1-k_1(T)/k_2]$, 
see the right panel in Fig.\ \ref{fig:force3}.
Therefore, a temperature cycle involving a temperature close to 
the melting temperature leads to a contact situation similar to 
the one obtained through some larger maximum compressive force. The
contact deformation/area and thus the force-free overlap become larger
due to the partial melting of the surface and also the stiffness is
increased accordingly.\\


The temperature dependence thus can lead to changes of the stiffness and 
increases the overlap (deformation) of the particles.  It is effective only 
if the temperature is changed -- the time dependence will be introduced
in the next subsection. 

\subsection{Temperature dependence with time}

In the material there are several time dependent processes taking place.
Since we are interested in the long time behavior of the material, we
assume that heat conduction and equilibration take place instantaneously,
as long as temperature changes are small and slow.
The realistic simulation of heat-conduction in the sample is far from
the scope of this study. Therefore, the particle size is adjusted
directly with the temperature according to Eq.\ (\ref{eq:aT}).

In addition to the direct effect of a temperature change on the particle
size and the stiffness, the material may change its internal, atomistic 
structure such that defects heal and disappear. 
This effect will occur mainly in the regime of high temperatures close 
to the melting point.  Therefore, in order to account for such slow 
microscopic processes, a time dependence is introduced that leads to 
a change of the material stiffness $k_1$ with time.  The change takes 
place extremely slowly with an algebraic time dependence, so that $k_1(T,t)$
lags behind when varying from its actual value to the desired, final value 
$k_1(T)$, as defined in the above equation (\ref{eq:k1T}). 

When the temperature is increased from a small value $T_0$ to $T$, 
then $k_1(T_0)$ changes to $k_1(T)$ following the law
\begin{equation}
k_1(T,t) 
   = k_1(T) \left [ 
              1 - \frac{1}{\frac{1}{1-k_1(T_0)/k_1(T)} - \frac{t}{t_0}}
            \right ] ~,
\label{eq:k1time}
\end{equation}
which corresponds to the rate of change
\begin{equation}
\frac{\partial k_1(T,t)}{\partial t} 
       = \pm \frac{[ k_1(T)-k_1(T,t) ]^2}{k_1(T) \, t_0} ~,
\label{eq:dk1time}
\end{equation}
with the time scale $t_0$ on which a typical change takes place. 
Note that $k_1$, $k_1(T)$, and $k_1(T,t)$ are different, in general,
and correspond to the maximum $k_1$, the temperature dependent
$k_1(T) < k_1$, and the time dependent $k_1(T,t)$ that tends towards
$k_1(T)$.  The sign in Eq.\ (\ref{eq:dk1time}) is chosen 
according to the sign of $[k_1(T)-k_1(T,t)]$ in Eq.\ (\ref{eq:k1time})
\footnote{\,In symbolic form this means: $\pm [y]^2 = y|y|$. 
For the numerical 
integration, we remark, that the value of the actual stiffness should be
$k_1(T,t-\Delta t)$, otherwise we obtained a numerically instable behavior 
in some situations}.

\begin{figure}[htb]
  \begin{center}
    \epsfig{file=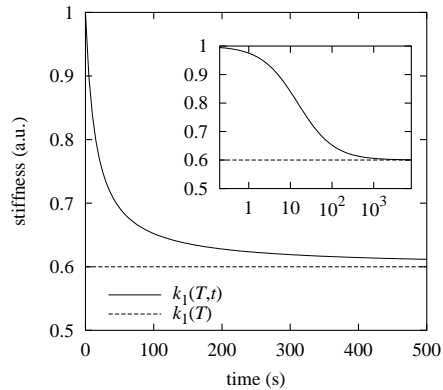,width=5.0cm,angle=-90}
  \end{center}
  \caption{Variation of the stiffness $k_1$ with time, see
           Eq.\ (\protect\ref{eq:k1time}). The inset shows the
           same function with a logarithmic time axis -- only after
           about $10^3$\,s the final stiffness is reached. }
  \label{fig:k1time}
\end{figure}

Assume $T_0=20^0$C and $T=118^0$C, so that $k_1(T_0)=1$
corresponds to $k_1(T)=0.6$.
The stiffness as a function of time is plotted in Fig.\ 
\ref{fig:k1time} for the time constant $t_0=10$s.

Note that, due to the factor $k_1(T)$ in the denominator of Eq.\ 
(\ref{eq:dk1time}), the change of stiffness is faster for higher
temperatures.  In the hot limit, changes take place very rapidly,
whereas in the cold limit changes are extremely slow.

Finally, we note that the adaption/relaxation of $k_1(T,t)$ to the
desired $k_1(T)$ value follows the above equations in all cases except when
the temperature is going down and the $k_1(T) > k_1(T,t)$.  In that
situation the contacts freeze rapidly and thus have to become strong
as fast as the system cools down.

\subsection{Cohesion dependence on stiffness and friction}
\label{sec:cohesion}

The cohesive properties of a particle contact depend on the 
temperature, in so far that a melted contact should have weak
tensile and compressive strength.  Therefore, we couple the
cohesive parameter $k_c$ to the magnitude of $k_1(T,t)$, see
Eq.\ (\ref{eq:k1time}), which 
decreases with temperature increasing. In addition,
in order to take into account a reduced tensile strength of a 
soft contact with weak deformation and thus small overlap, the 
cohesion is directly related to the stiffness
$k_2(\delta_{\rm max})$, see Eq.\ (\ref{eq:k2}). 
Thus, we propose
\begin{equation}
 k_c(T,t,\delta_{\rm max}) 
            = \frac{k_1(T,t)}{k_1}
              \frac{k_2(\delta_{\rm max})}{k_2} k_c ~.
\label{eq:kc}
\end{equation}
This is an arbitrary choice for the cohesive force factor $k_c$, 
but as long as no detailed experimental results are available, 
we stick to this empirical law, see Fig.\ \ref{fig:force2} for
a schematic picture.

\subsection{Tangential Contact Model}
\label{sec:forcetangential}

The force in tangential direction is implemented in the spirit
of Cundall and Strack \cite{cundall79} who introduced a tangential
spring in order to account for static friction.  Various authors
have used this idea and numerous variants were implemented, see
\cite{brendel98} for a summary and discussion.

Since we combine cohesion and friction and introduce time and temperature
dependencies, it is convenient to repeat
the model and define the implementation.  The tangential force is
coupled to the normal force via Coulomb's law, i.e.\ $f^t \le \mu f^n$,
where for the limit case one has sliding friction and for the case
of small forces $f^t$, one has static Coulomb friction.  The latter situation
requires an elastic spring in order to allow for a restoring force and
a non-zero remaining tangential force in static equilibrium due to 
activated Coulomb friction.

As a consequence of the cohesion force in normal direction,
attractive forces are possible so that Coulomb's law has to be modified
\begin{equation}
 f^t \le \mu (f^n - f_{\rm min}) ~,
\label{eq:ftC0}
\end{equation}
with the minimum (maximum attractive) force 
\begin{equation}
 f_{\rm min} 
               = -   \frac{k_2(\delta_{\rm max})-k_1(T,t)}
                          {1+k_2(\delta_{\rm max})/k_c(T,t,\delta_{\rm max})}
                     \delta_{\rm max} ~.
\end{equation}
The equal '$=$' in Eq.\ (\ref{eq:ftC0}) corresponds to (fully) activated,
sliding friction, while the smaller '$<$' corresponds to static friction.
Thus the tangential force is related to the normal force relative
to the point of cohesion-failure.  For normal forces larger than
$f_{\rm min}$, friction is always active and the amplitude is
proportional to $f^n - f_{\rm min}$ and $\mu$.  This can lead to 
a stable equilibrium of the solid also at $f^n \approx 0$ and activated
tangential static friction.  Note that $f_{\rm min}$ tends towards
zero for vanishing overlap, so that the difference from the original
model also vanishes for vanishing overlap.  

If $f^n - f_{\rm min}>0$, the tangential force is active, and we
project the tangential spring $\vn{\xi}$ into the actual tangential plane
by subtracting a possible (small) normal component.
\footnote{This is necessary, since the frame of reference of the contact
may have rotated since the last time-step}.
\begin{equation}
  \vn{\xi} = \vn{\xi} - \vn{n} ( \vn{n} \cdot \vn{\xi}) ~.
\label{eq:chimap}
\end{equation}
This action is relevant for an already existing spring, if the 
spring is new, the tangential force is zero anyway, however, the 
change of the spring-length defined below, is well defined.
Next, the tangential velocity is computed 
\begin{equation}
 \vn v_t = \vn v_{ij} - \vn{n} ( \vn{n} \cdot \vn{v}_{ij}) ~,
\end{equation}
with the total relative velocity 
\begin{equation}
\vn v_{ij} = \vn v_i - \vn v_j + a_i \vn n \times \vn \omega_i 
                                  + a_j \vn n \times \vn \omega_j  ~.
\end{equation}
In the next step we calculate the tangential test-force as the sum
of the tangential spring and a tangential viscous force (similar
to the normal viscous force)
\begin{equation}
 \vn f^{\rm o}_t = - k_t \, \vn \xi - \gamma_t \vn v_t ~, 
\label{eq:f_static}
\end{equation}
with the tangential spring stiffness $k_t = \alpha k_2(\delta_{\rm max})$, 
with a typical 
stiffness ratio $\alpha=0.2$, see \cite{luding97c}, and the tangential 
dissipation parameter $\gamma_t$.  If $|\vn f^{\rm o}_t| \le f_C$, with
$f_C = \mu (f^n-f_{\rm min})$, one has static friction and,
on the other hand, if $|\vn f^{\rm o}_t| > f_C$, sliding friction 
is active.  In the former case, the tangential test force is incremented
\begin{equation}
 \vn \xi' = \vn \xi + \vn v_t  \Delta t_{\rm MD} ~,
\end{equation}
to be used in the next iteration in Eq.\ (\ref{eq:chimap}),
and the force $\vn f^t=\vn f^{\rm o}_t$ from 
Eq.\ (\ref{eq:f_static}) is used.
In the latter case, the tangential spring is adjusted to a length
which is consistent with Coulomb's condition
\begin{equation}
 \vn \xi' = - \frac{1}{k_t} \left ( 
                f_C \, \vn t  \right ) ~,
\label{eq:xinew}
\end{equation}
with the tangential unit vector, $\vn t=\vn f^{\rm o}_t/|\vn f^{\rm o}_t|$, 
defined by the tangential spring, and the Coulomb force is used.
Inserting $\vn \xi'$ into Eq.\ (\ref{eq:f_static}) leads to 
$\vn f^t \approx f_C \vn t$ in the next 
iteration \footnote{
Note that $\vn f^{\rm o}_t$ and $\vn v_t$ are not necessarily
parallel in three dimensions. However, the mapping in Eq.\ (\ref{eq:xinew})
works always, rotating the new spring such that the direction of the
frictional force is unchanged}.
In short notation this reads
\begin{equation}
 \vn f^t = + {\rm min} \left ( f_C, |\vn f^{\rm o}_t| \right ) \vn t ~.
\end{equation}

Note that
the tangential force described above is identical to the classical
Cundall-Strack spring only in the limit $\gamma_t=0$ and $k_c=0$.
Besides the combination of the cohesive and the frictional
force, also the tangential dissipation is non-standard.  
Furthermore, we remark that the cohesion could also be coupled to
friction in the sense that a broken contact looses its tensile
strength when it is assumed brittle, so that $k_c=0, {\rm ~(if~sliding)}$,
i.e.\ if one has a sliding contact with $f^t = \mu (f^n-f_{\rm min})$
\footnote{
On the other hand, if the particles are very
small, attractive forces could still be present so that $k_c$
would not be affected by the type of the contact being either
sliding or sticking.  In this study we assume, as an arbitrary,
possibly inconsistent choice, $f^n \ge 0, {\rm ~(if~sliding)}$, 
thus disregarding cohesion in the sliding situation. The effect of
this choice has to be examined in more detailed elsewhere.}.

\subsection{Temperature dependence in tangential direction}

In parallel to the change of normal stiffness, the tangential stiffness
is always kept in a constant ratio to $k_2$ so that
\begin{equation}
   k_t = \alpha k_2(\delta_{\rm max}) ~,
\end{equation}
since the stiffness in tangential direction is based
on the same arguments as the material stiffness in normal direction.

The friction is coupled to the temperature dependent value of the 
stiffness $k_1(T,t)$, because friction should not be present in a 
liquid at large enough temperatures, so that
\begin{equation}
  \mu(T,t) = \frac{k_1(T,t)}{k_1} \mu ~.
  \label{eq:mu}
\end{equation}
Thus friction is modified together with the changes in normal direction.
No new ideas are introduced for the tangential forces.

\section{Results}
\label{sec:results}

\begin{figure}[htb]
  \begin{center}
    \epsfig{file=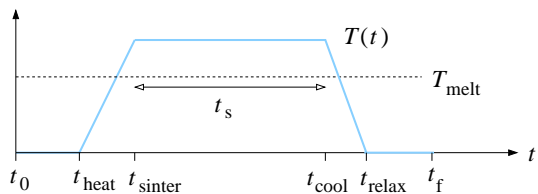,width=7.0cm,angle=0}
  \end{center}
  \caption{Schematic plot of the temperature variation during simulation.
           }
  \label{fig:schemNT}
\end{figure}

In this section, the sintering model is applied to the 
sintering process of a particulate material sample.  The material
is initially a loose powder and first has to be prepared at
low temperature from time $t_0$ to time $t_{\rm heat}$, see Fig.\
\ref{fig:schemNT}.  The preparation takes place with a system 
as described in section\ \ref{sec:model} with isotropic 
external pressure $p:=p_{\rm w}=p_{x} = p_{z} = 10$ or $100$.  
Pressure is measured in units of N\,m$^{-1}$ due to the two-dimensional
nature of the model, i.e.\ $p$ has the same units as the spring stiffness
$k_2$. The correct units of the pressure can be obtained by
division through the length of the cylinders (particles).  Since
the length is rather arbitrary, we drop the unit of pressure in the
following for the sake of simplicity. Due to the linearity of the
model, some stress $\sigma$ can be rescaled/non-dimensionalized with the 
stiffness $k_2$ and has thus to be read as dimensionless
quantity $2.10^5 \sigma / k_2$.  However, a detailed study
of the scaling behavior of the stresses is far from the scope of this study.

Note that the pressure is therefore related to the typical overlap of two 
particles: A pressure of $p=100$ corresponds to 
$100=2.10^5 p/k_2=2.10^5 f^n/(2ak_2)=2.10^5(\delta-\delta_0)/(2a)$
or $(\delta-\delta_0)/a \approx 10^{-3}$, while $p=10$ corresponds to 
approximately $(\delta-\delta_0)/a \approx 10^{-4}$. This can also
be interpreted as mean normal force $f^n \approx 2ap$.
However, these are rough estimates only, since the contact forces
and the overlaps are strongly varying in magnitude for one situation,
as will be discussed in more detail below.

The other system parameters are summarized in table\ \ref{tab:tabsys},
where multiple numbers mean that a series with the corresponding values 
was performed.
\begin{table}
\begin{tabular}{|l|l c|}
particle numbers $N$                  & 100, 300 & \\
\hline
mean particle radius $a_0$            & 1\,mm     &     \\
relative width of size distribution 
                                $w_0$ & $\pm 0.5$ &     \\
particle height $h$                   & 6\,mm     &     \\
particle density $\rho_p$             & 2000\,kg\,m$^{-3}$ & \\
\hline
particle stiffness $k_2$              & $2.10^5$\,N\,m$^{-1}$  &     \\
particle stiffness $k_1/k_2$          & 0.5                    &     \\
particle stiffness $k_c/k_2$          & 0.5                    &     \\
particle stiffness $k_t/k_2$          & 0.2                    &     \\
particle normal damping $\gamma_0$    & $0.2$\,kg\,s$^{-1}$    &     \\
particle tangential damping $\gamma_t$& $0.05$\,kg\,s$^{-1}$   &     \\
particle-particle friction $\mu$      & 0.5                 &     \\
particle-wall friction $\mu_w$        & 0                   &     \\
\hline
melting temperature $T_{\rm melt}$    & 120$^0$C  &     \\
temperature variation $T_{\rm var}$   & 10\,K     &     \\
density variation with temperature 
                         $\delta a_T$ & $10^{-4}$\,$K^{-1}$ &     \\
fluid overlap $\delta^{\rm fluid}$    & 0.2       &     \\
relaxation time $t_0$ (default)       & $10^3$\,s &     \\
relaxation time during sintering      & $10^{-1}$\,s &     \\
\hline
wall mass $m_{\rm w}$                 & 0.01      &    \\
wall damping $\gamma_{\rm w}$         & 0.2       &    \\
\end{tabular}
\caption{Summary of the system properties and material parameters as 
used in the simulations}
\label{tab:tabsys}
\end{table}
\subsection{Sample preparation}

The preparation of the sample consists of an initial relaxation
period at constant temperature $T=80^0$C until time $t_{\rm heat}$,
when the system is heated up to $T=140^0$C between time $t_{\rm heat}$
and $t_{\rm sinter}$.  During the sintering time $t_s$, the system is
allowed to sinter with a much shorter relaxation time $t_0=0.1$. This
`trick' allows for a long time sintering simulation while keeping the
simulation time small.  During sintering, the time axis should be stretched
by a factor of $10^{4}$ in order to obtain the real-time behavior.
At the end of the sintering process, at time $t_{\rm cool}$, the 
sample is slowly cooled down and, at time $t_{\rm relax}$ allowed
to relax at constant temperature until time $t_{\rm f}$.  With this
finished sample, tests will be performed later, but first we
discuss the preparation process.

For the preparation of the sample, we use the times $t_0=0$, 
$t_{\rm heat}=0.2$, $t_{\rm sinter}-t_{\rm heat}=0.1$, 
different sintering times $t_s:=t_{\rm cool}-t_{\rm sinter}$,
$t_{\rm relax}-t_{\rm cool}=0.1$, and $t_{\rm f}-t_{\rm relax}=0.1$.
Note that the time unit is arbitrary, since a slower preparation
procedure did not lead to noteably different results, i.e.\ the
process takes place in the quasi-static limit.  Only the long-time
sintering is affected by a change of the relaxation time $t_0$.

\subsubsection{Temperature and stiffness}

\begin{figure}[htb]
  \begin{center}
   \epsfig{file=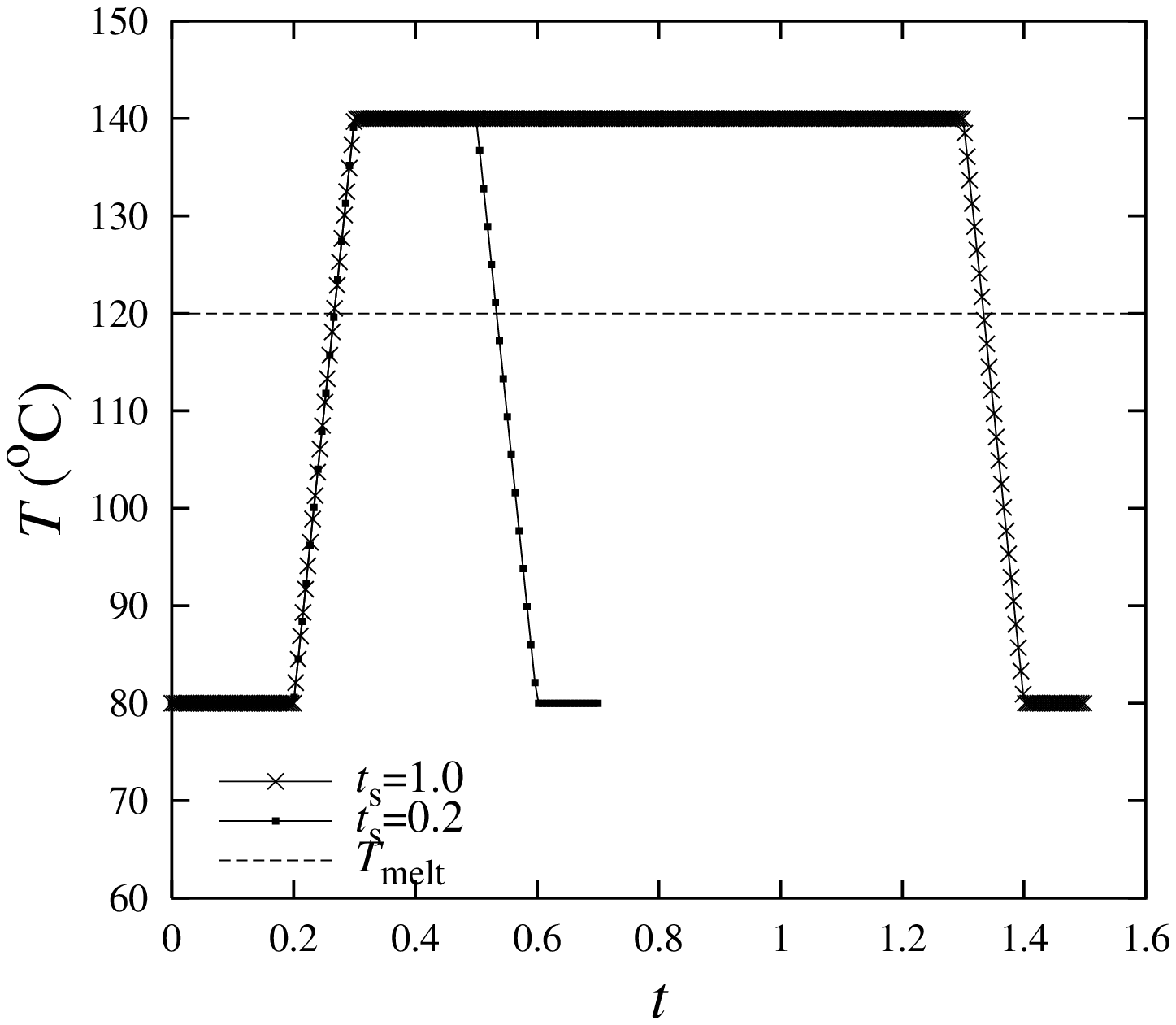,width=6.6cm,angle=-90} \hfill
   \epsfig{file=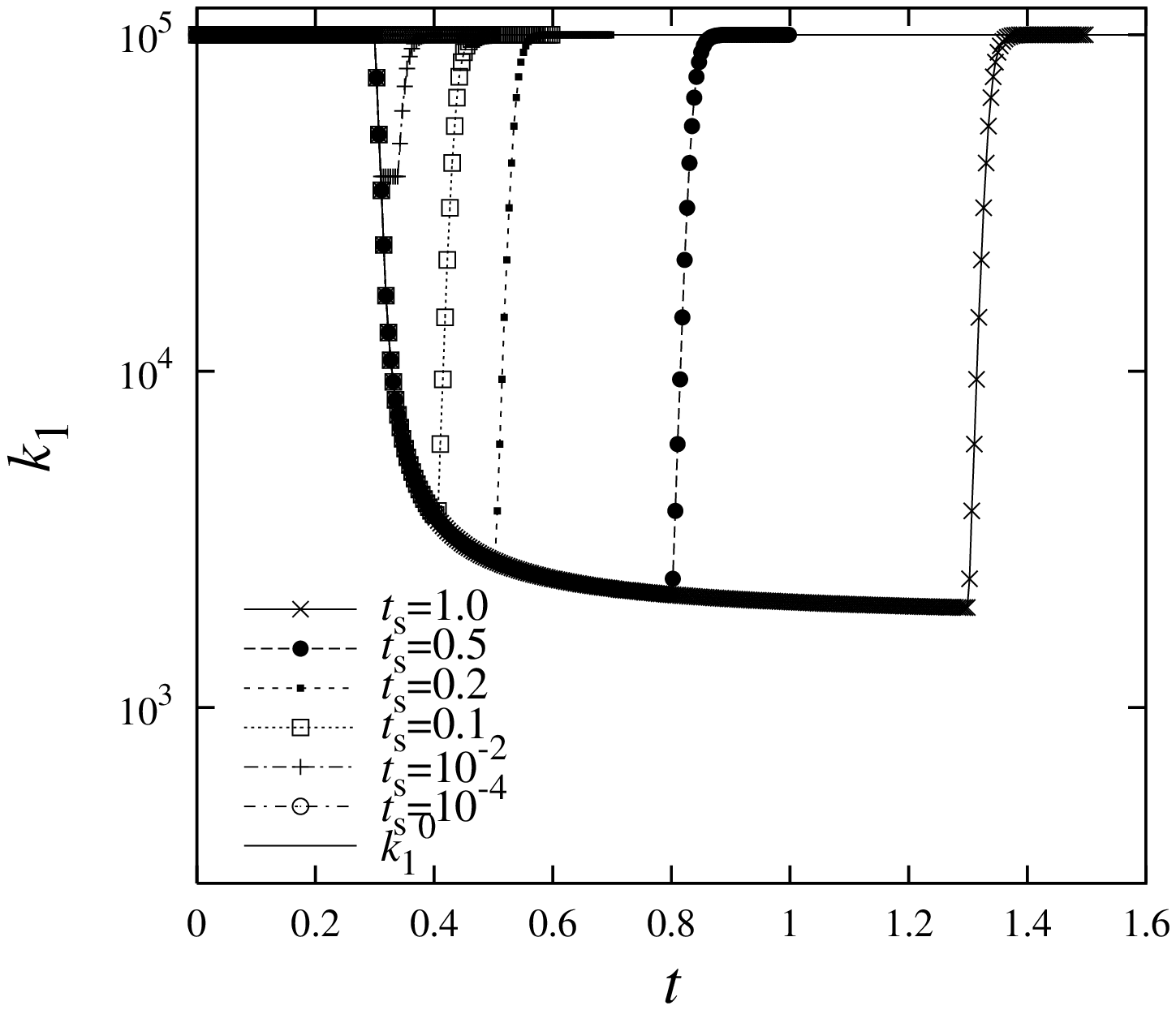,width=6.6cm,angle=-90} \hfill
  \end{center}
  \caption{(Top) Temperature during the preparation procedure and 
           (bottom) material stiffness $k_1$ as function of time for
           `experiments' with different sintering times $t_s$.}
  \label{fig:k1_prepare}
\end{figure}

The variation of $k_1(T,t)$ with time is plotted in Fig.\ \ref{fig:k1_prepare},
where the algebraic, slow decay of the stiffness with time becomes evident.  
The number of particles was $N=300$, the sidestress $p=100$, and the other 
parameters are given in table\ \ref{tab:tabsys}.

\subsubsection{Density}

\begin{figure}[htb]
  \begin{center}
   \epsfig{file=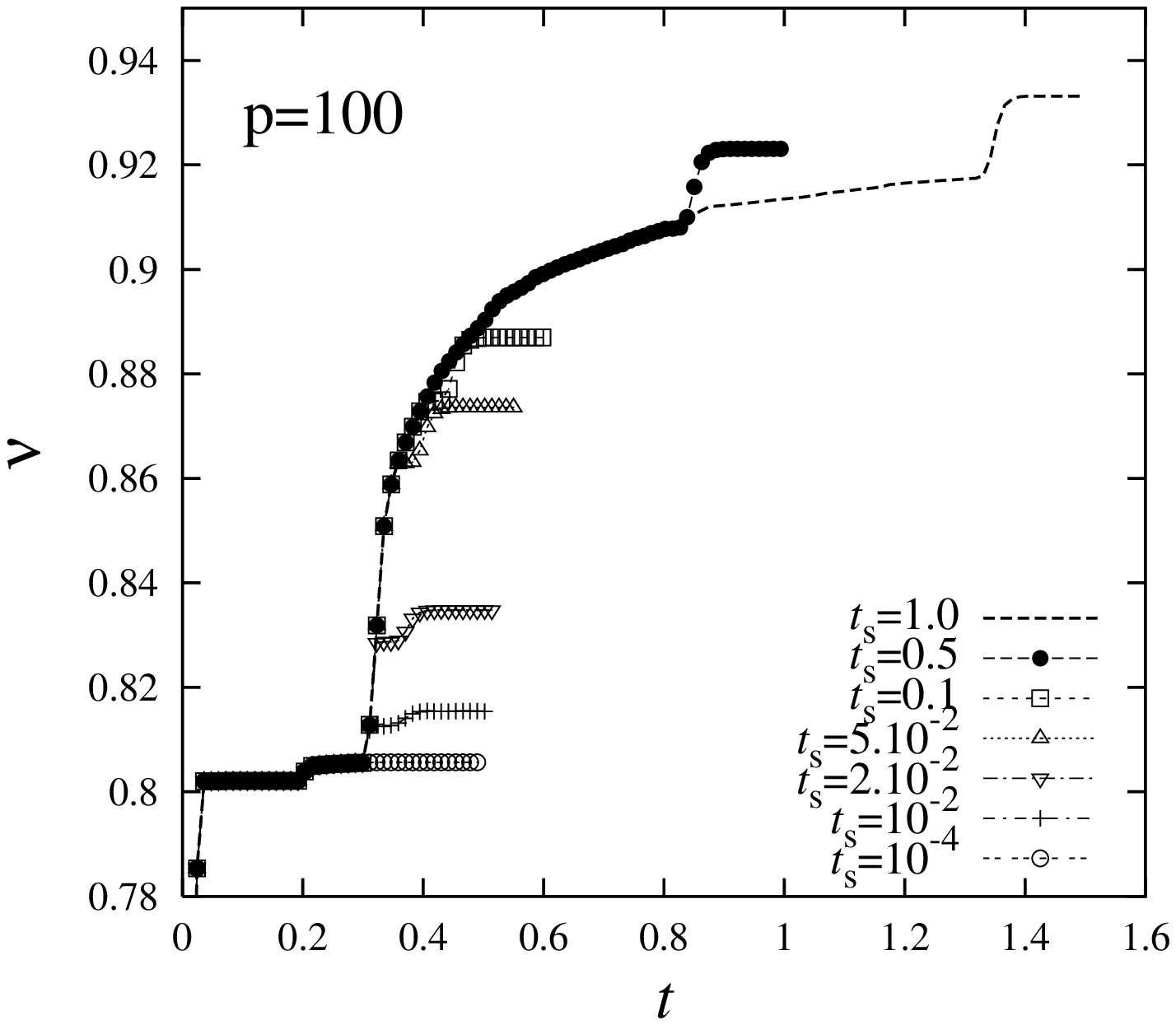,width=7.0cm,angle=-90}
   \epsfig{file=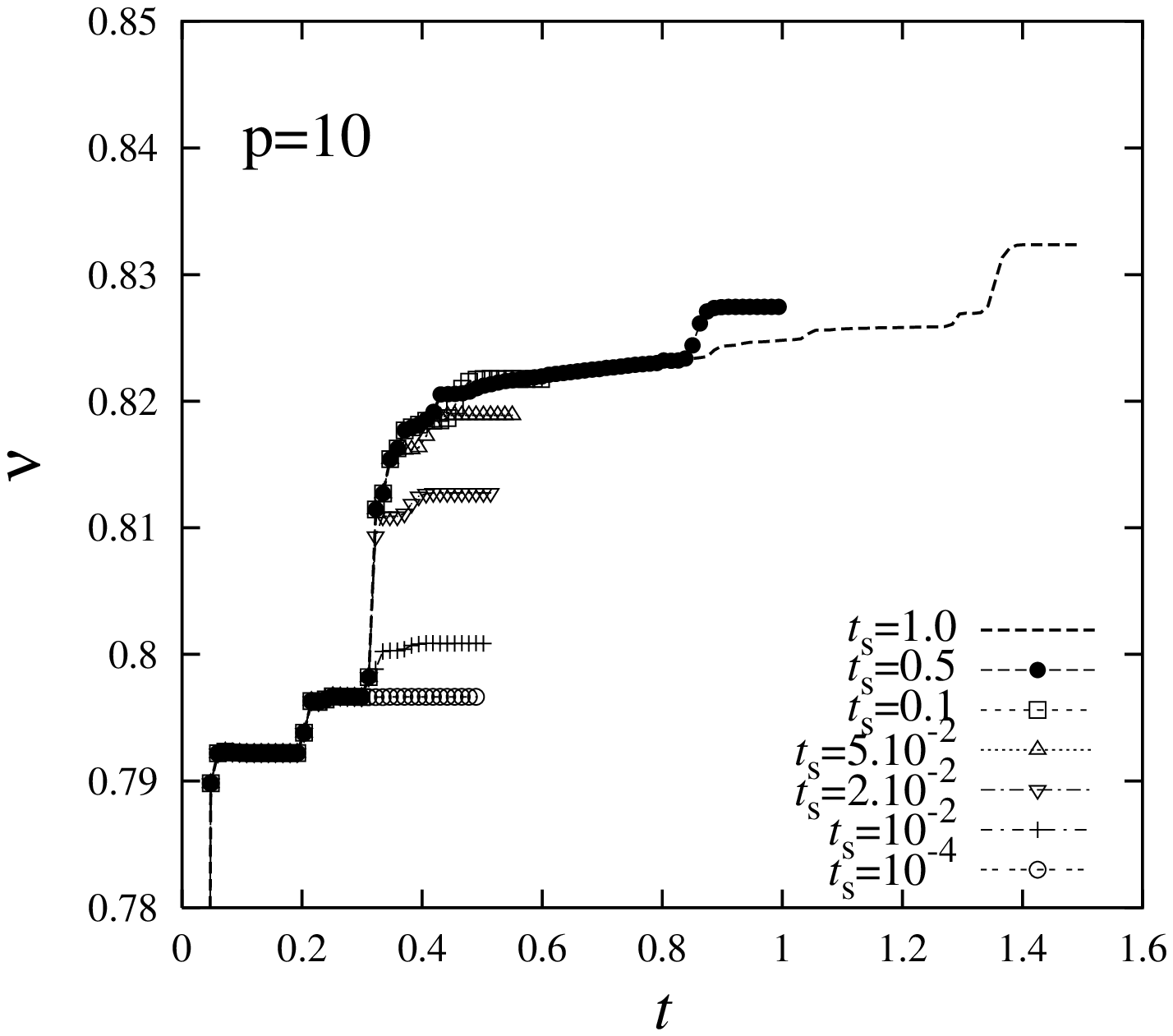,width=7.0cm,angle=-90}
  \end{center}
  \caption{Material density $\nu$ as function of time for simulations
           with $N=300$ particles and side stress $p=100$ (top) and
           $p=10$ (bottom). Note the different vertical axis scaling.}
  \label{fig:nu_prepare}
\end{figure}

The longer the sintering time $t_s$, the lower the value of $k_1$ gets.  
For short sintering time, the lowest values are never reached, because the
system is cooled down before the sintering is finished.
At the end of the sintering time, $k_1$ is increasing during the cooling
process of the sample and reaches its inital value.  However, the 
melting and sintering of the contacts is {\em not reversed}, as becomes
evident when plotting the volume fraction $\nu$ in Fig.\ \ref{fig:nu_prepare}.

The initial preparation step leads to a rather low density of $\nu \approx 0.8$.
At time $t_{\rm heat}$, the relaxed sample is heated up and, at the
same time, the wall friction is switched off.  The latter has an immediate 
effect, a slight increase in density is possible due to reorganizations.
The increased temperature becomes only effective at time $t_{\rm sinter}$,
when the relaxation time $t_0$ is decreased in order to accelerate the 
evolution of the system. Note that the increase in density due to the 
sintering seemingly appears quite rapid -- due to the quenched time axis
during sintering. Up to the beginning of the sintering process all
densities are equal due to an identical preparation of the sample.
For increasing duration of sintering, $t_s$, the density successivly 
increases with $t_s$.  This effect is strongest at the beginning, but
continues also for longer times.  Below, we will discuss different reasons
for this increase in density in more detail.  At the end of the sintering process,
the system is cooled down, corresponding to the final, small increase 
in density, and then relaxes to the final configuration.  

\subsubsection{Special cases}

As a control simulation, also three particles were simulation and sintered
in the same way as the larger samples with $N=100$ and $N=300$ above.  
A sample with only three particles has almost no possibility for a rearrangement
after the initial load is applied (actually, the three particles were either 
arranged on a triangle or on a line and did not change their configuration
for the cases observed).  This artificial test-case shows that the strong
increase in density observed for larger samples is caused by re-arrangements
of the packing and {\rm not only} due to the force model.  
More quantitatively, the sintering process leads to a densification of
about 5\% and 16\% for $p=10$ and $p=100$, respectively, whereas the
densification due to the contact law, without rearrangements of the
packing, can be estimated to the respective magnitudes of about
0.2\% and 8\% (data not shown here).  Thus, for low confining pressure,
reorganizations are much more important than in the case of large
external pressure -- at least for the parameters used here.
In this spirit, network-models or DEM simulations with fixed arrangements 
\cite{delo99,heyliger01,redanz01} can not account for the reorganization of 
the packing, which is of eminent importance during the sintering process.

Another control simulation without thermal expansion of the particles,
$\delta a_T=0$, gave no new insights and did not change the outcome of
the simulations after cool-down, even though the values of the density 
were slightly different from the values obtained with a thermal expansion.
Therefore, we conclude that the qualitative results are not dependent
on detailled parameter values and the sintering behavior, macroscopic
as well as microscopic, is generic within the framework of this model.

\subsubsection{Summary}

In summary, we obtain that longer sintering and larger confining pressures 
lead to higher densities of the sintered sample. As an example,
after the end of the longest sintering process (dashed line) at $t=1.3$,
a density of almost $0.92$ is reached for $p=100$. Only after cool-down,
the maximum density of almost $0.94$ is reached due to negative thermal 
expansion.

Through test simulations with different parameters, we verified that
the increase in density is only partially due to the contact model, but
also is caused by reorganizations in the sample.  The thermal expansion
of the particles, on the other hand, does not affect the results in 
a drastic way.

\subsection{Microscopic picture}

\begin{figure}[htb]
  \begin{center}
   \epsfig{file=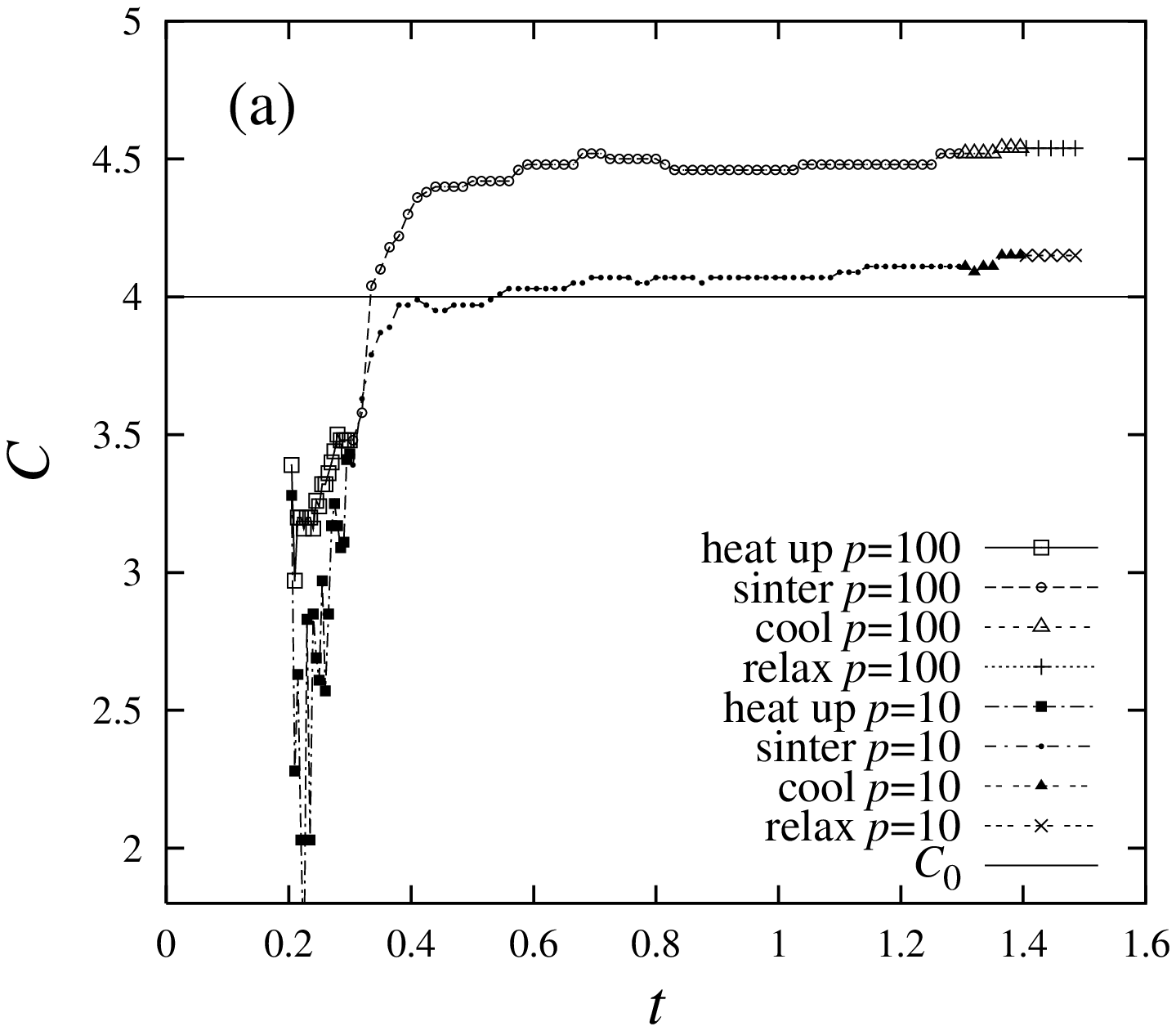,width=7.0cm,angle=-90}
   \epsfig{file=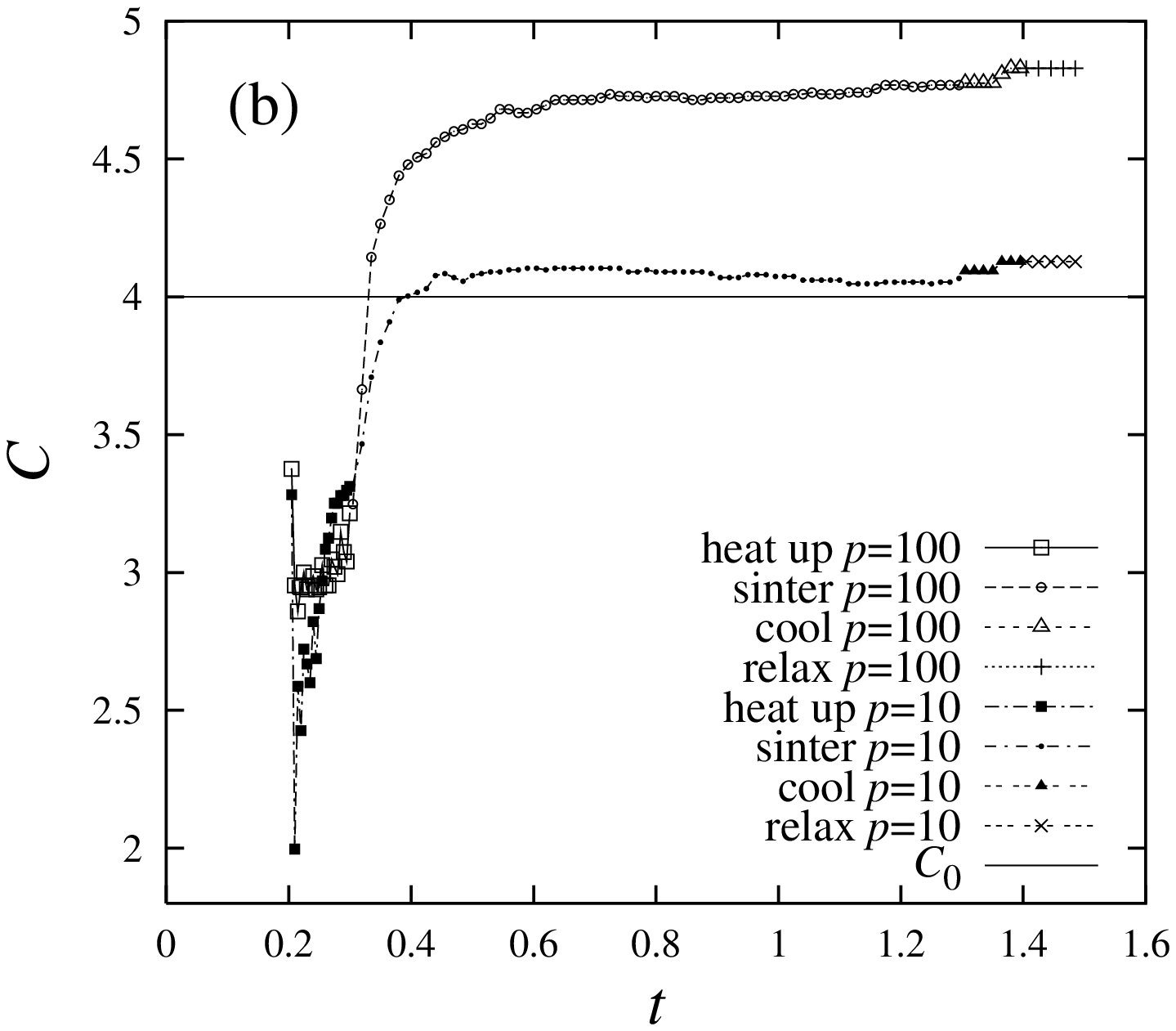,width=7.0cm,angle=-90}
  \end{center}
  \caption{Coordination number as function of time during the 
           preparation and sintering ($t_s=1$\,s) of the sample. Compared 
           are samples of different size, $N=100$ (a) and $N=300$ (b),
           and with different confining stress $p$.}
  \label{fig:cont1}
\end{figure}

In order to understand the behavior of the material during sintering,
we take a look at some microscopic quantities of the system, like the
coordination number, in Fig.\ \ref{fig:cont1}.

We note two major issues: First, the larger confining stress, $p=100$, leads 
to a larger number of contacts per particle, whereas the samples with small
confining stress, $p=10$, approach a coordination number of $C_0 \approx 4$,
as can be expected for a frictionless, isostatic arrangement of disks.
Second, the coordination number sometimes decays, whereas the density 
continuously increases.  We attribute this to the fact, that inside the
sample, reorganizations can take place, which save space and thus increase the
density, but at the same time can lead to a smaller $C$.

\begin{figure}[htb]
  \begin{center}
   \epsfig{file=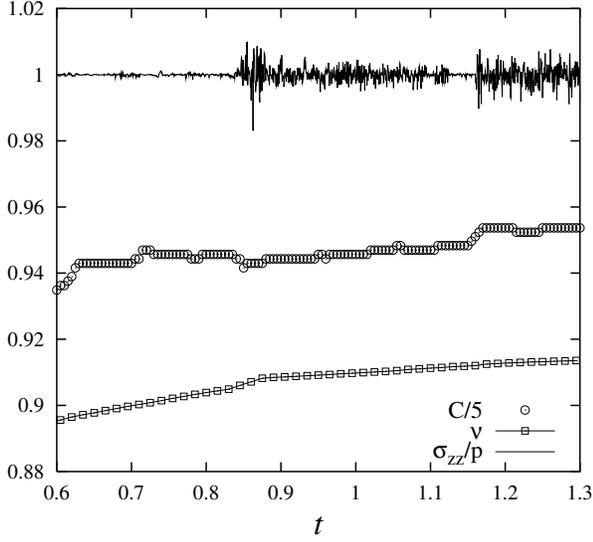,width=7.0cm,angle=-90}
  \end{center}
  \caption{Density (squares), coordination number (circles), and normalized 
           stress on the top-wall (line), as function of time during the 
           sintering ($t_s=1$\,s) of the sample.  The coordination number
           is scaled by a factor 0.2 in order to allow a comparison with the
           other quantities.
           }
  \label{fig:cont2}
\end{figure}

In order to deepen this insight, the coordination number, the density and
the measured wall-stress $\sigma_{\rm zz}/p$ are plotted in Fig.\ \ref{fig:cont2}.
The continuous increase in density is sometimes accompanied by a decrease
of the coordination number and large stress fluctuations.  Large scale 
re-organizations of the packing thus may be responsible for a detectable
fluctuation of the stress at the boundaries of the sample, possibly 
connected to sound emission.

\subsection{Contact statistics}

\subsubsection{Compression/deformation/overlap probabilities}

A very specific microscopic property of the sintered sample is the 
statistics of the contacts, i.e.\ the probability distribution function
(pdf) to find a certain overlap $\delta$.  After the initial preparation of
the powder sample at low temperature, the probability for larger 
overlaps decays rapidly -- no large deformations have taken place. 

\begin{figure}[htb]
  \begin{center}
   \epsfig{file=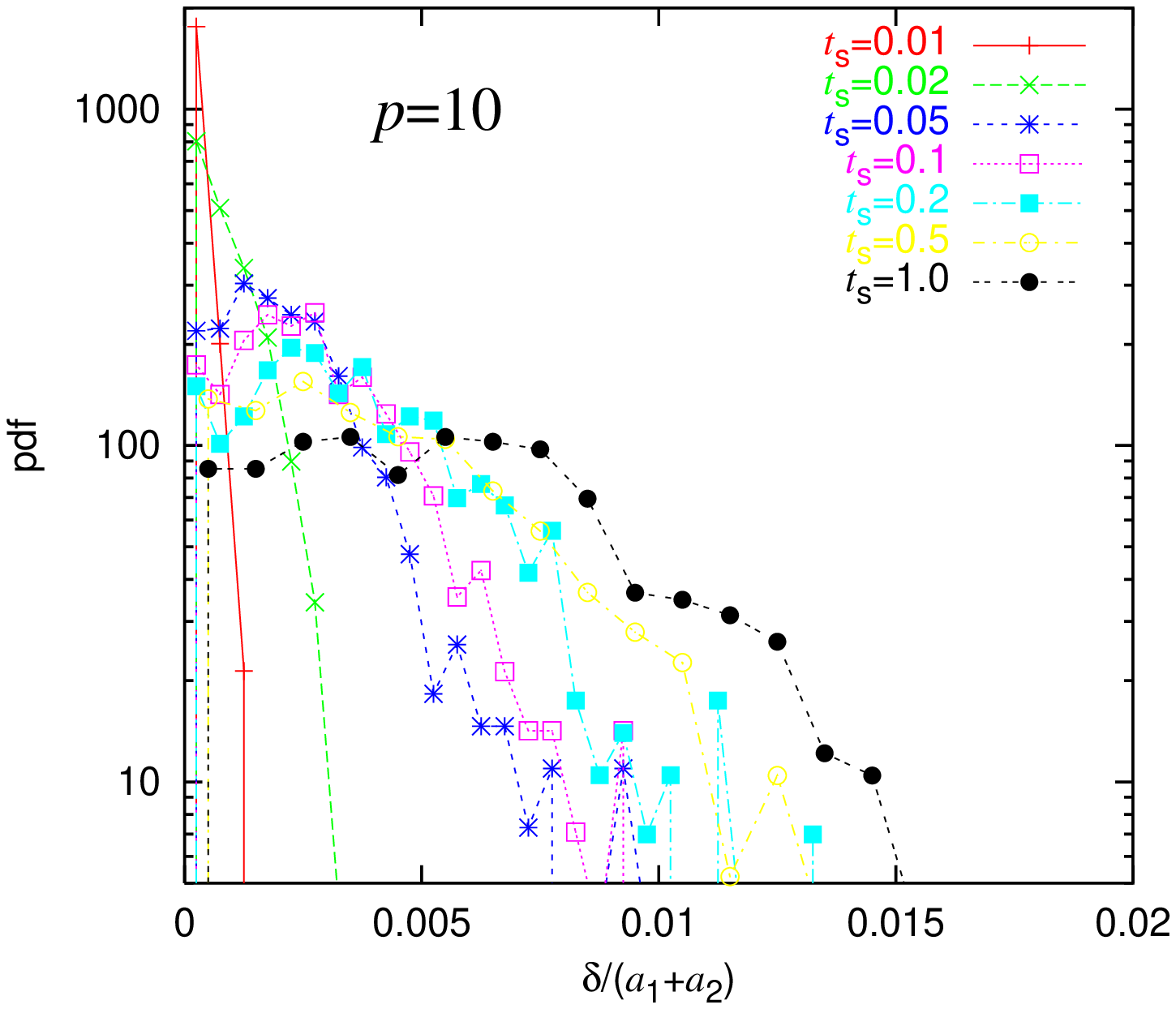,width=7.0cm,angle=-90}
   \epsfig{file=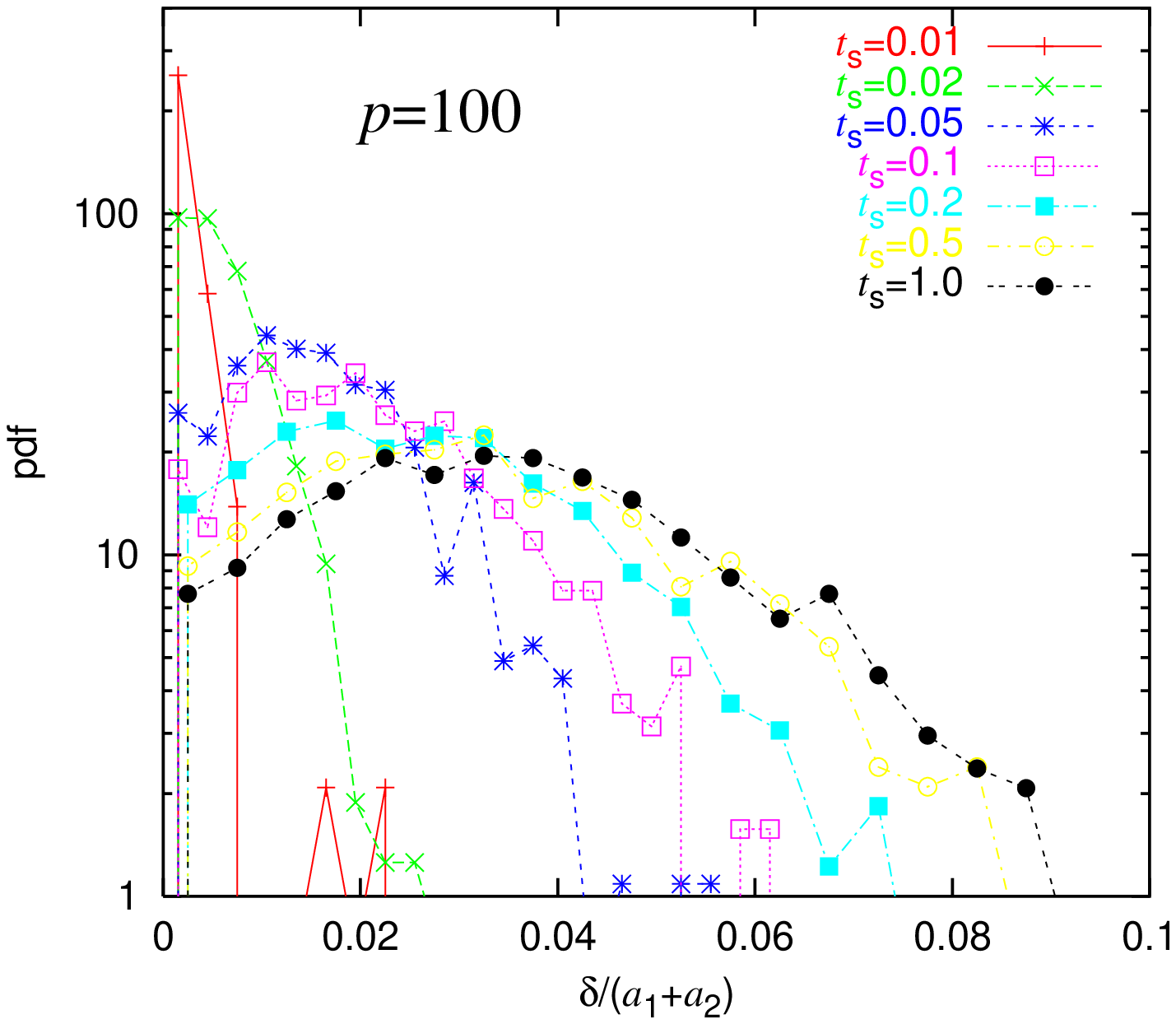,width=7.0cm,angle=-90}
  \end{center}
  \caption{Overlap probability for overlaps, rescaled with the sum
           of the radii $a_1$ and $a_2$ of the contacting particles, after
           different sintering durations. The data are taken from the simulations
           with $N=300$ at different confining pressure $p$.
           }
  \label{fig:overlap3}
\end{figure}

The overlap probabilities at the final stage of the sintering
process are plotted in Fig.\ \ref{fig:overlap3} for $N=300$, and both
side pressures $p=10$ and $p=100$.  The probability distributions
show many contacts with small overlaps and a rapidly decaying
probability for large overlaps at short sintering times. 
The mean overlap increases, as expected, with the side stress 
and the sintering time.  The larger side stress leads to an underpopulation
of small overlaps, because {\em all} particles are compressed quite strongly.

\begin{figure}[htb]
  \begin{center}
   \epsfig{file=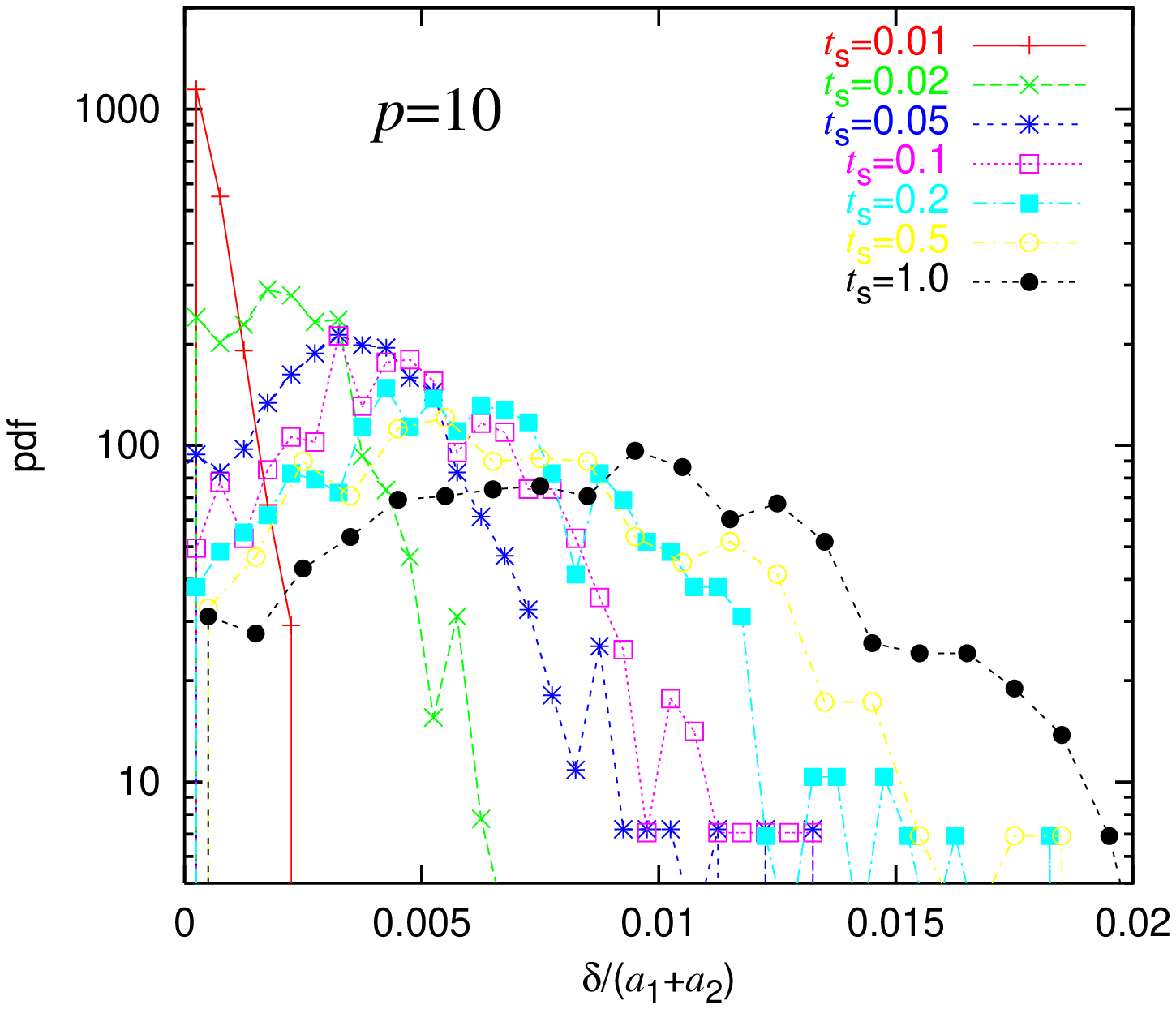,width=7.0cm,angle=-90}
   \epsfig{file=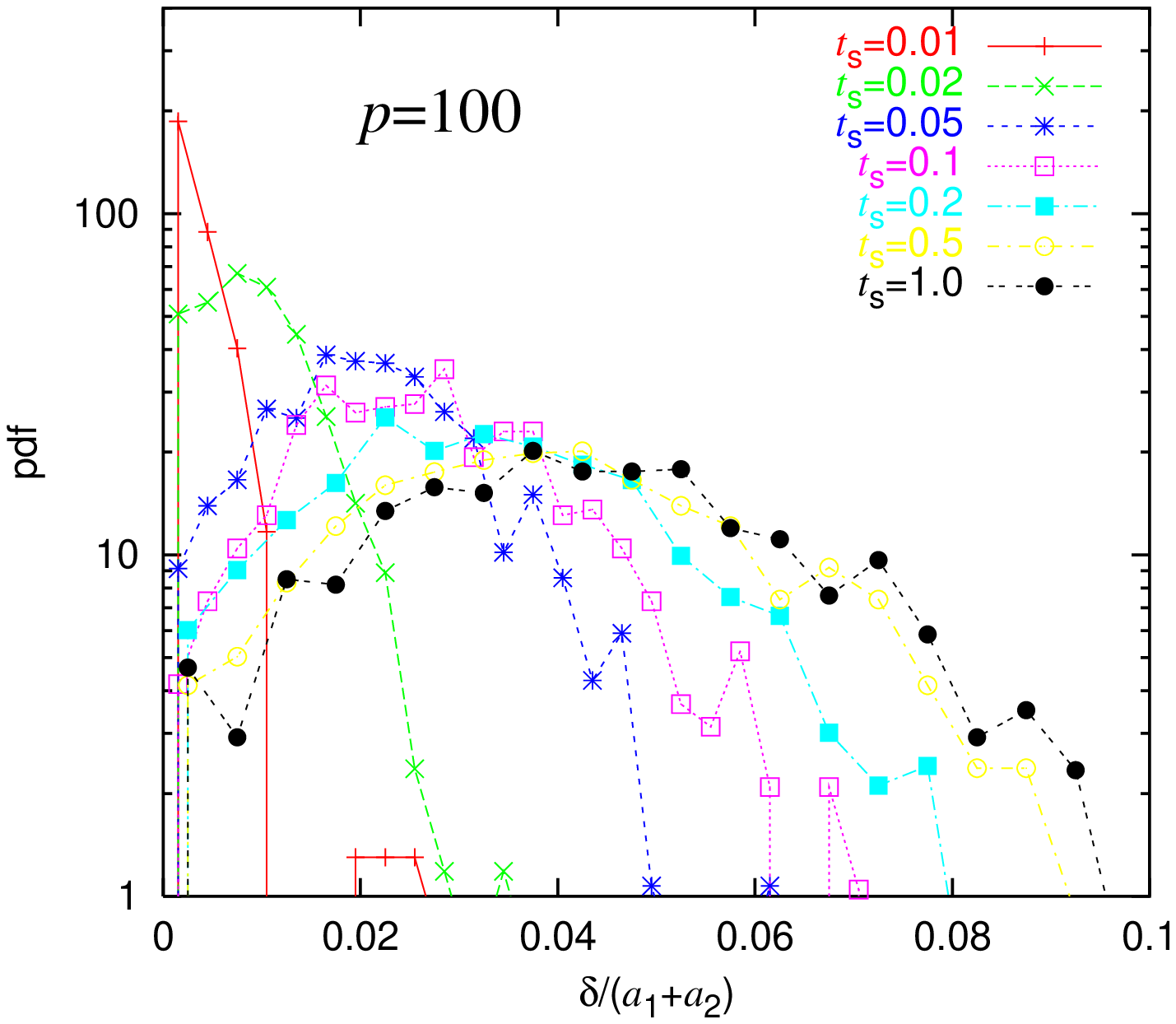,width=7.0cm,angle=-90}
  \end{center}
  \caption{Overlap probability for overlaps, rescaled with the sum
           of the radii $a_1$ and $a_2$ of the contacting particles, after
           different sintering durations. The data are taken from the simulations
           with $N=300$ at different confining pressure $p$.
           }
  \label{fig:overlap5}
\end{figure}

The probability distributions after the cool-down and the subsequent
relaxation is plotted in Fig.\ \ref{fig:overlap5} for the same simulations.
The remarkable difference is not the width of the distribution -- that is
only slightly wider. The most striking difference is the shape, showing
that after cool-down contacts with small overlap are rarefied, while
contacts with a the typical overlap are overpopulated as compared
to the situation before cool-down.  The probability for the largest 
overlaps is still decreasing rapidly.


\subsubsection{Normal contact force probabilities}

\begin{figure}[htb]
  \begin{center}
   \epsfig{file=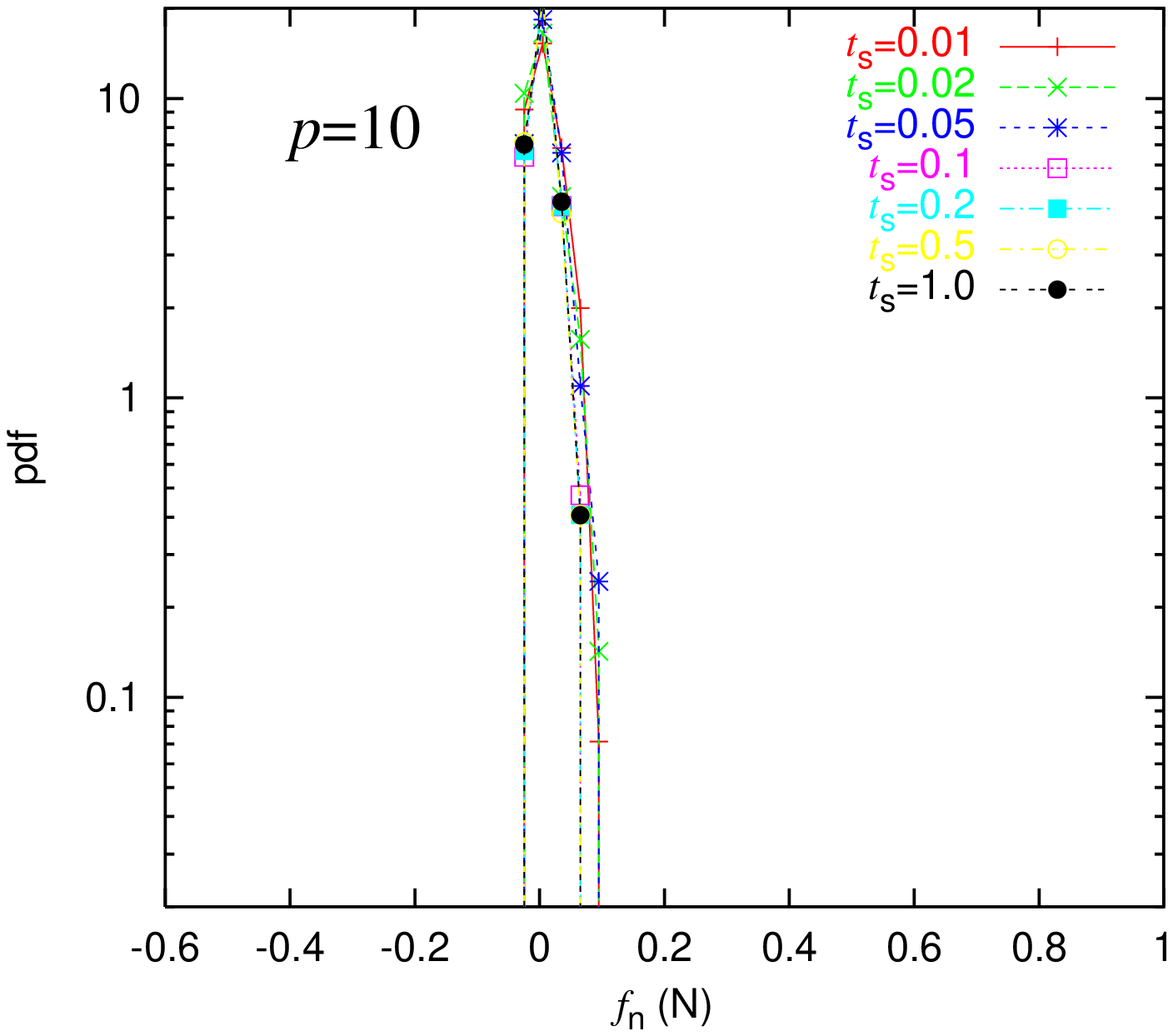,width=6.0cm,angle=-90}
   \epsfig{file=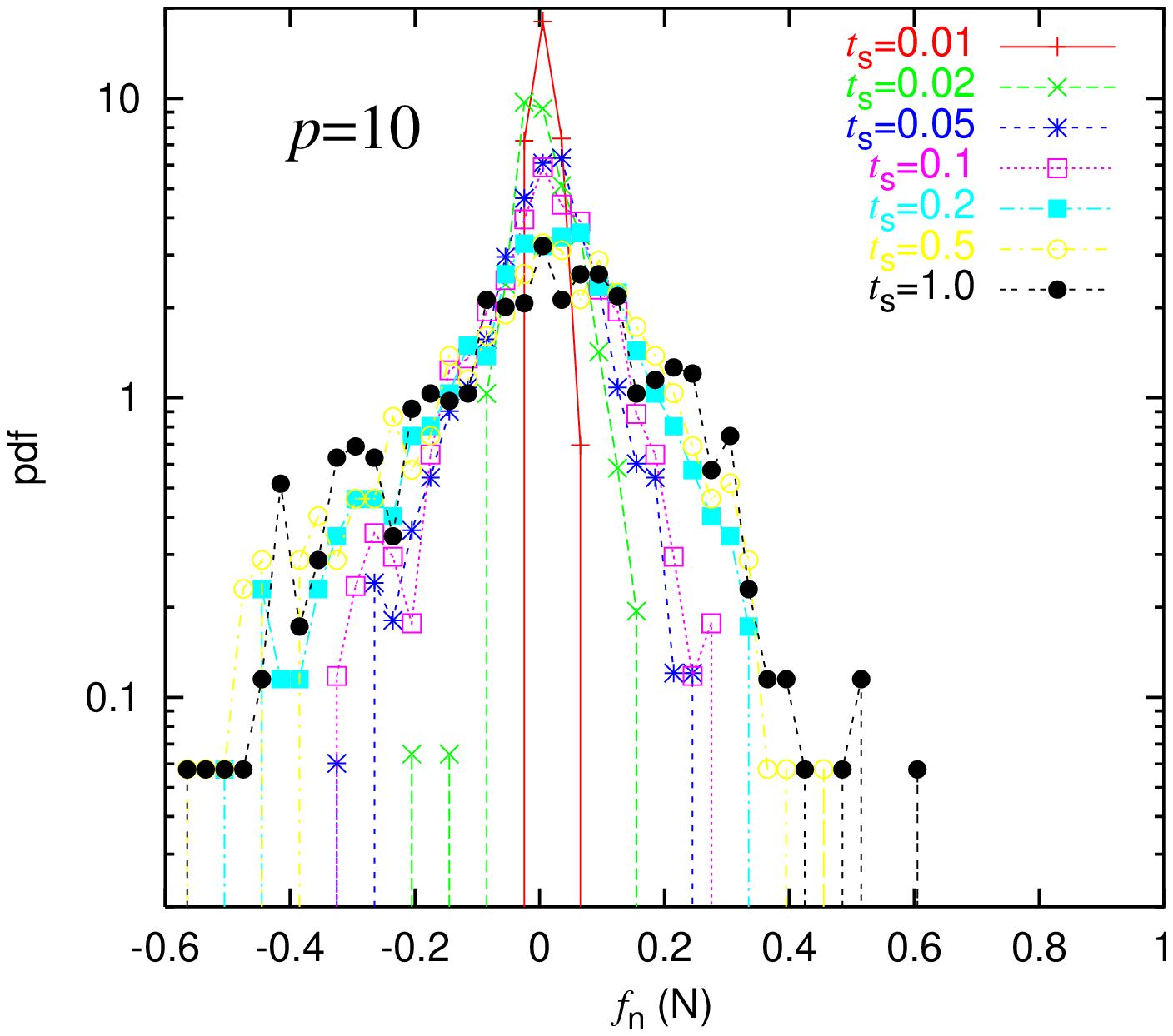,width=6.0cm,angle=-90}
  \end{center}
  \caption{Contact force probability distribution after the sintering (top)
           and after the cool-down and relaxation (bottom), for $N=300$ and
           $p=10$.
           }
  \label{fig:force10}
\end{figure}

\begin{figure}[htb]
  \begin{center}
   \epsfig{file=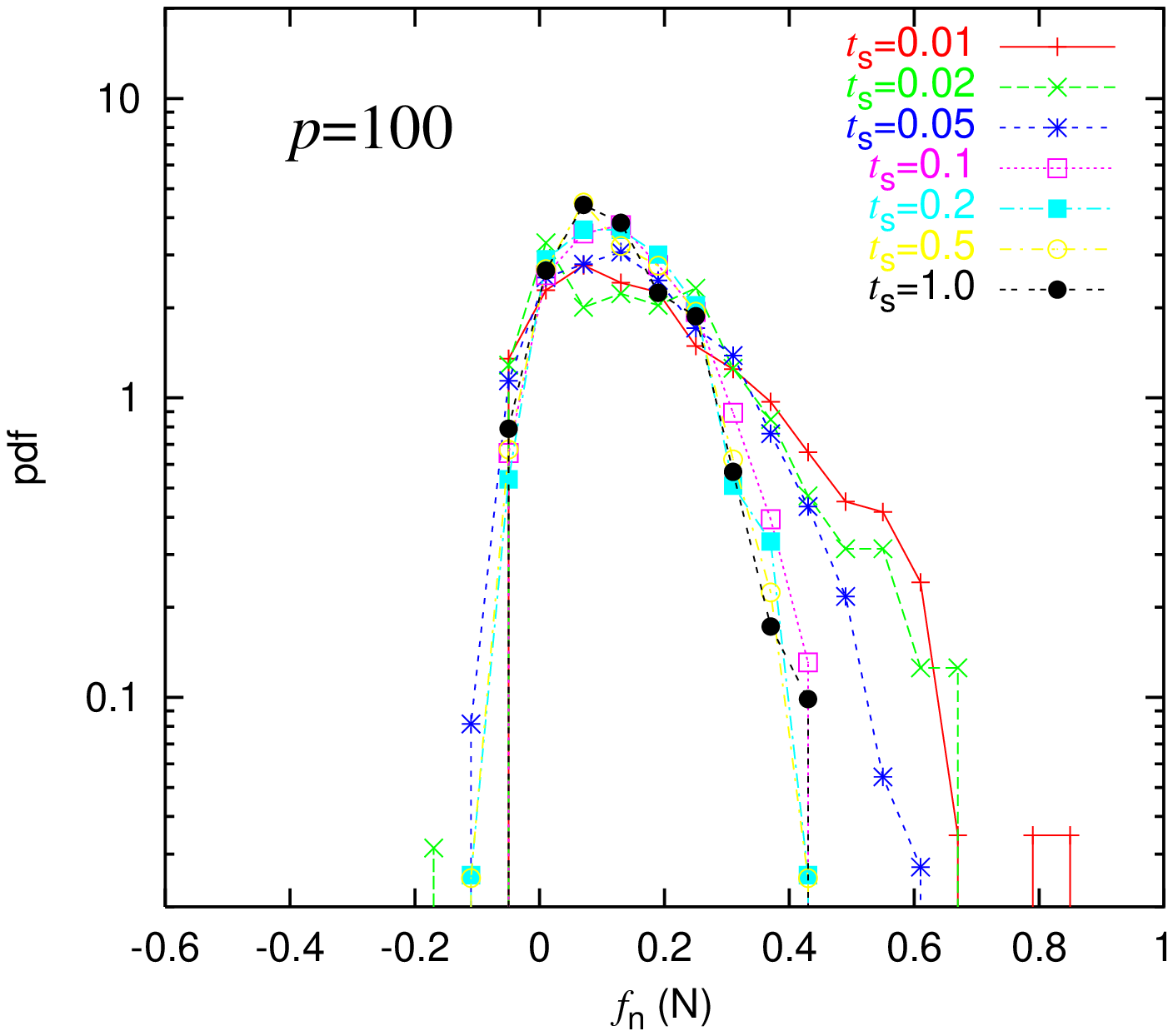,width=6.0cm,angle=-90}
   \epsfig{file=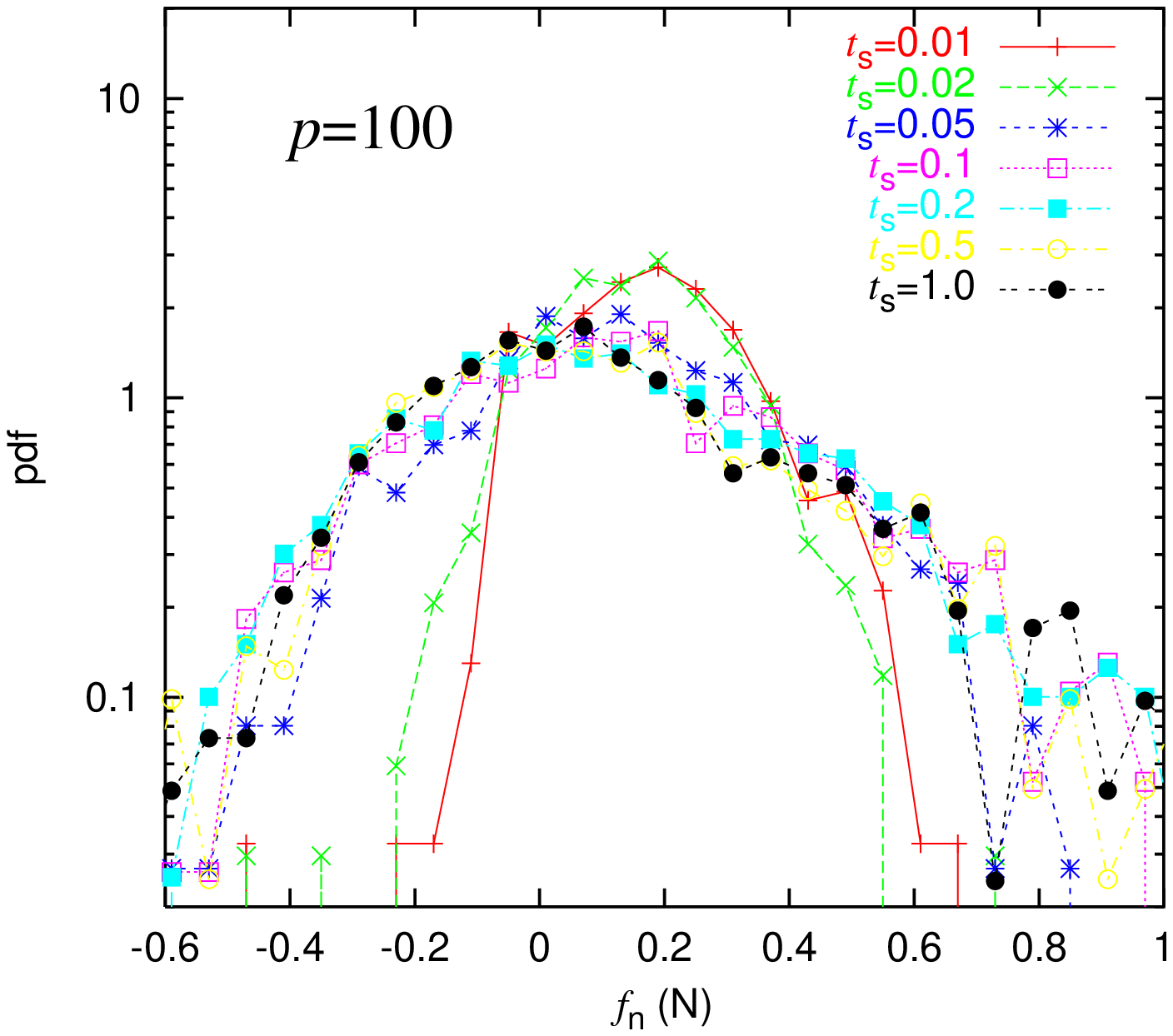,width=6.0cm,angle=-90}
  \end{center}
  \caption{Contact force probability distribution after the sintering (top)
           and after the cool-down and relaxation (bottom), for $N=300$ and
           $p=100$.
           }
  \label{fig:force100}
\end{figure}

Due to the advanced contact force law, as introduced and used in this stude,
it is not possible to directly obtain the contact force from the overlap.
Therefore, the pdf for the normal contact forces is plotted in 
Figs.\ \ref{fig:force10} and \ref{fig:force100} for small and large
side-stresses $p=10$ and $p=100$, respectively, and for different
sintering times, $t_s$ as given in the inset.

Comparing the pdf for the two situations, after sintering and
after cool-down, the astonishing outcome of the simulation is
the fact that the contact forces are mostly repulsive and rather
small in the hot situation, just before cool-down.  This situation
changes during cooling down: The cooling goes ahead with a broadening
of the distribution towards both positive and negative forces.
This qualitative result is independent of the side pressure, however,
the width of the distribution increases with increasing side stress
and increasing sintering time.

Another interesing and unexpected result is the observation that 
the force distribution {\em becomes narrower} with sintering.
In other words, the extremely large forces are ``destroyed'' due
to long time sintering and the distribution becomes more homogeneous
in the sense that its width becomes smaller, see top panel in 
Fig.\ \ref{fig:force100}.

\subsubsection{Tangential forces}

From the tangential force distribution (data not shown), there are less 
clear observations to be made.  The force distribution shows that during
sintering the tangential forces become weaker -- as can be expected from
the force model, see Eq.\ (\ref{eq:mu}), since $k_1(T,t)$ decays with 
increasing temperature and time.  Thus longer sintering leads to
weaker tangential forces.  However, after cool-down, the tangential
force distribution is almost independent of the side stress and the sintering
time, if both are sufficiently large.  With other words, only for very
small side stresses and sintering times, the tangential forces are 
weakly activated, for larger $p$ or $t_s$, the tangential forces
reach a saturation distribution that is decaying almost exponentially for
large $f_t$.



The second possibility to look at the tangential forces is to measure the
amount of friction that is activated, namely $\mu_t:=f_t/(f_n-f_{\rm min})$.
This quantity should not become larger than $\mu(T,t)$, as is 
consistently observed from the data.  Contacts where $\mu_t \approx \mu(T,t)$
are referred to as contacts with fully activated friction.  These
become less probable for longer sintering times since, as mentioned
before, the attractive forces become stronger after long sintering and
the cooling down. Larger attractive forces correspond to a larger magnitude
of $f_{\rm min}$, so that $\mu_t$ becomes smaller.

\subsection{Material properties}

The samples prepared via the procedure described above are now
tested via two test experiments, a compression test, where the vertical
confining pressure is slowly increased, and a vibration test, where the 
confining stress is removed, and the sample is vibrated vertically
on a flat bottom. The compression test is performed for two sample
sizes, namely with $N=100$ and $N=300$, in order to judge the effect
of a rather small particle number on the outcome of the simulations.

\subsubsection{Compression test (100 particles)}

The compression test is a variant of the bi-axial compression 
frequently used in soil- or powder mechanics. A given 
compressive stress in one direction ($p=\sigma_{xx}$) is kept constant 
by moving the right, vertical wall, if neccessary. The other 
confining stress $\sigma_{zz}$ is increased by moving the vertical
wall down in a defined way -- see also section\ \ref{sec:model}. 
This vertical motion is thus strain controlled, where the vertical strain is 
defined as $\epsilon_{zz}(t)=1-z(t)/z_0$, with the vertical position of the
top wall $z(t)$ and $z_0=z(0)$. Note that a compressive vertical
strain is defined positive for convenience here.

\begin{figure}[htb]
  \begin{center}
   \epsfig{file=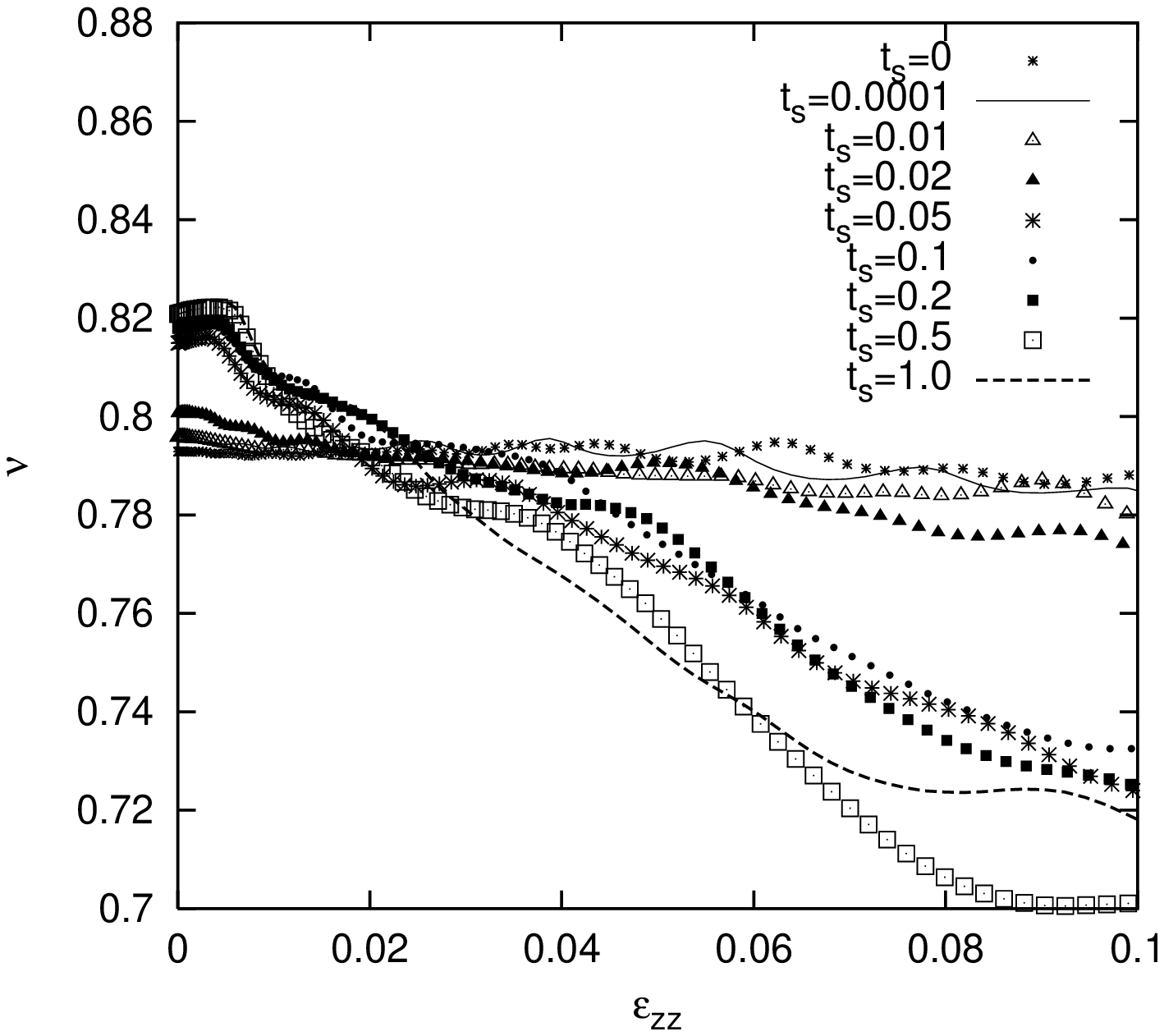,width=7.0cm,angle=-90} \hfill
   \epsfig{file=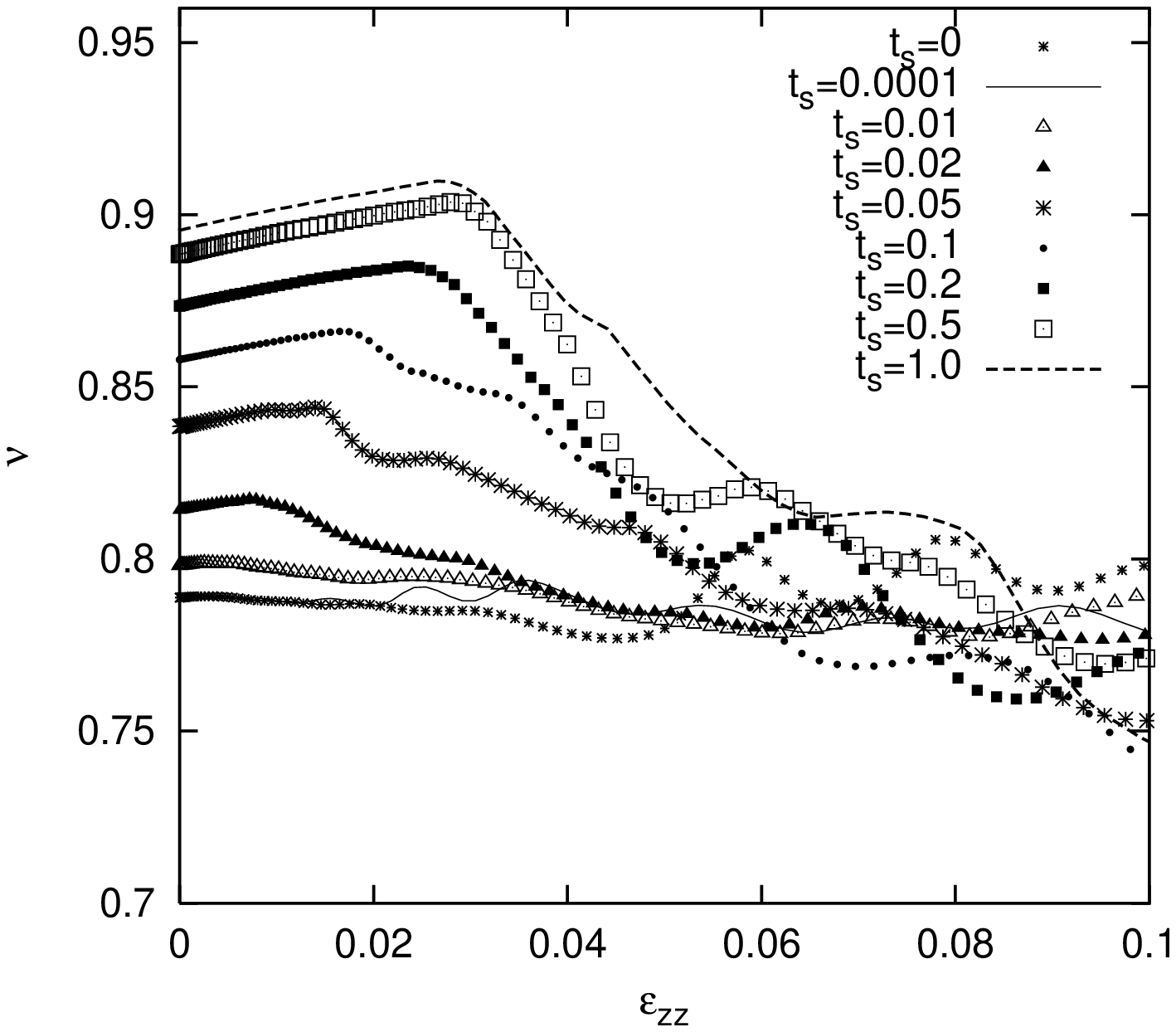,width=7.0cm,angle=-90}
  \end{center}
  \caption{Material density as function of the vertical strain
           for samples prepared with different sintering times $t_s$
           and different pressure $p_{\rm w}=10$ (Top) and $p_{\rm w}=100$
           (Bottom).}
  \label{fig:S1nu}
\end{figure}

For the compression test, the previously prepared samples are used and the top
wall is displaced slowly with the vertical strain $\epsilon_{zz}(t)$.
The density of a sample with $N=100$ particles is plotted against the
strain for various sintering times and for the two pressures $p=10$ and
$p=100$ in Fig.\ {\ref{fig:S1nu}}. The results show again that
samples, which sintered longer have a higher density.  Furthermore, 
the compressive pressure also affects the sample density in magnitude
and also in the variation, i.e.\ the higher compressive pressure
leads to higher densities and also to a broader variation in densities
between.

During compression, the density slightly increases first and then
decreases strongly.  The former is due to compression, the latter is 
due to dilatancy that is neccessary for the material to shear.
The longer the sample was sintered, the stronger is
the change in density.  We relate this to the existence of the attractive
contact forces.  For longer sintering, stronger attraction is activated,
holding together parts of the sample, so that the compression test leads
to fragements with increasing size for increasing sintering duration.
For the shortest sintering, the density change due to compression is
negligible and the sample fragments into single grains.

\begin{figure}[htb]
  \begin{center}
   \epsfig{file=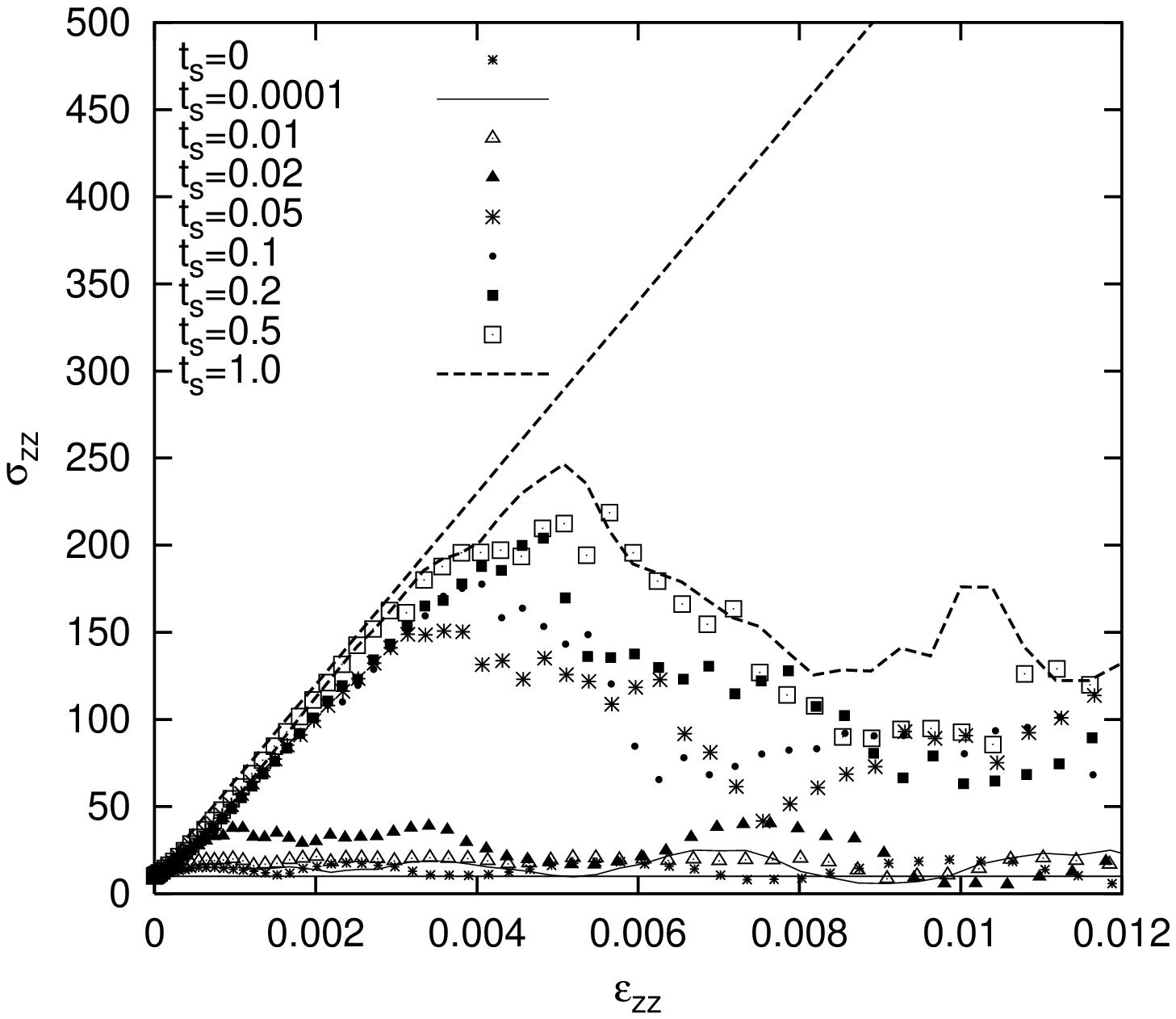,width=7.0cm,angle=-90} \hfill
   \epsfig{file=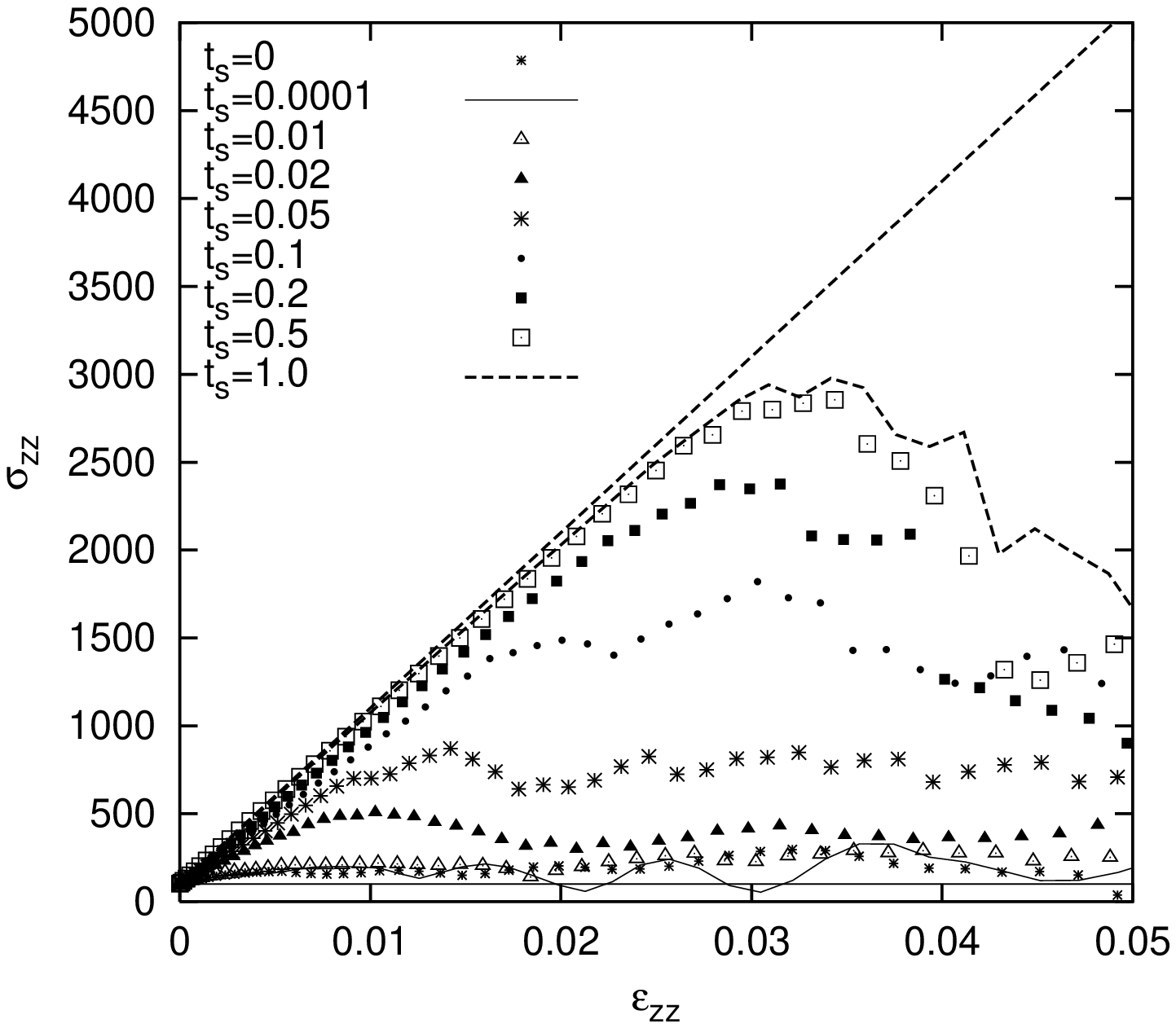,width=7.0cm,angle=-90}
  \end{center}
  \caption{Vertical stress as function of the vertical strain
           for samples prepared with different sintering times $t_s$
           and different pressure $p_{\rm w}=10$ (Top) and $p_{\rm w}=100$ 
           (Bottom). The slope of the dashed line (material stiffness)
           is about double for the higher pressure.}
  \label{fig:S1sxx}
\end{figure}

The behavior of the material, as for instance its stiffness, is better 
described in terms of the vertical stress that the material can sustain
under load, as plotted in Fig.\ {\ref{fig:S1sxx}}.  The vertical stress
in the sample increases and it fails typically at some strain and at some 
magnitude of stress. 
The failure stress increases with increasing sintering time and
increasing external stress.  The material stiffness (the slope in this 
representation) is increased by a factor of about two when the confining 
stress is increased by a factor of ten.  Moreover, the critical strain where 
the material fails increases with increasing confining pressure. Finally, a 
rather large jump in material strength is observed for sintering times 
between $t_s=0.02$ and $0.05$ for the small pressure, not paralleled
by a similar outcome for the large stress. Thus the combination of
sintering time, compressive pressure and test mode appears quite
non-linear and not straightforward.

\subsubsection{Compression test (300 particles)}

\begin{figure}[htb]
  \begin{center}
   \epsfig{file=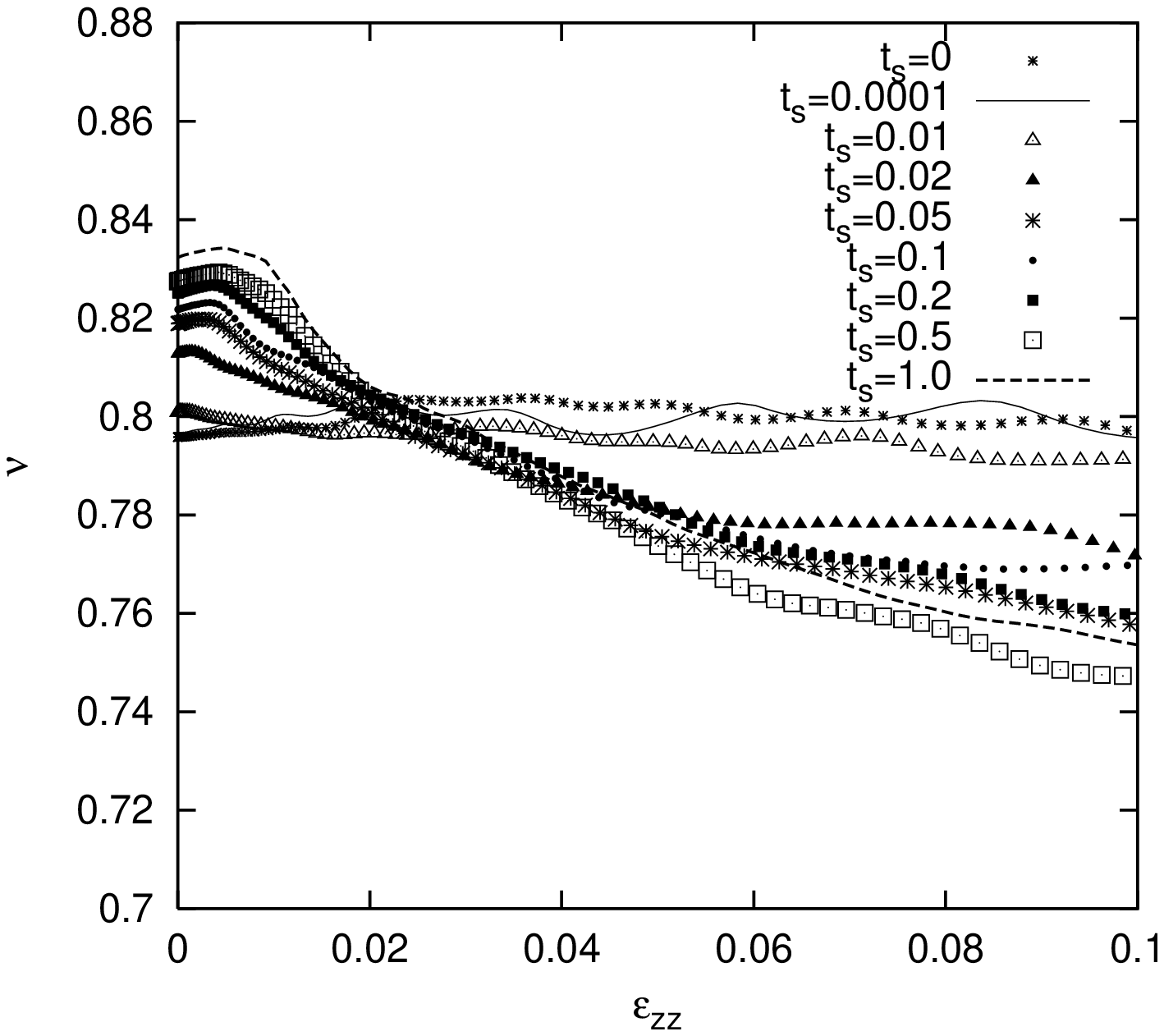,width=7.0cm,angle=-90} \hfill
   \epsfig{file=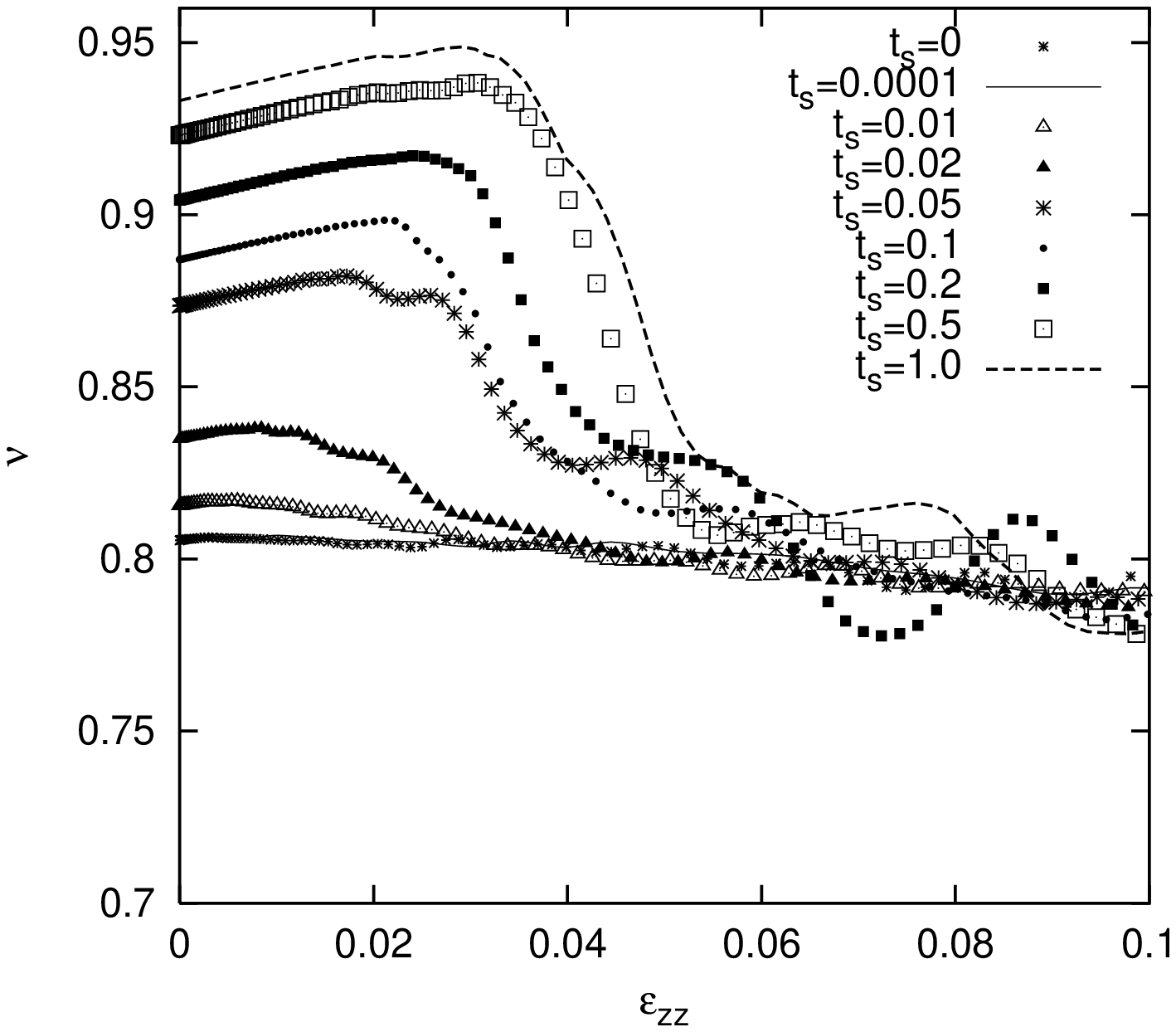,width=7.0cm,angle=-90}
  \end{center}
  \caption{Material density as function of the vertical strain
           for samples prepared with different sintering times $t_s$
           and different pressure $p_{\rm w}=10$ (Top) and $p_{\rm w}=100$
           (Bottom).}
  \label{fig:S3nu}
\end{figure}

For the compression test of the larger samples, also the prepared 
samples are used and the top wall is displaced slowly with the 
vertical strain.  The density of a sample with $N=300$ particles 
is plotted against the strain for various sintering times and for 
the two pressures $p$ in Fig.\ {\ref{fig:S3nu}}. The results are 
qualitatively similar, only that the larger sample shows somewhat
larger densities and a weaker decay of density after failure.

\begin{figure}[htb]
  \begin{center}
   \epsfig{file=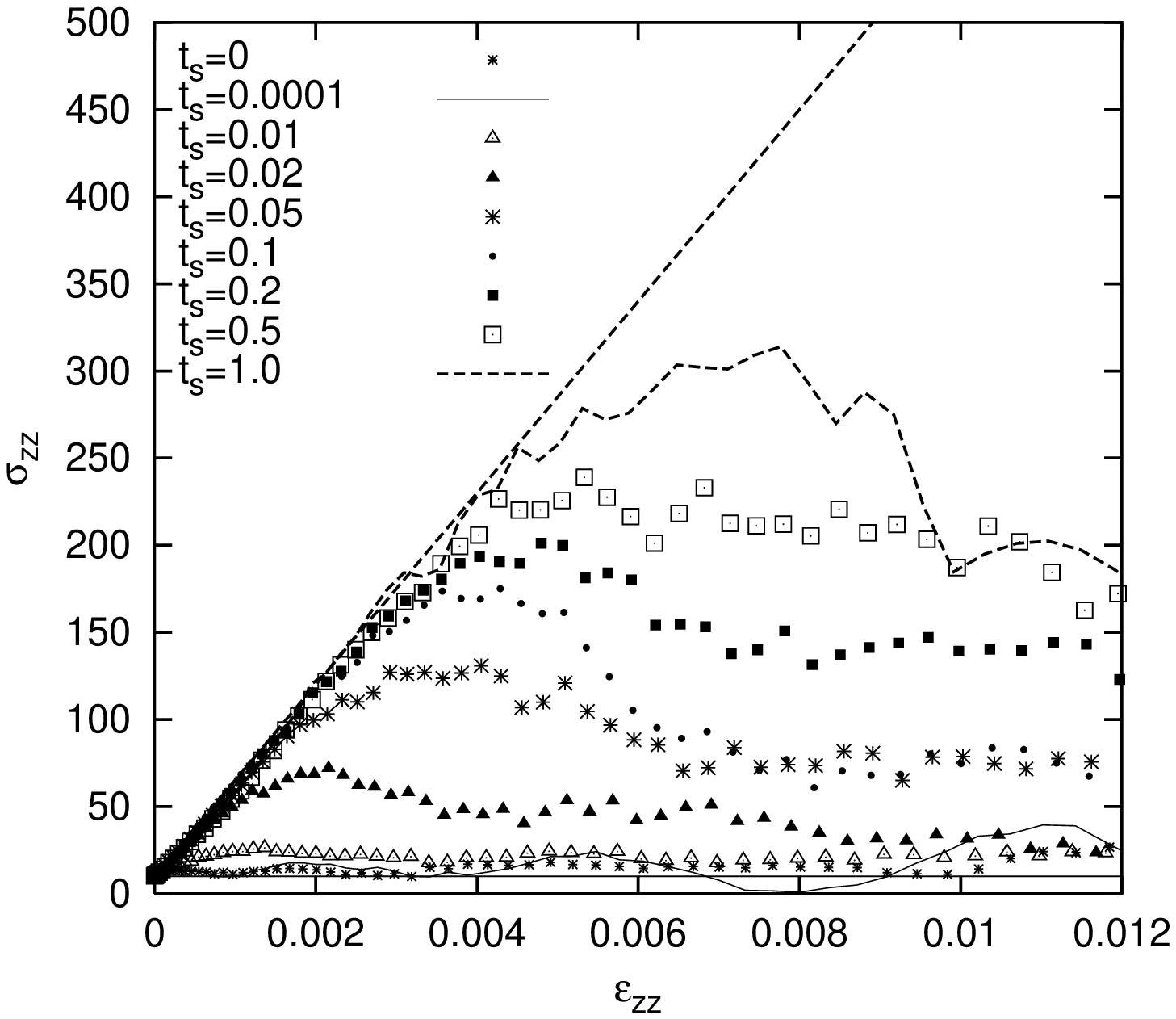,width=7.0cm,angle=-90} \hfill
   \epsfig{file=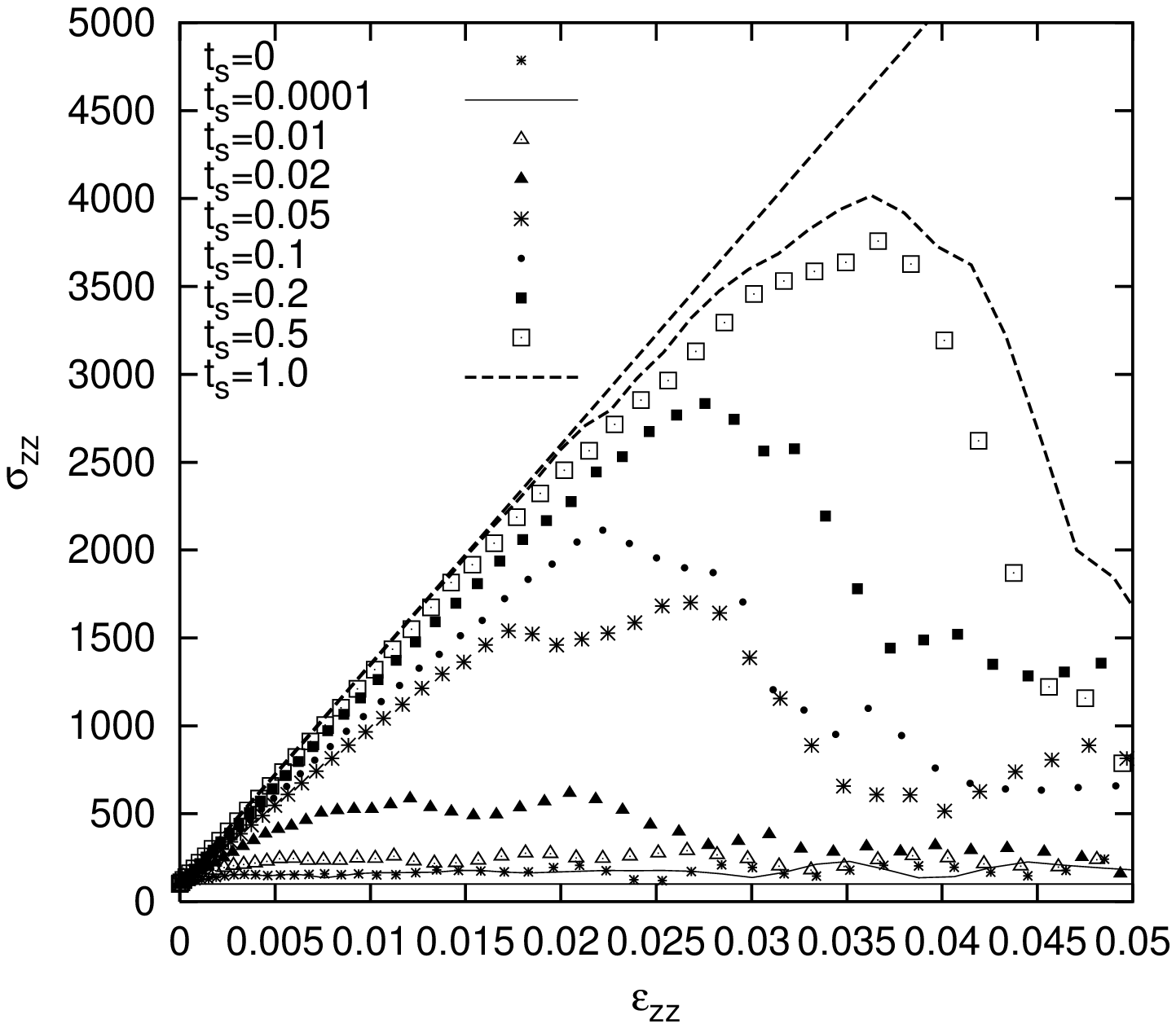,width=7.0cm,angle=-90}
  \end{center}
  \caption{Vertical stress as function of the vertical strain
           for samples prepared with different sintering times $t_s$
           and different pressure $p_{\rm w}=10$ (Top) and $p_{\rm w}=100$ 
           (Bottom). The slope of the dashed line (material stiffness)
           is about double for the higher pressure.}
  \label{fig:S3sxx}
\end{figure}

The vertical stress is plotted against the strain in Fig.\ {\ref{fig:S3sxx}}. 
With increasing strain, the vertical stress in the sample increases and it 
fails typically at some slightly larger strain and stress as in the small
sample.  The failure stress increases with increasing sintering time,
increasing external stress, and increasing sample size.  However, the latter
observation is not clearly a size-effect, since the sample size is
so small that a non-negligible fraction of the particles is in contact
with the walls and thus leads to different outcome.
The material stiffness (dashed line in Fig.\ {\ref{fig:S3sxx}}) is again
increased by about a factor of two when the confining stress is increased 
by a factor of ten.  Moreover, the critical strain where the
material fails increases with increasing confining pressure, sintering
time and system size.

\subsubsection{Compression test snapshots}

\begin{figure}[htb]
  \begin{center}
   \epsfig{file=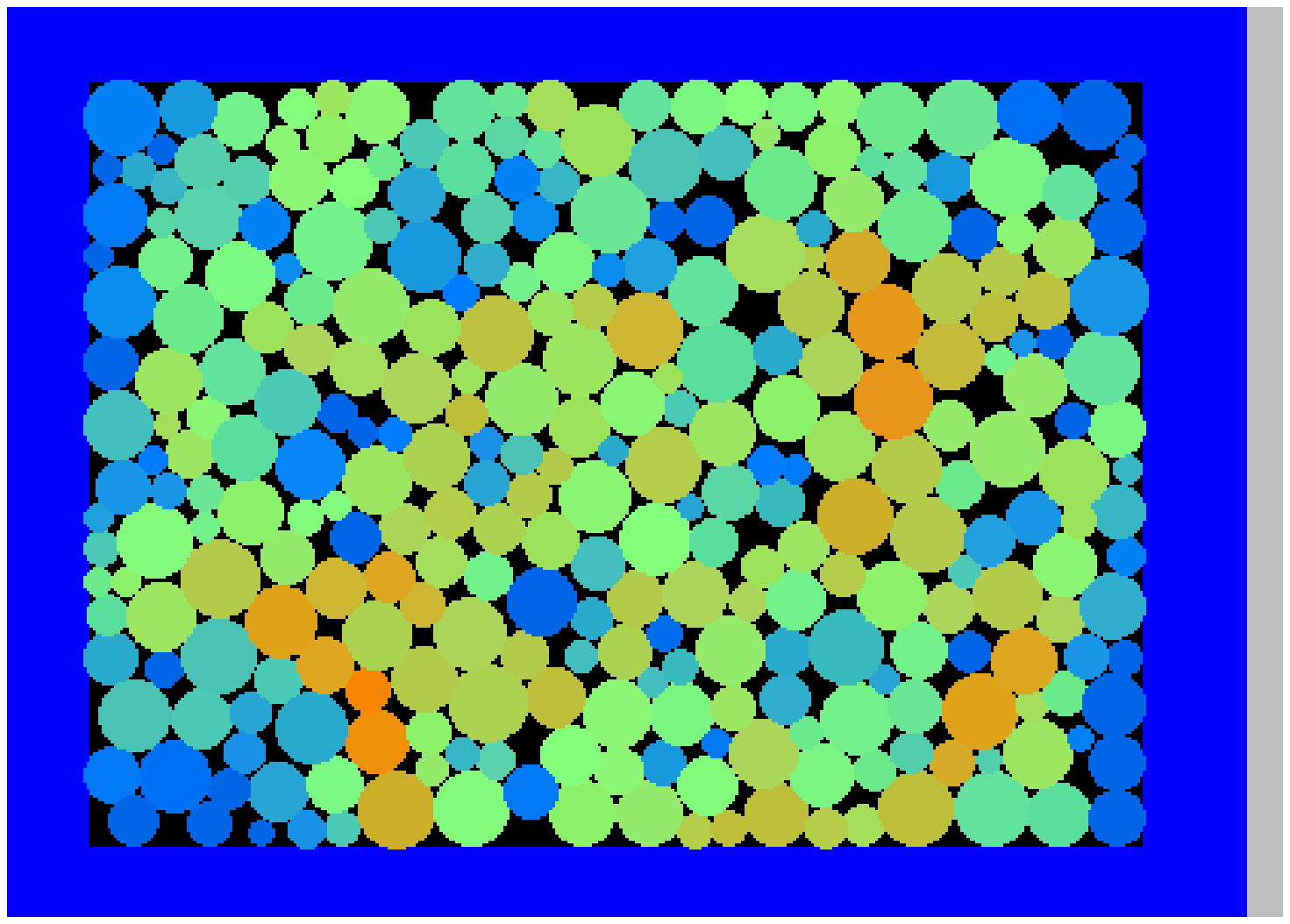,width=4.2cm,angle=0} \hfill
   \epsfig{file=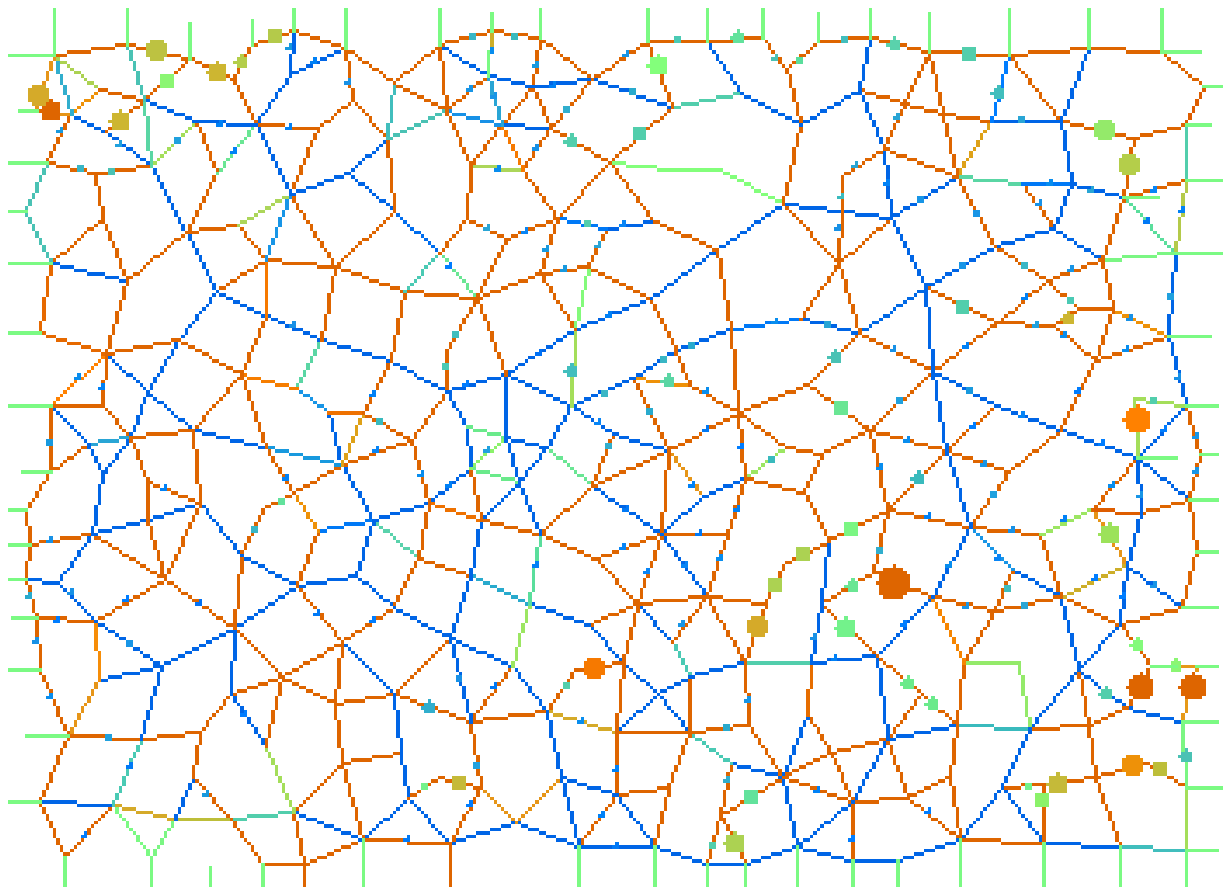,width=4.2cm,angle=0}
   \epsfig{file=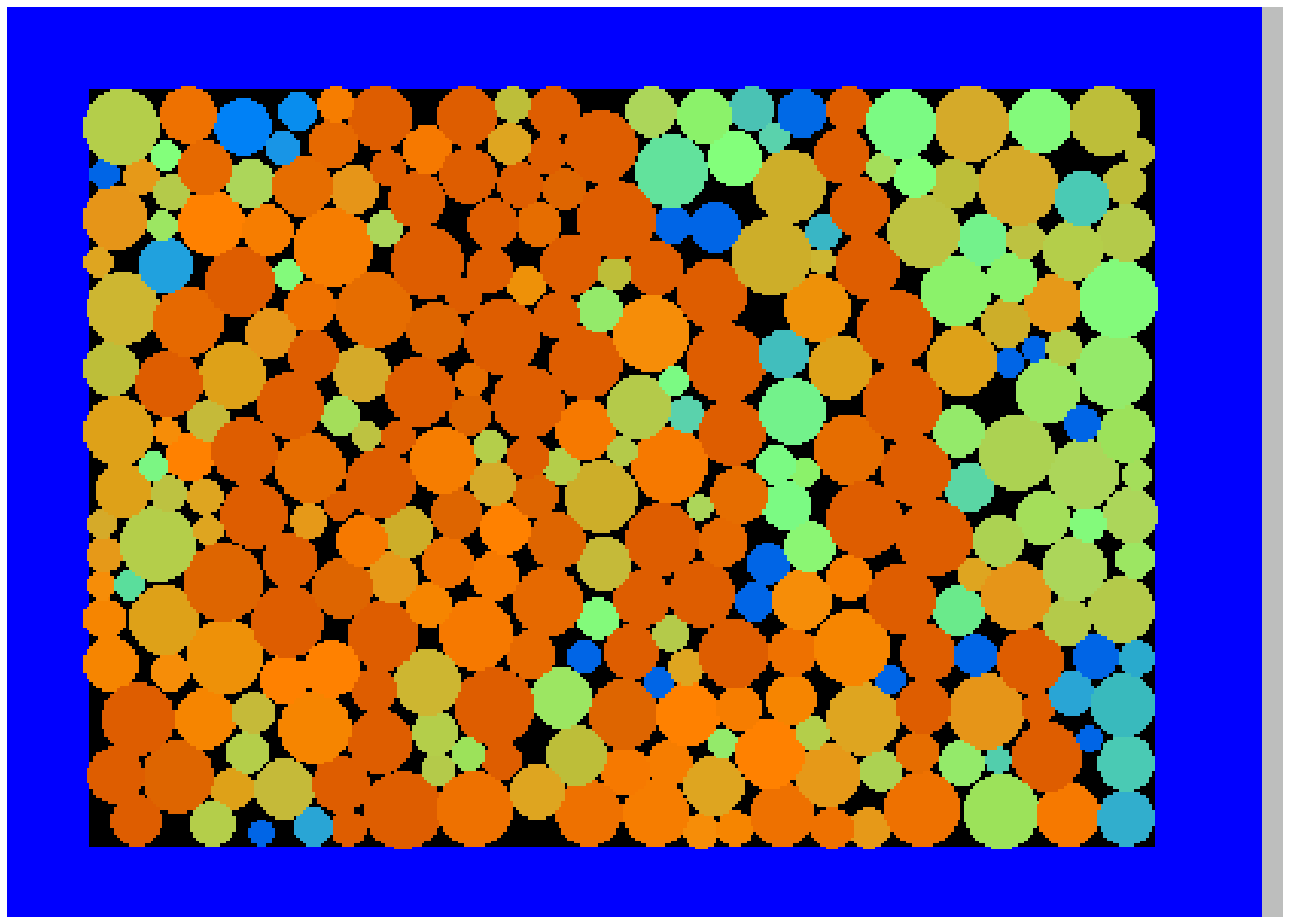,width=4.2cm,angle=0} \hfill
   \epsfig{file=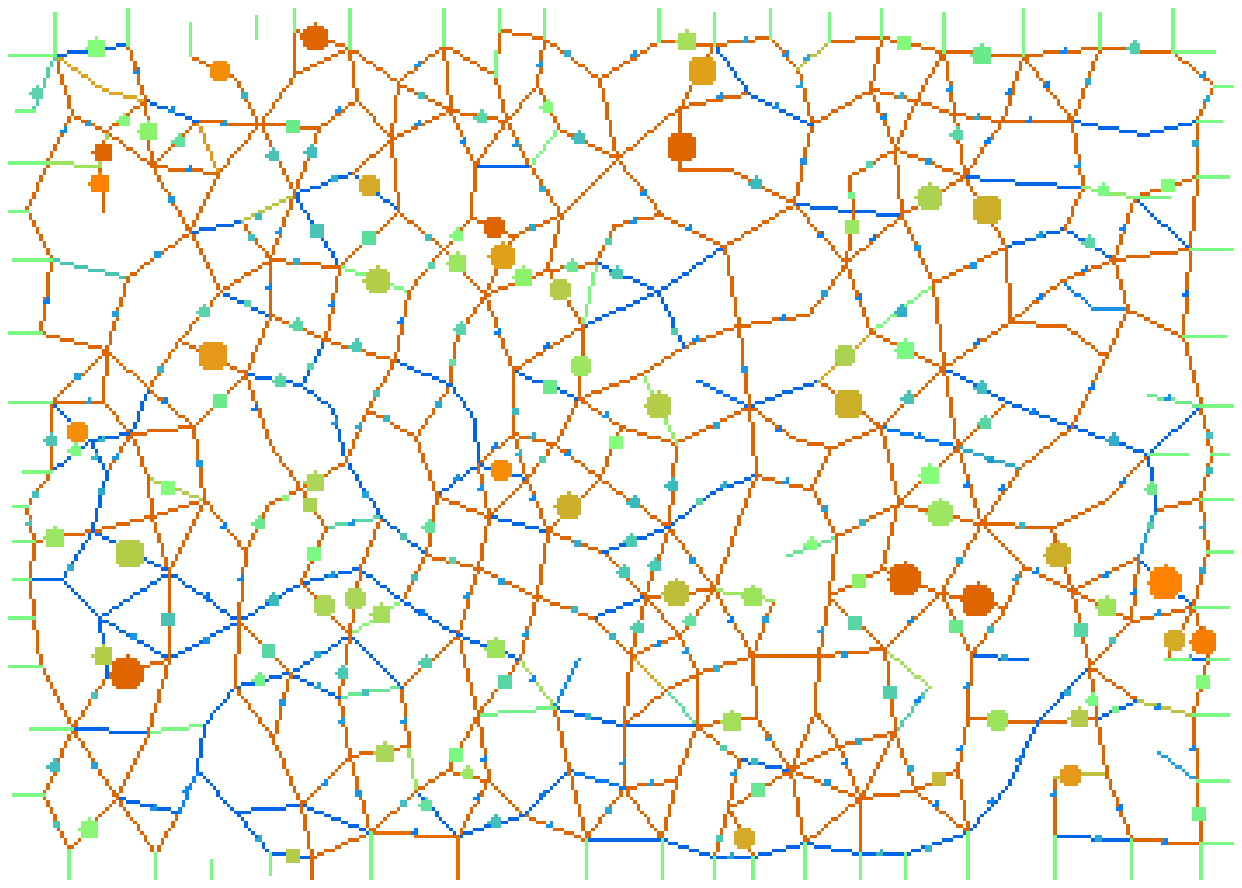,width=4.2cm,angle=0}
   \epsfig{file=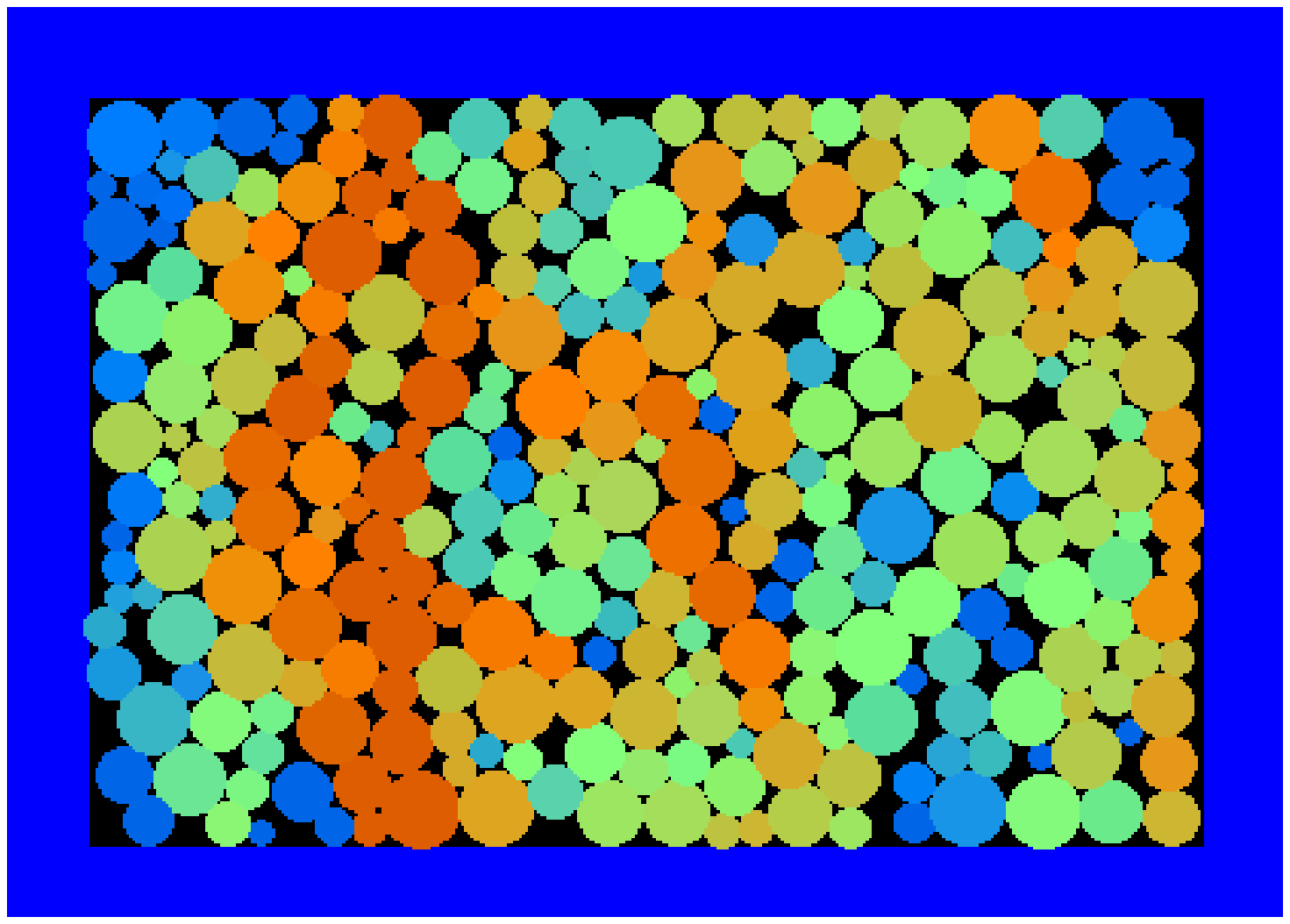,width=4.2cm,angle=0} \hfill
   \epsfig{file=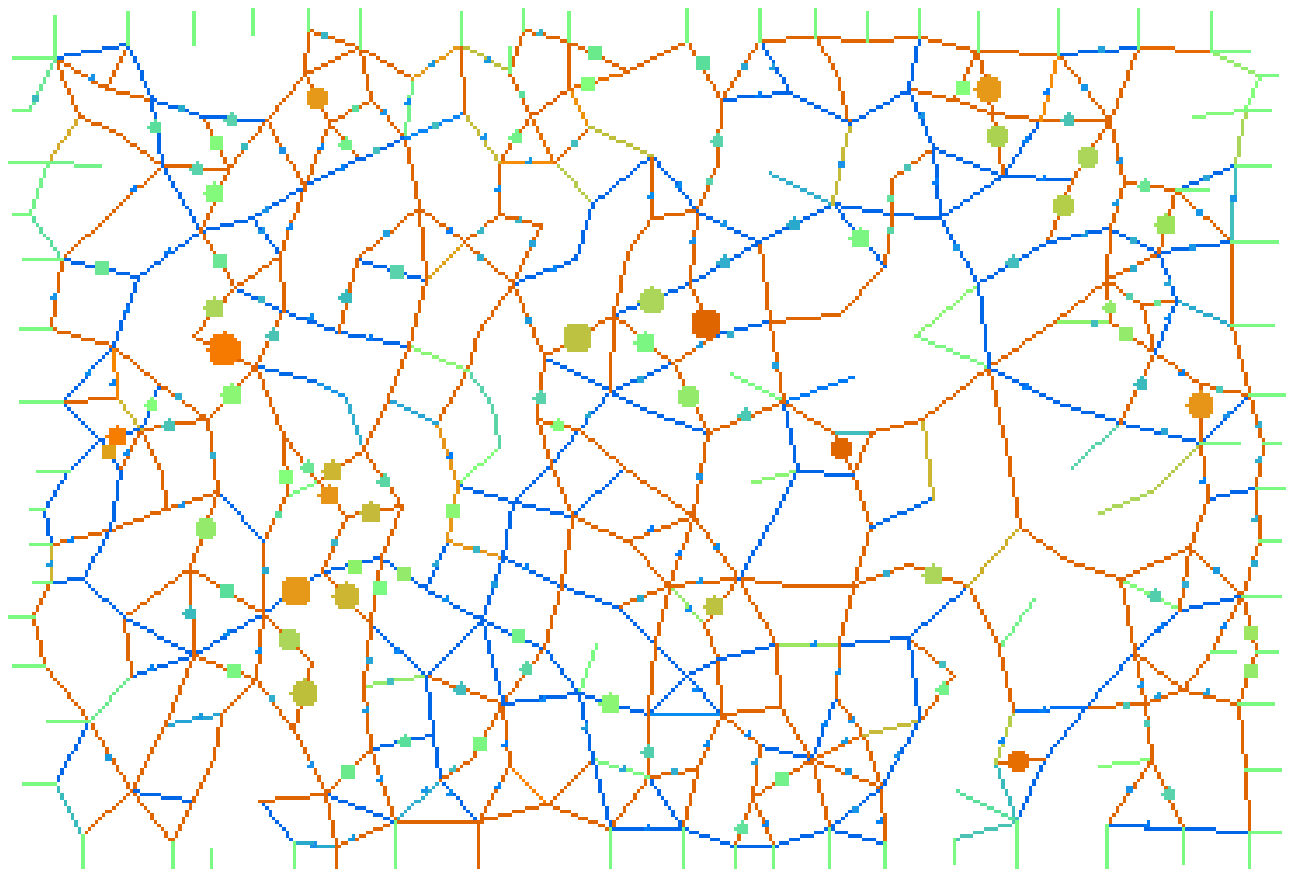,width=4.2cm,angle=0}
   \epsfig{file=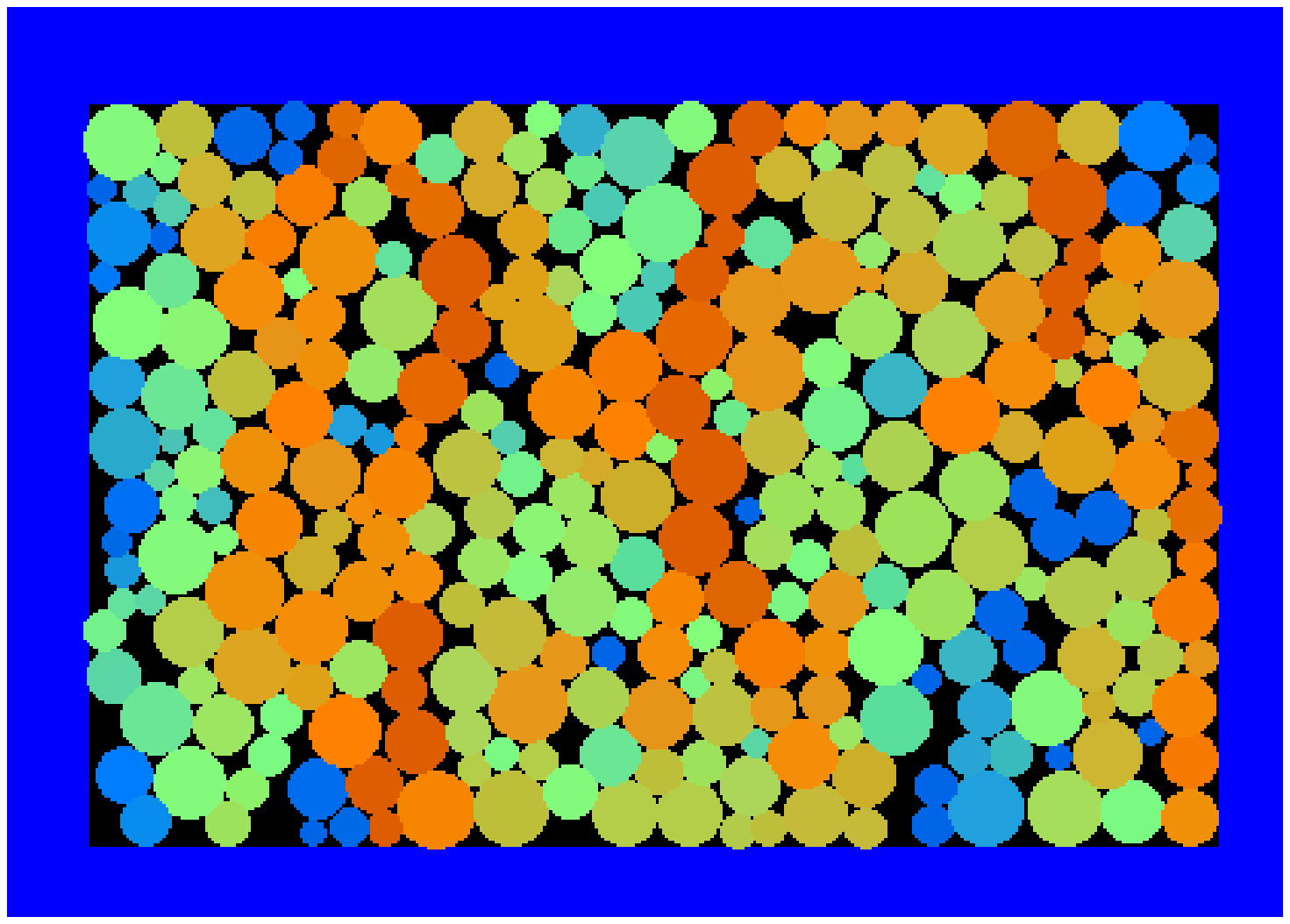,width=4.2cm,angle=0} \hfill
   \epsfig{file=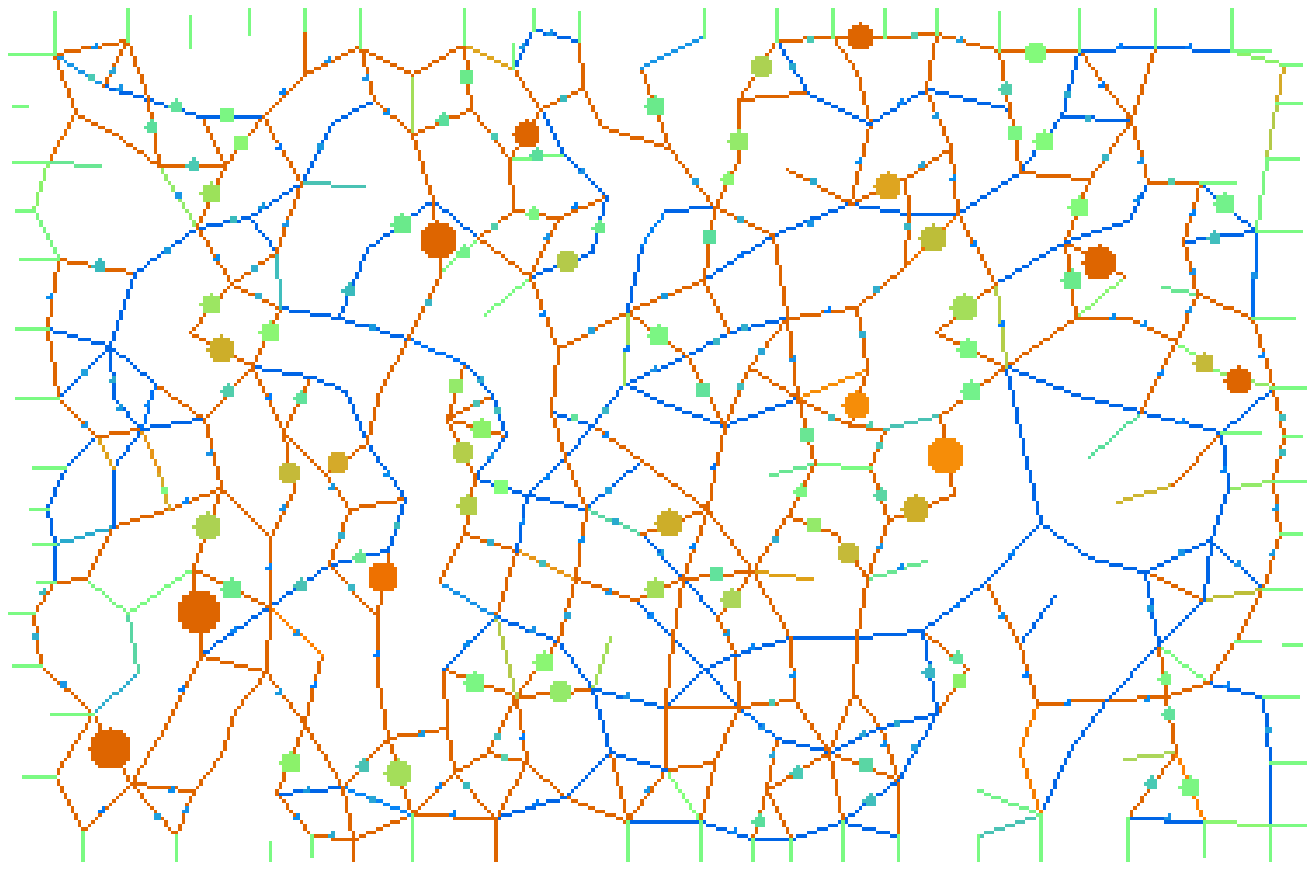,width=4.2cm,angle=0}
   \epsfig{file=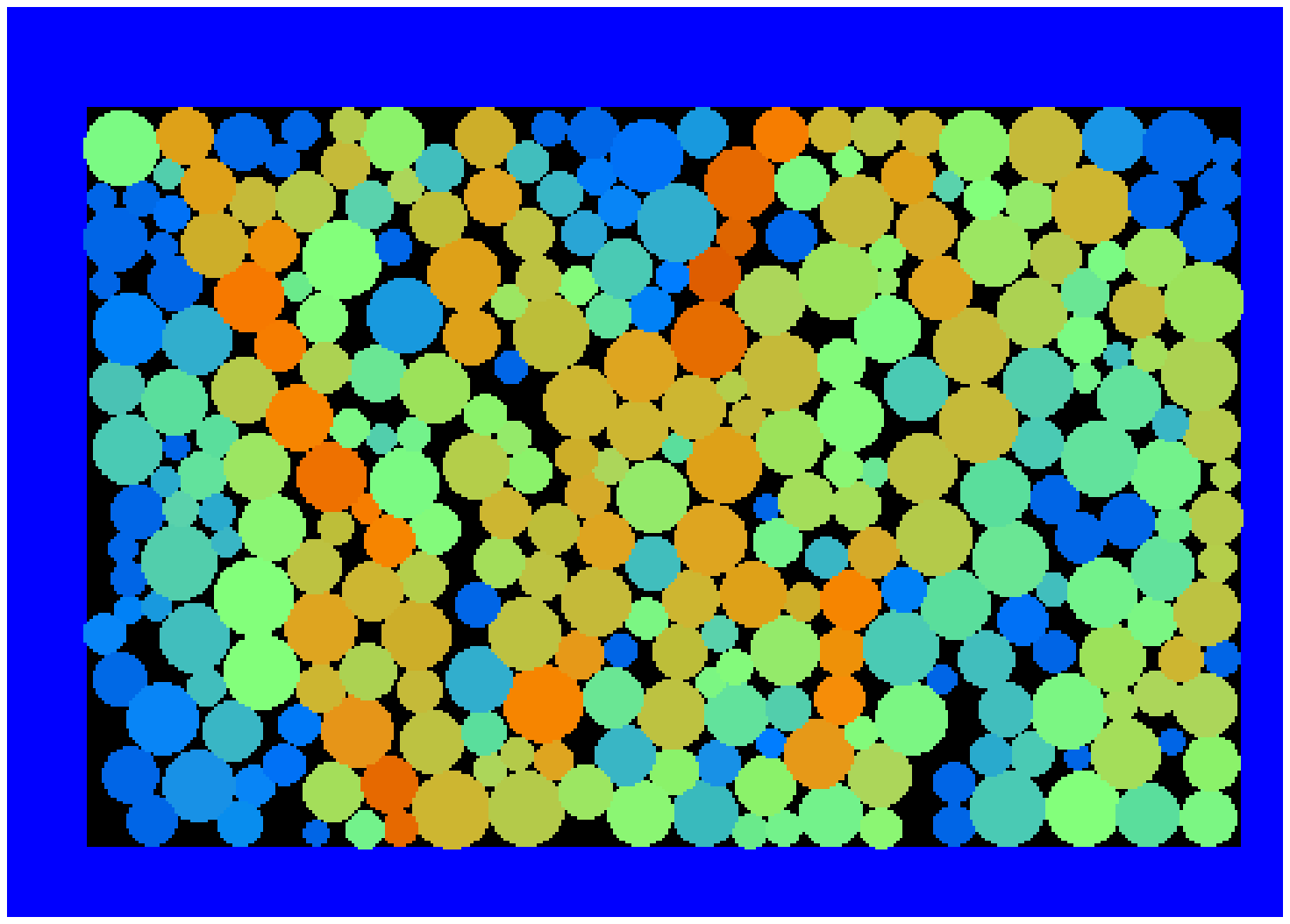,width=4.2cm,angle=0} \hfill
   \epsfig{file=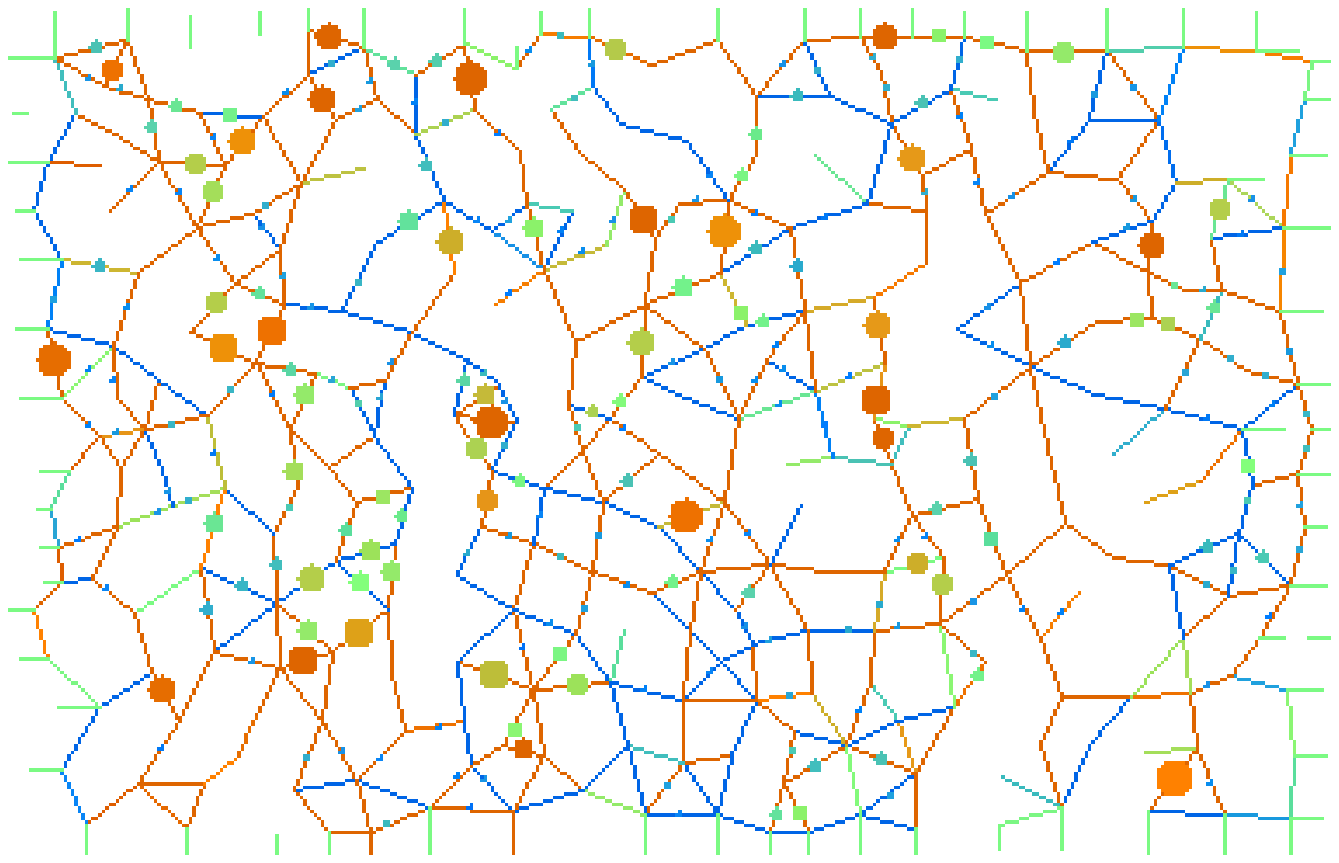,width=4.2cm,angle=0}
  \end{center}
  \caption{Snapshots of the compression test with $N=300$ and $p=10$.
           at times $t=0.005$, $0.02$, $0.03$, $0.035$, and $0.04$ 
           (from top to bottom); the material sample was sintered for 
           $t_s=1.0$.  The blue bars are the walls, the circles are the 
           particles with their colors coding the average stress, blue, 
           green and red correspond to low, medium and large stresses. 
           The lines are the contacts with their colors coding attractive 
           (blue) or repulsive (red) normal forces. The small solid circles 
           denote the tangential forces, with their size proportional to the
           magnitude of the tangential force.}
  \label{fig:300comp10}
\end{figure}

One compression test for a long sintering time, $N=300$ and $p=10$ is
presented in Fig.\ \ref{fig:300comp10}. During compression (from top to
bottom), the lines of strong attraction (so to say the backbone of the sample)
are desctroyed and. at the same time, more and more frictional contacts
occur due to local shear. Moreover, gaps between the parts of the sample
open and it fragments into pieces.

This is only a representative example for the compression test; a more
detailed study of the fracture behavior, sample-size- and 
sintering-time-dependence is far from the scope of this study.

\subsection{Vibration test}

\begin{figure}[htb]
  \begin{center}
   \epsfig{file=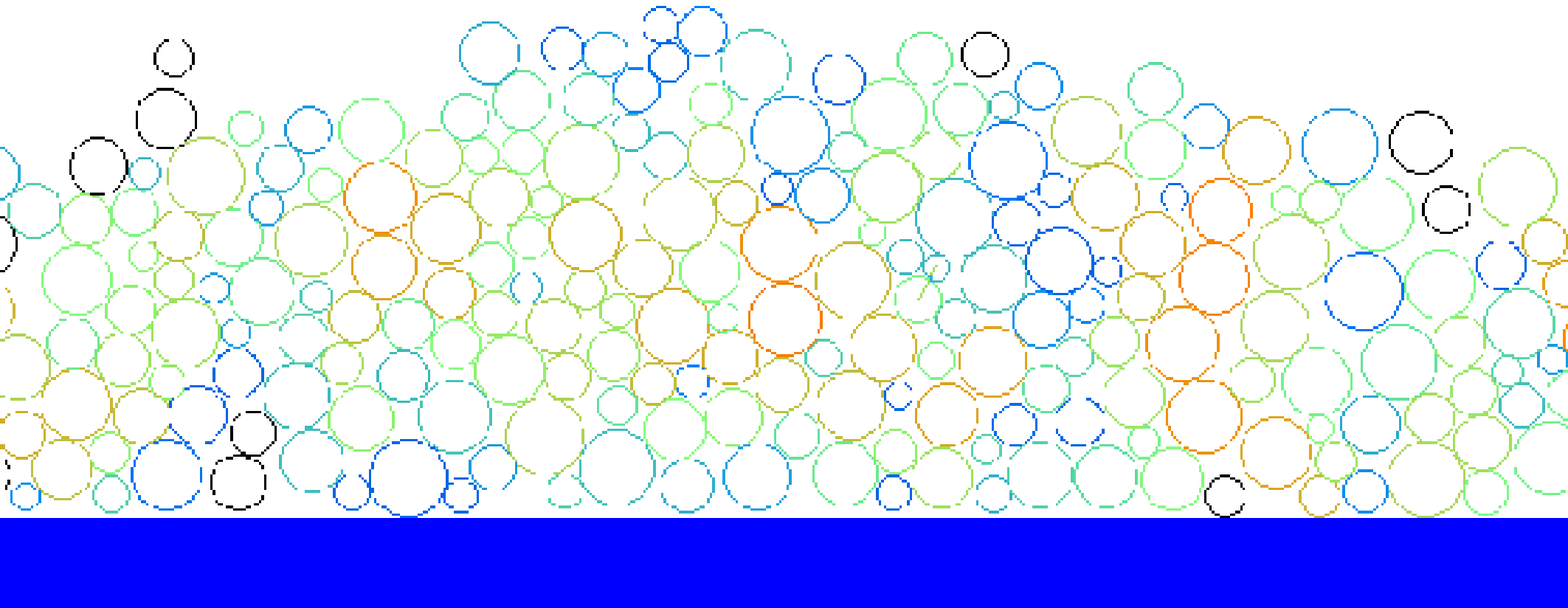,width=7.0cm,angle=0} 
   \epsfig{file=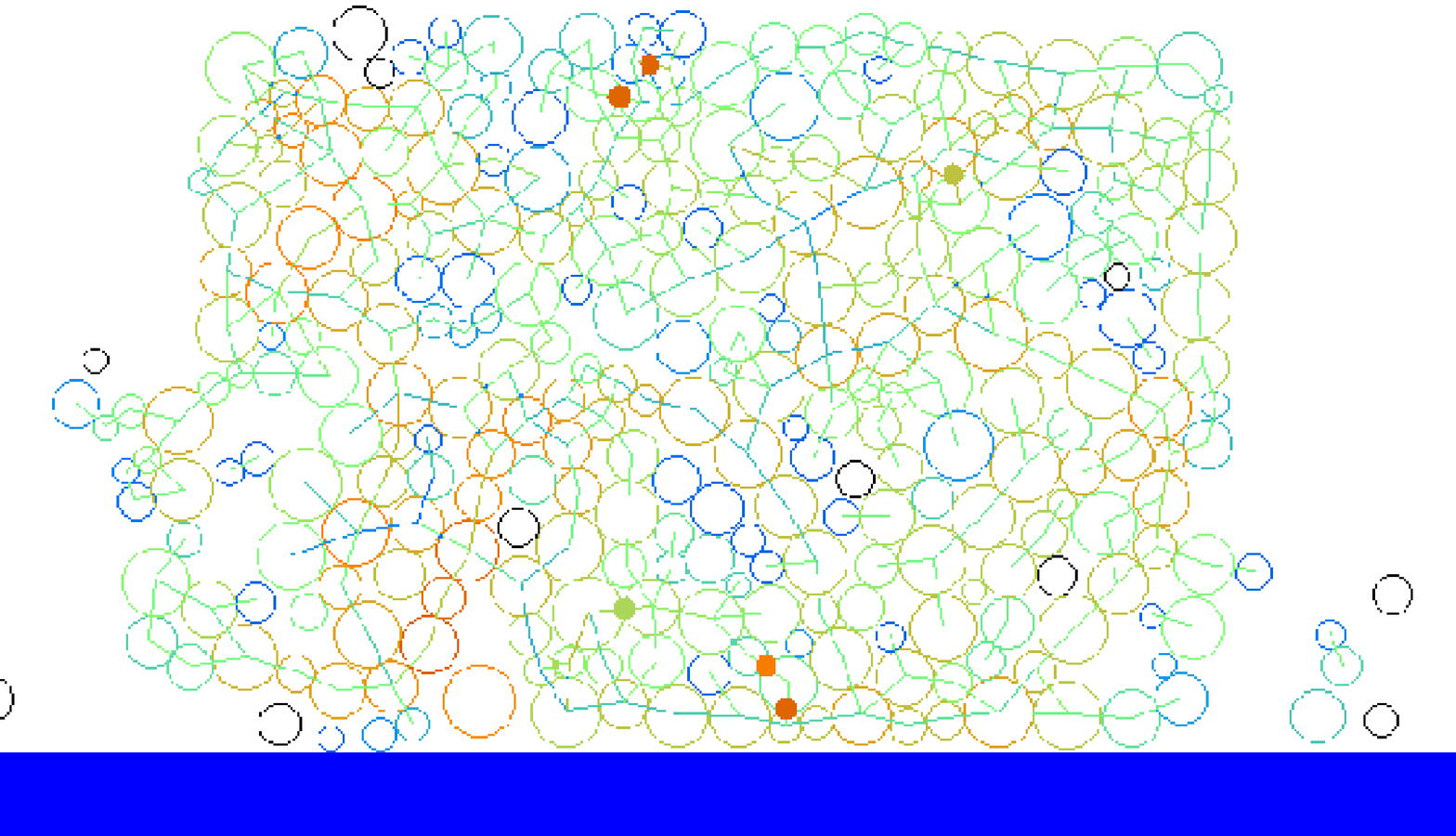,width=7.0cm,angle=0} 
   \epsfig{file=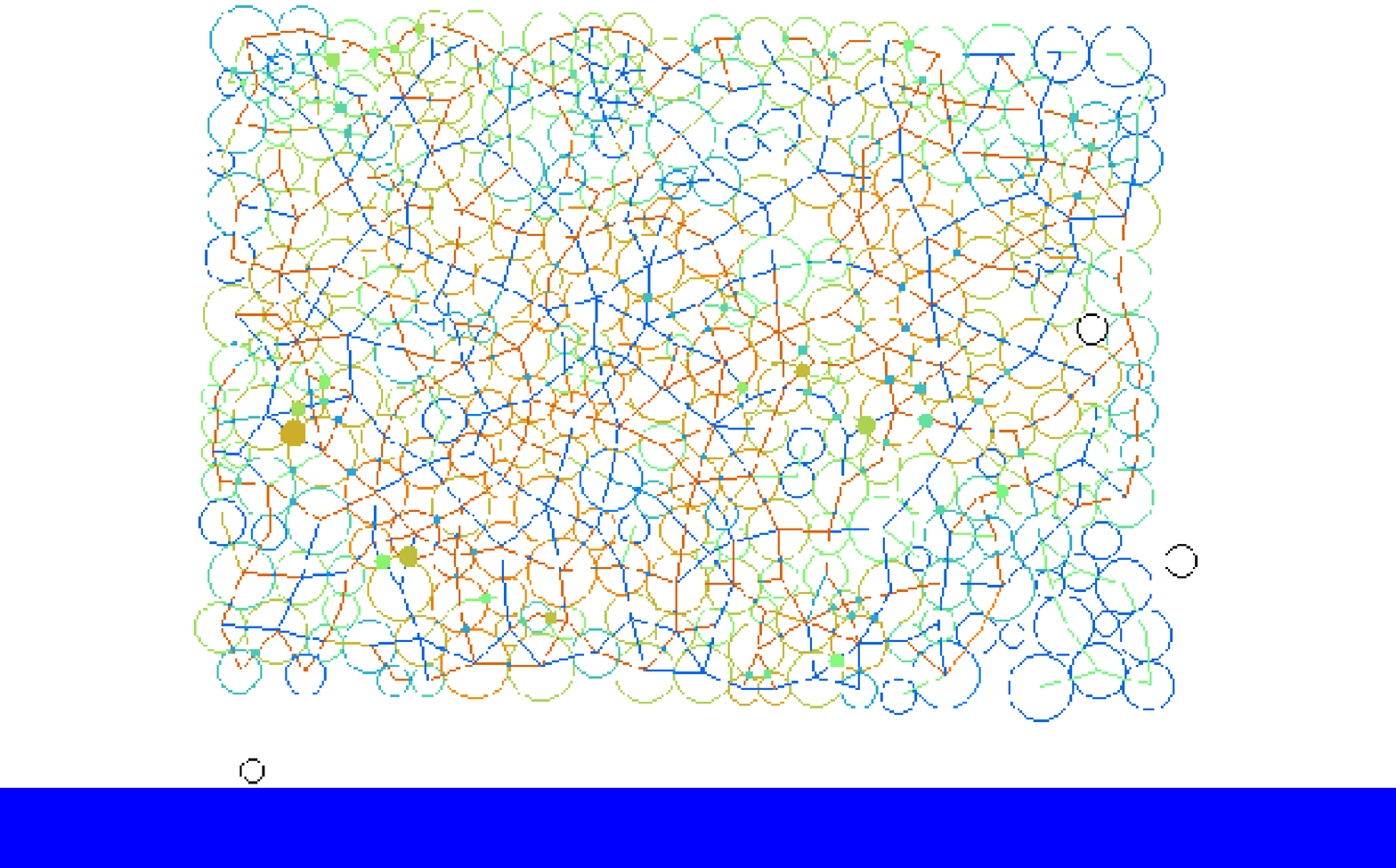,width=7.0cm,angle=0} 
  \end{center}
  \caption{Snapshots after a vibration test of $t=0.1$, with frequency 
           $f=100$\,Hz and amplitude $a=0.2$\,mm.  The material sample
           was sintered (from top to bottom) for $t_s=10^{-4}$, 
           $0.02$, and $1.0$ with $p=10$. The open circles are particles 
           with their colors coding the average stress, blue, green and red 
           correspond to low, medium and large stresses. The lines are
           the contacts with their colors coding attractive (blue) or
           repulsive (red) normal forces. The small solid circles denote
           the tangential forces, with their size proportional to the 
           magnitude of the tangential force.}
  \label{fig:vibrate10}
\end{figure}

\begin{figure}[htb]
  \begin{center}
   \epsfig{file=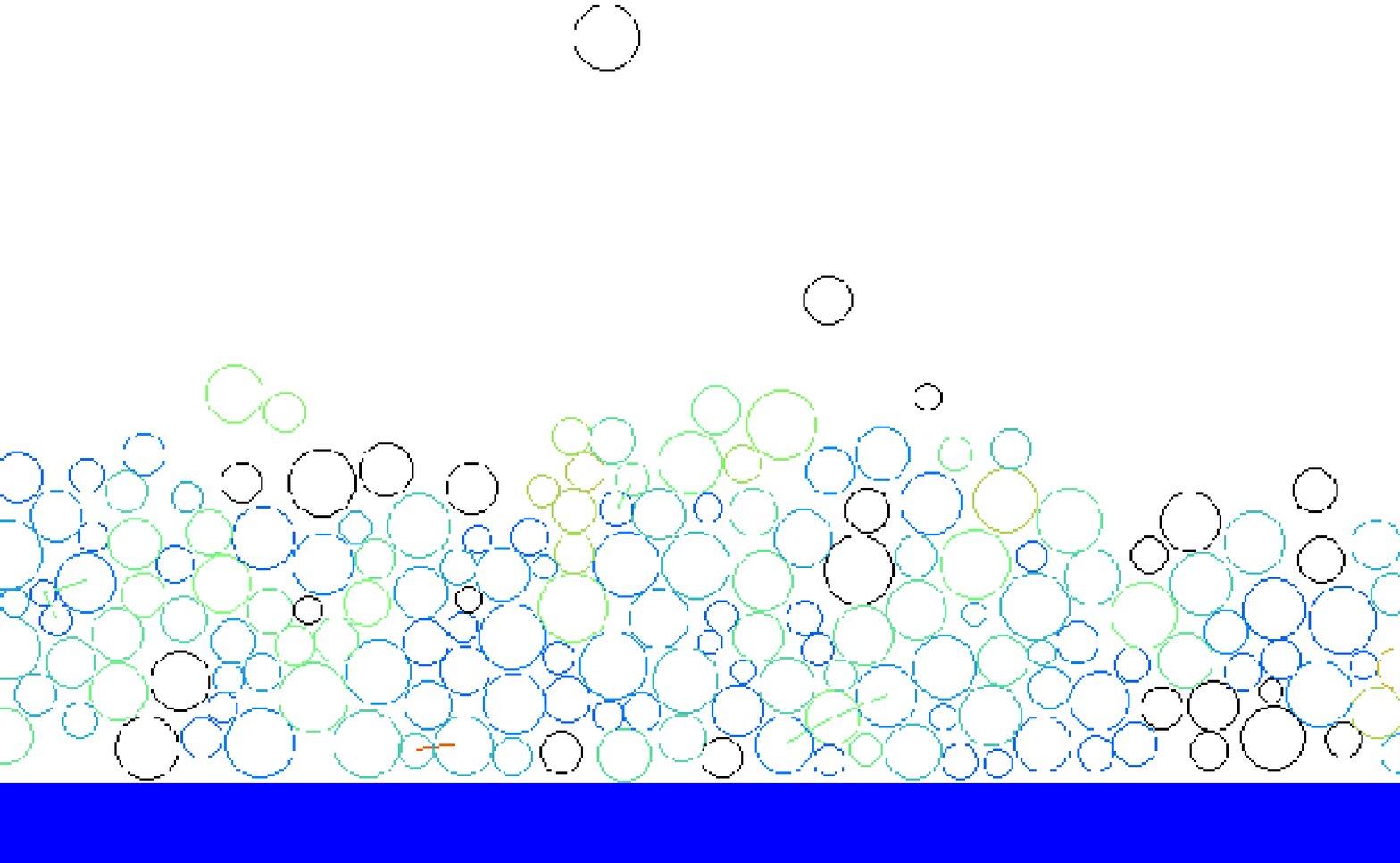,width=7.0cm,angle=0} 
   \epsfig{file=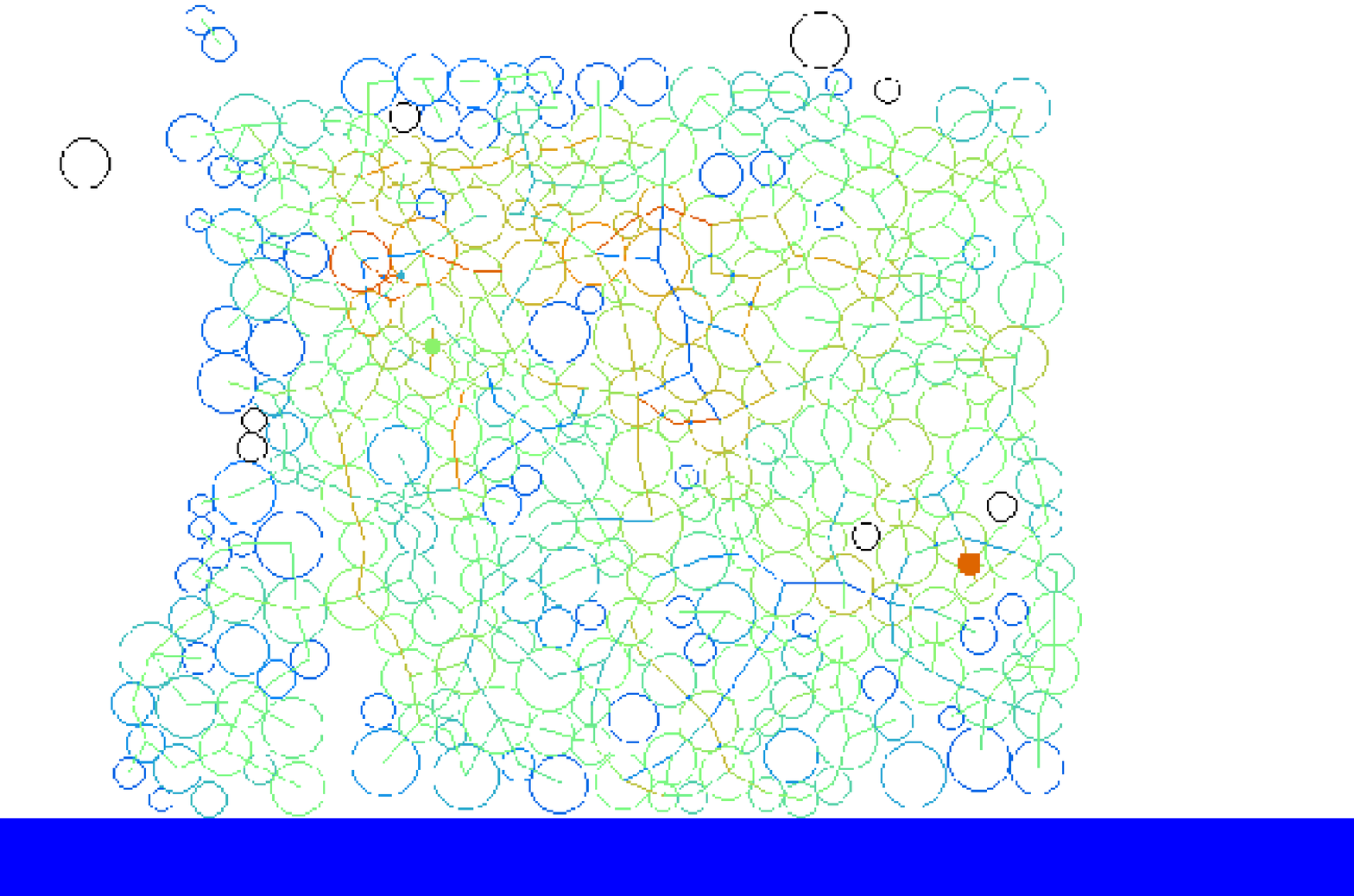,width=7.0cm,angle=0} 
   \epsfig{file=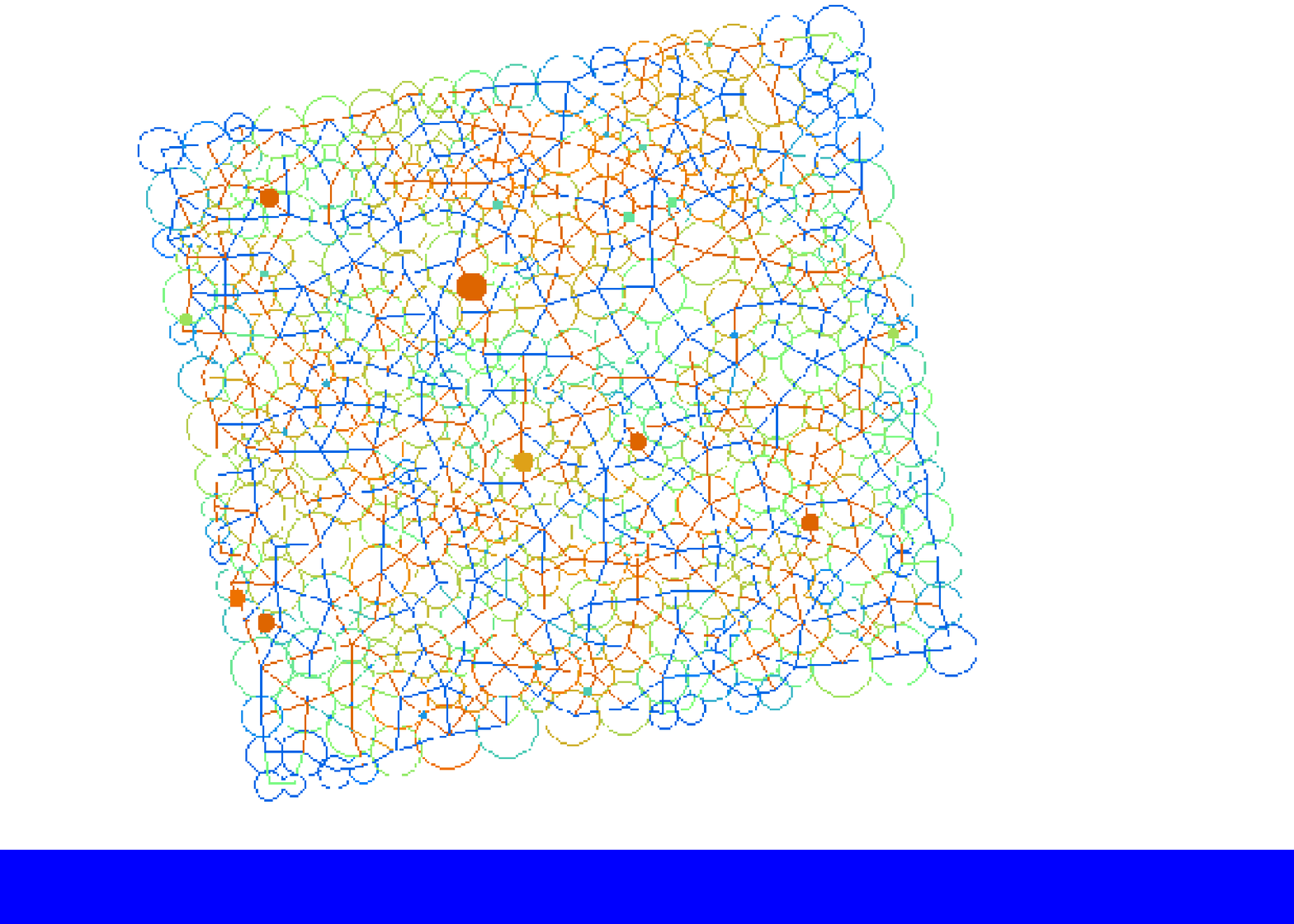,width=7.0cm,angle=0} 
  \end{center}
  \caption{Snapshots after a vibration test of $t=0.1$, with frequency 
           $f=100$\,Hz and amplitude $a=0.2$\,mm.  The material sample
           was sintered (from top to bottom) for $t_s=10^{-4}$, 
           $0.01$, and $1.0$ with $p=100$. The color coding and 
           figure meaning is the same as in Fig.\ \protect\ref{fig:vibrate10}.}
  \label{fig:vibrate100}
\end{figure}

The sintered samples can also be vibrated (in the gravitational
field) in order to probe their stability.  The results of the previous
tests are paralleled by a vibration test, see Figs.\
\ref{fig:vibrate10} and \ref{fig:vibrate100}.

Short sintering times leads to unstable material samples, whereas the
sample becomes more and more stable with increasing sintering time and
confining pressure.  The sample with the longest sintering time
$t_s=1.0$ is almost perfectly stable even under strong shaking --
some corner- or boundary particles sometimes break off.
For shorter sintering time the sample is less stable and
fragments into pieces. For very short sintering time, the sample
consists of single particles only.

\section{Conclusion}

In summary, a discrete model for the sintering of particulate materials
was introduced and simple material samples were sintered for different
times and confining pressures.  Then they were tested with respect to 
their anisotropic load strength:  Longer sintering and
stronger confining pressure systematically increases the density and
the strength of the material.  Depending on the sintering duration, either
isolated particles, fragments or a single solid block of material could be
produced.  

Besides this macroscopic point of view, also the microscopic picture
was examined.  The series of astonishing observations includes:
(i) The coordination number may slightly decrease due to reorganizations
while the density monotonously increases. (ii) Sintering leads to a broadening
overlap distribution, but to a narrowing force distributon, and thus
to a homogenization of the sample. 
(iii) The normal forces become strongly attractive during the cooling 
down of the sample below the melting temperature.
Besides these facts, a lot of open questions concerning the 
sintering process remain, especially concerning the connection
between the microscopic contact-model parameters and the macroscopic
material parameters.

The research to be done is an accurate testing of the model via a 
comparison with experimental data.  Since these are only available in
three-dimensional systems, the 2D model presented here may be not
helpful. However, the model can easily be extended to three dimensions,
where `only' more particles are needed. Note that the model presented
here increases the amount of computation necessary for each contact
by a large factor, so that the number of particles possible to simulate
becomes rather small for a standard computer.  Thus an extension to
three dimensions requires a proper tuning of the implementation of the
force-model and possibly the view of particles as blocks of material
that cannot fragment, rather than isolated particles.  The last missing
ingredient in the model is a rolling resistance which accounts for a
torque resistance of the contacts.

\section{Acknowledgements}

The authors thank M. L\"atzel, J. Tomas, S. Diebels, H. Besserer,
and G. A. D\'{}Addetta for discussions and acknowledge financial
support by the Deutsche Forschungsgemeinschaft (DFG) and by the
DaimlerChrysler research division.


\end{document}